%% file: FermionPairScalar.tex
\newcommand{\version}{December 19, 2014}
\documentclass[letterpaper,twoside,11pt,pdftex]{article}
\pdfoutput=1
\usepackage[bindingoffset=0.2cm,textheight=22.4cm,hdivide={2.7cm,*,2.7cm}, vdivide={*,22cm,*}]{geometry}
\usepackage{amsmath,amsfonts}

\IfFileExists{dsfont.sty}
	{\usepackage{dsfont}
         \let\mathbb=\mathds
         \newcommand{\id}{\mathds{1}}}
	{\typeout{Package dsfont.sty was not found, using alternative macros.}
         \let\mathds=\mathbb
         \newcommand{\id}{\mbox{1 \kern-.59em {\rm l}}}}

\usepackage[pdftex]{graphicx}
\usepackage{caption,subcaption} 
\usepackage[pdftex,bookmarksnumbered=true,breaklinks=true]{hyperref}

\usepackage[numbers,square,comma,sort&compress]{natbib}
\usepackage{slashed}

\newcommand{\cD}{\mathcal{D}}

\newcommand{\bPhi}{\boldsymbol{\Phi}}
\newcommand{\bL}{\mathbf{L}}
\newcommand{\ba}{\mathbf{a}}
\newcommand{\bR}{\mathbf{R}}
\newcommand{\bQ}{\mathbf{Q}}
\newcommand{\eqnref}[1]{Eq.~(\ref{#1})}		
\newcommand{\figref}[1]{Fig.~\ref{#1}}			
\newcommand{\secref}[1]{Sec.~\ref{#1}}		
\newcommand{\appref}[1]{Appendix~\ref{#1}}		
\newcommand{\co}[2]{\left[#1,#2\right]}					
\newcommand{\aco}[2]{\left\{#1,#2\right\}}				
\newcommand{\var}[2]{\frac{\d #1}{\d #2}}				
\newcommand{\pa}{\partial}						
\newcommand{\diff}[2]{\frac{\pa #1}{\pa #2}}				
\newcommand{\sgn}[1]{\text{sgn}(#1)}

\renewcommand{\a}{\alpha}
\renewcommand{\b}{\beta}
\newcommand{\g}{\gamma}
\renewcommand{\d}{\delta}
\newcommand{\e}{\epsilon}

\newcommand{\z}{\zeta}

\renewcommand{\th}{\theta}

\renewcommand{\l}{\lambda}
\newcommand{\m}{\mu}
\newcommand{\n}{\nu}

\renewcommand{\r}{\rho}
\newcommand{\s}{\sigma}
\renewcommand{\t}{\tau}

\newcommand{\G}{\Gamma}

\newcommand{\inv}[1]{\frac{1}{#1}}				

\newcommand{\Z}{\mathds{Z}}

\newcommand{\bpsi}{\overline{\psi}}

\def\nbox#1#2{\vcenter{\hrule \hbox{\vrule height#2in
\kern#1in \vrule} \hrule}}
\def\sq{\,\raise.5pt\hbox{$\nbox{.09}{.09}$}\,}
\def\sqb{\,\raise.5pt\hbox{$\overline{\nbox{.09}{.09}}$}\,}

\newcommand{\bea}{\begin{eqnarray}}
\newcommand{\eea}{\end{eqnarray}}
\newcommand{\be}{\begin{equation}}
\newcommand{\ee}{\end{equation}}
\newcommand{\bes}{\begin{subequations}}
\newcommand{\ees}{\end{subequations}}
\def\lag{\langle}
\def\rag{\rangle}

\newcommand{\nn}{\nonumber\\}

\makeatletter

 \@addtoreset{equation}{section}
 \makeatother

\newcommand{\Appendix}[1]{
  \refstepcounter{section}
  \section*{Appendix \thesection:\hspace*{1.5ex} #1}
  \addcontentsline{toc}{section}{Appendix \thesection}
}

\title{\vspace{-2.5cm}\begin{flushright}
        {\small LA-UR-13-29168\\[-2ex]LA-UR-14-25252}
       \end{flushright}\vspace{3em}
Fermion Pairing and the Scalar Boson of \\the 2D Conformal Anomaly}
\author{Daniel N. Blaschke\footnote{\ttfamily{dblaschke@lanl.gov}}~, Ra\'ul Carballo-Rubio\footnote{\ttfamily{raulc@iaa.es}}~, and Emil Mottola\footnote{\ttfamily{emil@lanl.gov}}}

\date{\version}

\graphicspath{{./}{./Figures/}}
\begin{document}
\maketitle

\thispagestyle{empty}

\begin{center}
\renewcommand{\thefootnote}{\fnsymbol{footnote}}
\vspace{-0.3cm}
\footnotemark[1]\footnotemark[3]\textit{Theoretical Division, Los Alamos National Laboratory\\
Los Alamos, NM, 87545, USA}\\[0.3cm]
\footnotemark[2]\textit{Instituto de Astrof\'{\i}sica de Andaluc\'{\i}a (IAA-CSIC)\\
Glorieta de la Astronom\'{\i}a, 18008 Granada, Spain}\\[0.9cm]
\end{center}

\begin{abstract}

We analyze the phenomenon of fermion pairing into an effective boson associated with anomalies and 
the anomalous commutators of currents, bilinear in the fermion fields. In two spacetime dimensions the chiral 
bosonization of the Schwinger model is determined by the chiral current anomaly of massless Dirac fermions. 
A similar bosonized description applies to the 2D conformal trace anomaly of the fermion stress-energy tensor.
For both the chiral and conformal anomalies, correlation functions involving anomalous currents, $j^{\m}_5$ or $T^{\m\n}$ of 
massless fermions exhibit a massless boson $1/k^2$ pole, and the associated spectral functions obey a UV finite 
sum rule, becoming $\d$-functions in the massless limit. In both cases the corresponding effective action of the 
anomaly is non-local, but may be expressed in a local form by the introduction of a new bosonic field, which 
becomes a \textit{bona fide} propagating quantum field in its own right. In both cases this is expressed in Fock space
by the anomalous Schwinger commutators of currents becoming the canonical commutation relations of the 
corresponding boson. The boson has a Fock space operator realization as a coherent superposition of massless 
fermion pairs, which saturates the intermediate state sums in quantum correlation functions of fermion currents. 
The Casimir energy of fermions on a finite spatial interval $[0,L]$ can also be described as a coherent scalar 
condensation of pairs, and the one-loop correlation function of any number $n$ of fermion stress-energy tensors 
$\lag TT\dots T\rag$ may be expressed as a combinatoric sum of $n!/2$ linear tree diagrams of the scalar boson.

\end{abstract}

\newpage
\tableofcontents

\section{Introduction}

In many-body physics it is well-known that gapless fermion excitations in the vicinity
of a Fermi surface can pair up into effective bosonic degrees of freedom. The formation of such fermion
Cooper pairs is the basis for the BCS theory of superconductivity and the superfluidity of $^3$He \cite{Leggett:2006}. 
This amounts to a reorganization of the ground state of the system from weakly interacting fermions to interacting
effective bosons, themselves consisting of bound fermion pairs.

In this paper we study the mechanism of fermion pairing in relativistic quantum field
theory, emphasizing that the pairing is a direct result of quantum anomalies in otherwise classically conserved
currents that are bilinear in the fermion fields. The particular focus of the paper is the 2D conformal anomaly 
of the stress-energy tensor \cite{Davies:1976ei,Brown:1976wc,Birrell:1982ix} and the bosonized
description it leads to. By studying this case in detail, our aim is to lay the groundwork for the
extension of our considerations of anomaly induced pairing and corresponding bosons in four (and higher) 
dimensions with the appropriate modifications.

The best known example of fermion pairing in a relativistic quantum field theory is provided by the
Schwinger model, \textit{i.e.}\ quantum electrodynamics of massless fermions in two spacetime
dimensions \cite{Schwinger:1962tn,Schwinger:1962tp}. The study of this model has a long history, 
and over the years has been solved by a number of different techniques \cite{Brown:1963,Lowenstein:1971fc,Casher:1974vf,
Halpern:1975jc,Manton:1985jm,Wolf:1985,Hetrick:1988yg,Link:1990bc,Sachs:1991en,Smilga:1992hx,Durr:1996im,Azakov:2005qn,Hosotani:1998za}.
We begin in Sec. \ref{sec:chiral-review} by reviewing the Schwinger model, and emphasizing that its main feature 
of fermion pairing into an effective massive boson may be understood by both functional integral and operator methods 
most simply and directly as a consequence of its chiral current anomaly.

The basic signature of an anomaly is the existence of a bosonic excitation in correlators
involving anomalous currents, which becomes an isolated $1/k^2$ pole
when the underlying fermions are massless \cite{Dolgov:1971ri,Horejsi:1992tx,Giannotti:2008cv,Armillis:2009im}. 
In $d=2$ QED, the one-loop current polarization tensor $\lag j^{\m}(x) j^{\n}(x')\rag$ has an imaginary part 
and spectral representation which obeys an ultraviolet finite sum rule and becomes a $\d$-function 
in the limit of massless fermions, indicating the presence of a massless boson intermediate state 
composed of fermion pairs. Correspondingly, the real part of the same correlation function exhibits 
a $1/k^2$ pole, which is the propagator of a dynamical massless boson. The residue of 
this massless pole in the correlation function of non-anomalous currents is proportional to the 
coefficient of the chiral current anomaly.

In the functional integral approach, the non-local effective action of the anomaly may be expressed
in a local form describing a massless boson, which becomes a \textit{bona fide} propagating
field, inheriting its dynamics from that of the underlying fermions, even in the absence 
of other interactions. In the Schwinger model the effect of the interaction with the gauge potential
leads to the boson acquiring a mass $M^2=e^2/\pi$, but the fermion pairing into a boson
state occurs even in the limit $e\to 0$ of infinitesimally small coupling to the gauge field.

In the Fock space operator approach to the Schwinger model, after careful definition of the
fermion vacuum and normal ordering, there is a non-zero anomalous equal time commutator 
(Schwinger term) of currents \cite{Schwinger:1959xd}. It is worth noting that although the Schwinger 
term occurs in the commutator $[j^0, j^1]$ of apparently non-anomalous vector current components, 
this non-zero commutator can be viewed as a direct result of the chiral anomaly, and the 
boson degree of freedom it describes by fermion pairing. Indeed since the fermion
current is linear in the chiral boson field, the Schwinger commutator term in the currents 
may be mapped precisely into the canonical equal time commutator 
of the boson, itself composed of fermion pairs, showing that the boson field is
a true propagating degree of freedom.

Having reconsidered and rederived the standard results of the Schwinger model 
from the vantage point of its chiral anomaly, we proceed to apply the same
methods to the conformal trace anomaly of the fermion stress-energy tensor.
We show that most of the same features associated with the chiral anomaly
reappear in the case of the conformal anomaly. In particular, correlation
functions involving the stress-energy tensor of massless fermions again exhibit an 
isolated massless $1/k^2$ pole, with finite residue determined by the anomaly.
There is again an ultraviolet sum rule for the corresponding spectral function
which becomes a $\d$-function in the massless limit \cite{Giannotti:2008cv}.
The $\sq^{-1}$ propagator in the non-local Polyakov action of the conformal anomaly
\cite{Polyakov:1981rd} can be rewritten in terms of a local scalar field $\varphi$ that 
becomes a \textit{bona fide} propagating scalar, inheriting its dynamics from the underlying
fermions of which it is composed. The anomalous equal time commutators
or Schwinger terms of the stress-energy tensor, \textit{i.e.}\ the central terms 
in the Virasoro algebra, may also be recognized as equivalent to canonical
equal time commutation relations of the scalar composite boson, which
therefore must be treated as a dynamical degree of freedom in its own right.

A new feature of the conformal trace anomaly of the stress-energy tensor
is that the local scalar field $\varphi$ has its own quantum stress-energy
tensor containing terms that are both linear and bilinear in $\varphi$, and hence 
its own quantum anomaly. This would shift the anomaly coefficient of $N$ fermions 
from $N$ to $N + 1$, if uncompensated. Complete equivalence with the original 
$N$ fermion theory can be achieved in one of two different ways. If $\varphi$ is 
treated as a full quantum field participating in internal loops, then a compensating shift 
of $N$ to $N-1$ in the $\varphi$ effective action must be introduced, so that the one 
less fermion degree of freedom is replaced by one boson degree of freedom. 
Alternately, if one is interested only in the correlation functions of the fermions,
one can treat the unshifted anomaly induced effective action of the $N$ fermions 
as a purely tree-level effective action for the scalar $\varphi$ in a gravitational field. 
By thereby forbidding the scalar $\varphi$ from participating in internal loops, 
there is no shift in $N$ to be compensated for, and it is possible to identify
the fermion stress-energy tensor at the operator level with only those terms in the bosonic 
stress-energy tensor linear in the quantum $\varphi$ field, analogous to the chiral
bosonization of the Schwinger model \cite{Klaiber:1967jz,Coleman:1974bu,Coleman:1975pw}. 

This second approach also makes it possible to prove a remarkable theorem relating 
the correlation functions of an arbitrary number of fermion stress-energy tensors 
$\lag T^{\m_1\n_1}(x_1)\dots T^{\m_n\n_n}(x_n)\rag$ 
at one-loop level to pure linear tree diagrams of the boson $\varphi$. The two-fermion 
intermediate states in the quantum correlation functions of the stress-energy tensor of the
fermions are therefore identical to the single boson states of the corresponding bosonic 
tree diagrams for any number of stress-energy tensor insertions. This amounts to an alternative 
bosonization scheme for coupling to gravity, different from the usual chiral 
bosonization coupling to electromagnetism in the Schwinger model.

Once fermions are paired into an effective boson field, the bosons can condense
and non-vanishing condensates are possible. In the simplest case of free fermions
with anti-periodic boundary conditions on the finite spatial interval $[0,L]$, 
the quantum Casimir energy of the fermions can be described as a 
finite condensate $\bar \varphi = \lag \varphi\rag$ of the boson field. The value of this
boson condensate can be obtained from simple geometric considerations
of a conformal transformation from flat ${\mathbb R}^2$ (assumed to have
vanishing vacuum energy) to ${\mathbb R} \times {\mathbb S}^1$, appropriate
for the periodically identified finite interval. This shows an interesting
connection of the anomalous action of the boson and its condensate to
the topology of spacetime. In the axial case there is a corresponding 
relationship to the topological winding number of the gauge field
and its vacuum structure \cite{Manton:1985jm,Witten:1978bc,Veneziano:1979ec,Azakov:2000eq}.

This paper is organized as follows. After reviewing the path integral, dispersive and
Fock space approaches to the Schwinger model in \secref{sec:chiral-review}, we proceed
in \secref{sec:gravity} to couple the fermion theory to gravity. The effective action of the
conformal anomaly is first found by the functional integral method in \secref{subsec:GravFunctInt}.
The two-point correlation function of stress-energy tensors, spectral function, UV finite sum rule
and $1/k^2$ pole is considered in \secref{sec:correlation-gravity}, the Fock space operator representation
and Schwinger terms in \secref{TFermion}, the boson condensate and Casimir energy in \secref{sec:casimir},
the canonical field representation in \secref{sec:canonical}, and saturation of the intermediate 
state sum in $\lag TT\rag$ by the boson in \secref{sec:IntermedTT}. 
In \secref{sec:correlation-functions} we show that the fermion pairing into the
boson $\varphi$ associated with the conformal anomaly implies a complete
equivalence between one-loop quantum correlation functions of arbitrary numbers of
fermion stress-energy tensors $\lag TT\dots T\rag$ to a set of bosonic linear tree diagrams, in 
which the intermediate boson states are precisely the fermion pairs. 
\secref{Sec:Conclusion} contains a summary of our conclusions and discussion of
how these results may extend to $d > 2$ dimensions. There are two Appendices. The
Fock space algebra of currents is computed first for the standard chiral bosonization of 
charge currents in Appendix  \ref{sec:commutators}, and then secondly for the Virasoro algebra 
of stress-energy tensor moments in Appendix \ref{sec:virasoro}.

\section{Fermion Pairing and Bosonization in the Schwinger Model}
\label{sec:chiral-review}
\subsection{Covariant Path Integral and Effective Action}

Perhaps the best known example of the phenomenon of fermion pairing associated with an anomaly in relativistic QFT
is the Schwinger model, \textit{i.e.}\ massless QED in $d=2$ dimensional flat spacetime  \cite{Schwinger:1962tn,Schwinger:1962tp}. 
We generalize this slightly and consider the action for $N$ identical Dirac fermion species (flavors), and rescale the coupling 
$e^2 \rightarrow e^2/N$, so that we consider the theory described by the classical action
\be
S_{\rm cl}= S_f[\psi, \bpsi; A]  + \frac{N}{e^2}\, S_g [A] =
i\!\int\!d^2x \sum_{j=1}^N \bpsi_j\,\g^\m (\! \stackrel{\leftrightarrow}{\ \pa_\m}-iA_\m)\psi_j 
- \frac{N}{4e^2}\! \int\!d^2x\, F_{\m\n}F^{\m\n}\,.
\label{Schwmod}
\ee
As usual $F_{\m\n} = \pa_\m A_\n - \pa_\n A_\m$ is the field strength tensor, whose only
non-vanishing component in $d=2$ is the electric field $F_{01} = -E = -F^{01}$. At certain points
below we shall extend the model to include a fermion mass term $S_{\rm cl} \to S_{\rm cl} + \int d^2x\, m\bpsi\psi$,
but our primary focus is on the massless case.  

The Dirac matrices obey the anti-commutation relations
\be
\gamma^\m \gamma^\n + \gamma^\n\gamma^\m = 2\,{\rm diag}\, (+,-) = -2\, \eta^{\m\n} 
\label{Dirmat}
\ee
in the flat metric $\eta^{\m\n}$ with $\eta^{00}=-1$. In two dimensions, these are satisfied in the $2 \times 2$ 
chiral representation in terms of the Pauli matrices by
\begin{align}
&\gamma^0=\sigma_1=
\begin{pmatrix}
 0&1\\1&0
\end{pmatrix}\,,&
&\gamma^1=-i\sigma_2=
\begin{pmatrix}
 0&-1\\1&0
\end{pmatrix}\,, &
&\gamma_5=\gamma^0\gamma^1=\sigma_3=
\begin{pmatrix}
 1&0\\0&-1
\end{pmatrix}
\,\label{gamdef}
\end{align}
with $\bpsi = \psi^{\dag} \gamma^0$, and $\g^0 = (\g^0)^{\dag}$ Hermitian. Thus a free Dirac fermion 
in $1+1$ dimensions may be represented as a two-component complex spinor
\be
 \psi=\begin{pmatrix}
        \psi_+ \\ \psi_-
       \end{pmatrix}
\,,  \qquad \psi_{\pm} = \tfrac{1}{2} (\id\pm\g_5)\,\psi
\ee
with $(\id \pm \gamma_5)/2$ projecting out the right- and left-handed (right moving and left moving) single component 
Dirac fields $\psi_{\pm}$ respectively. For massless fermions these two components propagate independently.

A special property of the Dirac matrices in two dimensions is 
\be
\g^{\m}\g_5 = \g_\n\,\e^{\n\m} 
\label{gamdual}
\ee
where $\e^{\m\n} = -\e^{\n\m}$ is the anti-symmetric symbol, defined so that $\e_{01} = +1 =- \e^{01}$, obeying
\be
\e^{\m\l}\e_{\n\r} = -\d^\m_{\ \n}  \d^\l_{\ \r} +  \d^\m_{\ \r} \d^\l_{\ \n}\,,\qquad \e^{\m\n}\e_{\n\r}  = \d^{\m}_{\ \r}  \,.
\label{epsident}
\ee
Because of (\ref{gamdual}), the vector coupling of the Dirac fermions can equally well be expressed 
as a chiral coupling to the \textit{dual} gauge field 
\be
\widetilde A_{\m} \equiv \eta_{\m\l}\e^{\l\n}A_{\n} \,,\qquad A_{\m} =  \eta_{\m\l}\e^{\l\n}\widetilde A_{\n}\,,\qquad
\textit{i.e.}\qquad  \widetilde A_0 = A_1\,,\quad \widetilde A_1 = A_0
\label{Adual}
\ee
according to
\be
\g^\m A_\m  = \g^\m  \e_{\m}^{\ \,\n}\widetilde A_{\n} =  \g^{\n} \g_5 \widetilde A_\n\,.
\label{chirB}
\ee
This also implies that the vector and axial vector currents 
\be
j^\m \equiv \var{S_f}{A_\m}=\sum_{i=1}^N \bpsi_i\g^\m\psi_i   \,,\qquad\qquad  
j^\m_5\equiv \var{S_f}{\widetilde A_\m} =\sum_{i=1}^N\bpsi_i\g^\m\g_5\psi_i 
\label{vecaxvec}
\ee
are related by
\be
j^\m =  j^{\n}_5\, \e^{\ \m}_{\n} \,\,, \qquad\qquad  \,j^{\m}_5 = j^{\n}\e^{\ \m}_{\n} \,.
\ee
For free fermions, $e=0$, each of these currents is classically conserved, corresponding to the 
invariance of fermion action $S_f$ in (\ref{Schwmod}) under both $U(1)$ vector gauge symmetry 
\bes
\bea
\psi_i &\rightarrow& e^{i\a}\psi_i\\
A_\m &\rightarrow& A_\m + \pa_\m \a
\eea
\label{U1vec}\ees
and the $U_A(1)$ axial transformation
\bes
\bea
\psi_i &\rightarrow& e^{i\b\g_5}\psi_i  \label{U1axpsi}\\
\widetilde A_\m &\rightarrow& \widetilde A_\m + \pa_\m \b\,.
\eea
\label{U1ax}\ees
The global versions of these symmetries would be sufficient to guarantee conservation of
the Noether currents (\ref{vecaxvec}) in the free theory. As is well known, the vector and axial 
symmetries (\ref{U1vec}) and (\ref{U1ax}) cannot both be preserved at the quantum level, and 
conservation of (at most) only one of the two classical currents (\ref{vecaxvec}) can be maintained
together with Lorentz invariance. Since as soon as $e\neq 0$, the gauge field action $S_g[A]$ is 
invariant only under the local vector symmetry (\ref{U1vec}), enforcing the choice of $U(1)$ vector 
gauge invariance
\be
\pa_\m j^\m=0
\label{currcons}
\ee
leads necessarily to a well-defined anomalous divergence of the axial current \cite{Johnson:1963vz}
\be
\pa_\m j_5^\m=\frac{N}{\pi}\,\widetilde{F} = -\var{\G_{\rm eff}[\b]}{\b}
\label{axanom}
\ee
where the pseudoscalar dual of the field strength tensor is
\be
\widetilde F \equiv \tfrac{1}{2}\, \e^{\m\n}F_{\m\n} =E\,.
\label{Fdual}
\ee
The second relation of (\ref{axanom}) indicates that through the chiral anomaly the effective action 
of the fermions $\G_{\rm eff}[\b]$ necessarily acquires a non-vanishing dependence
upon a $U_A(1)$ rotation (\ref{U1ax}) by $\b$.

In any number of dimensions the decomposition of a vector field into its parallel and transverse components is
\bes
\begin{align}
&A_{\m} = A^{\parallel}_{\m} + A^{\perp}_{\m} \\
&A^{\parallel}_{\m} = \pa_{\m}\sq^{-1}(\pa^{\n}A_{\n})\equiv \pa_{\m} \a\\
&\hspace{-1.5cm}A_{\m}^{\perp} = (\d_{\m}^{\ \n} - \pa_{\m}\sq^{-1}\pa^{\n})A_{\n}\,, \qquad\pa_{\m}A^{\perp\,\m} = 0
\label{Aperpdef}\end{align}
\label{Hdecomp}\ees
where $\sq^{-1}$ is the Green's function of the scalar wave operator $\sq$. The decomposition (\ref{Hdecomp})
is unique up to zero modes of $\sq$ (which we neglect for present purposes), and implies that the gauge invariant information 
resides in the transverse component $A^{\perp}_{\m}$. A special property of two dimensions is that the transverse component 
can be written as the Hodge dual of a scalar gradient $A^{\perp\,\m} = \e^{\m\n}\pa_\n \b$ \cite{Nakahara:2003}. Hence 
starting from $A_\m =0$, an arbitrary gauge potential is composed of a combined $U(1)$ and $U_A(1)$ transformation 
(\ref{U1vec}) and (\ref{U1ax}) in the form
\be
A_\m = \pa_\m\a + \eta_{\m\l}\e^{\l\n}\pa_\n \b
\label{Adecomp}
\ee 
and we obtain from this, (\ref{Adual}) and (\ref{Fdual}) that
\be
\widetilde F = \e^{\m\n}\pa_{\m}A_{\n}^{\perp} = \pa_{\m}\widetilde A^{\m} = \sq \b  \,.
\label{Fdbeta}
\ee
Inserting this relation into (\ref{axanom}) gives
\be
\var{\G_{\rm eff}[\b]}{\b}  = -\frac{N}{\pi} \sq \b
\label{varGam}
\ee
which being linear in $\b$, allows for immediate integration to the one-loop effective action 
\be
\G_{\rm eff} [\b] =-\frac{N}{2\pi} \int d^2x\, \b \sq \b
\label{Gameff}
\ee
quadratic in $\b$. This action is exact (up to zero modes and an additive constant which we may 
set to zero at $\b = 0$) and entirely determined by the chiral current anomaly. 

Using (\ref{Fdbeta}) again to formally solve for $\b = \sq^{-1} \widetilde F$ allows us to express the anomalous effective 
action (\ref{Gameff}) in the non-local gauge invariant form \cite{Jackiw:1983}
\be
\G_{\rm eff} [\b] =  -\frac{N}{2\pi}\int\!d^2x\,\int\! d^2x'\ 
\widetilde{F}_{\!x}\, (\sq^{-1})_{xx'}\, \widetilde{F}_{\!x'} = -\frac{N}{2\pi} \int\! d^2x\, A_{\m}^{\perp} A^{\perp\m}
\equiv S_{\rm anom}[A]
\,. \label{SeffA}
\ee
Thus the effect of integrating out the massless fermions in the functional integral \cite{Jackiw:1983,Burgess:1993np} is
\be
Z_f^{(N)}[A]= \int \prod_{i=1}^N [\cD \psi_i][\cD \bpsi_i] \exp\{iS_f[\psi, \bpsi; A]\} = [{\rm det}_F(\slashed{\pa})]^{N}
\exp\{ iS_{\rm anom}[A]\}
\label{intSeffA}
\ee
with $S_{\rm a nom}[A]$ given by (\ref{SeffA}). In the functional integral approach the breaking of chiral symmetry 
and the axial anomaly (\ref{axanom}) may be ascribed to the non-invariance of the fermionic functional 
measure $\prod_{i=1}^N[\cD \psi_i][\cD \bpsi_i]$ under the axial $U_A(1)$ transformation (\ref{U1axpsi})
\cite{Fujikawa:1979ay,Fujikawa:1980eg,Fujikawa:2003az,Bertlmann-book}. 

That all the gauge invariant information resides in $A^{\perp}_{\m}$ which,  owing to (\ref{Adecomp}), is generated by a $U_A(1)$ 
axial transformation by $\beta$ of the fermion determinant in two dimensions, and that this dependence upon 
$\beta$ in (\ref{varGam}) is only linear through the axial anomaly (\ref{axanom}) so that the effective
action (\ref{SeffA}) is purely quadratic in $A_{\perp}$ are the essential points leading to the Schwinger model 
being exactly soluble.

The appearance of the massless scalar propagator $ (\sq^{-1})_{xx'}$ in the one-loop effective action (\ref{SeffA}) 
is the first indication that an effective scalar boson degree of freedom is associated with the chiral anomaly. Indeed 
a pseudoscalar boson field $\chi$ may be introduced so as to rewrite the result for the non-local effective action in 
(\ref{SeffA})-(\ref{intSeffA}) in the form
\be
Z_f^{(N)}[A] =[{\rm det}_f(\slashed{\pa})]^{N} [{\rm det}_B(-\sq)]^{\frac{1}{2}}\!
\int [\cD\chi]\,\exp\left\{i S_{\rm anom}[\chi; A]\right\}
\label{Zchi}
\ee
with the local bosonic action
\be
S_{\rm anom}[\chi; A] \equiv \frac{N}{\pi}\!\int\!d^2x\left(\tfrac{1}{2}\chi \sq \chi - \widetilde{F}\, \chi \right)
\label{Seffchi}
\ee
accounting for the anomaly. By varying this action the local field $\chi$ satisfies the eq.\ of motion
\be
\sq\chi = \widetilde{F} = \sq\beta
\label{chieom}
\ee
while performing the Gaussian integral over $\chi$ simply reproduces (\ref{intSeffA}), when account
is taken of the $ [{\rm det}_B(-\sq)]^{\frac{1}{2}}$ prefactor in (\ref{Zchi}). Thus the pseudoscalar
boson field $\chi$ is a completely gauge invariant local field equivalent to the pre-potential $\b$ 
determining the transverse gauge invariant part of the vector potential, up to homogeneous solutions 
of the two dimensional wave equation.

Varying (\ref{Seffchi}) with respect to the vector potential $A_{\m}$ or its dual $\widetilde A_{\m}$ yields
the vector and axial vector currents (\ref{vecaxvec}) in terms of the effective boson field $\chi$ as
\be
j^\m = \var{S_{\rm anom}}{A_\m\ }=-\frac{N}{\pi}\, \e^{\m\n}\pa_\n\chi \qquad {\rm and} \qquad  j^\m_5 
=\var{S_{\rm anom}}{\widetilde A_\m\ }=\frac{N}{\pi}\,\pa^\m\chi\
\label{vecaxvecchi}
\ee 
so that the axial anomaly (\ref{axanom}) is recovered by the eq.\ of motion for $\chi$ (\ref{chieom})
\be
\pa_{\m}  j^\m_5  = \frac{N}{\pi}\,\sq \chi = \frac{N}{\pi}\,\widetilde{F}
\ee
and the Maxwell eq.\ is
\be
\pa_{\n}F^{\m \n} = -\e^{\m\n} \pa_{\n}(\sq\b) = \frac{e^2}{N\,}\,j^{\m} =-\frac{e^2}{\pi\,}\, \e^{\m\n}\pa_\n\chi
\label{Maxeq}
\ee
with current conservation (\ref{currcons}) becoming a topological identity, equivalent to the
single-val\-uedness of $\chi$. 

Because $\widetilde F $ is a total derivative, \textit{cf.}\ (\ref{Fdbeta}), its integral $\int d^2 x \,\widetilde F = 2 \pi \n$
is a topological invariant, $\n$ being the Pontryagin index, and the action (\ref{Seffchi}) is invariant up to a surface 
term under the shift of $\chi$ by a spacetime constant. This leads to the existence of a Noether current
\be
J^{\m}_5 \equiv  j^{\m}_5 - \frac{N}{\pi} \e^{\m\n}A_{\n}  \equiv j^{\m}_5 + 2N K^{\m}
\label{j5tilde}
\ee
which is gauge dependent but conserved, $\pa_{\m} j^{\m}_5 [A] = 0$ by (\ref{axanom}) and (\ref{Fdual}).  
Note also that (\ref{Maxeq}) can be immediately integrated and implies 
\be
E= \sq\b = \frac{e^2}{\pi}\, \chi + E_0
\ee
where $E_0$ is a spacetime independent integration constant, that can be regarded as an external constant electric field.
Since $E_0$ can be eliminated by shifting $\chi \rightarrow \chi - \pi E_0/e^2$, reference to (\ref{Seffchi}) shows that
this is equivalent to adding to the action a topological term $\theta N\n$, with the arbitrary $\theta$ 
vacuum parameter of the Schwinger model given by $E_0 = e^2\theta/2$ in the present units \cite{Coleman:1976uz}.

The complete solution of the Schwinger model is achieved by making use of (\ref{Zchi}) together with the 
classical action in (\ref{Schwmod}) to integrate over the inequivalent gauge orbits of the vector potential $A_{\mu}$ 
by means of the gauge invariant functional measure \cite{Mottola:1995sj}
\be
\frac{[\cD A]}{{\rm Vol}\, [U(1)]} = [{\rm det}_B(-\sq)]^{\frac{1}{2}} [\cD A^{\perp}] = {\rm det}_B(-\sq) \, [\cD \b]
\label{Ameas}
\ee
again up to zero modes. In these relations the determinants in the functional measure are Jacobians of the
transformations from $A_{\m}$ to $A_{\m}^{\perp}$ to $\beta$. These Jacobians and in particular the last determinant
in (\ref{Ameas}) are responsible for cancelling the contributions of the apparent ghost in the gauge field action
$S_g[A] = -1/4 \int d^2 x F_{\m\n}F^{\m\n} = 1/2 \int d^2 x (\sq \b)^2$ in the latter higher derivative form expressed
in terms of the gauge invariant chiral pre-potential $\b$. The Gaussian integral over 
$A^{\perp}_{\m}=\e_{\m}^{\ \n}\pa_\n \b$ has a saddle point at $\widetilde F = e^2 \chi/\pi$ and yields 
\begin{align}
Z^{(N)} &= \int \frac{[\cD A]}{{\rm Vol}\, [U(1)]} \exp\left\{\frac{iN}{e^2}S_g[A]\right\}\,  Z_f^{(N)}[A] \nn
&= [{\rm det}_f(\slashed{\pa})]^{N} [{\rm det}_B(-\sq)]^{\frac{1}{2}} \!
\int [\cD\chi]\,\exp\left\{\frac{iN}{2\pi}\!\int\!d^2x\left(\chi \sq \chi - \frac{\,e^2}{\pi}\, \chi^2 \right)\right\}\,.
\label{ZNf}
\end{align}
We have retained the determinants in (\ref{intSeffA}) and (\ref{Zchi}) in order to keep track of the number of
local degrees of freedom, starting with $N$ local fermionic degrees of freedom (and none originally in the vector potential, 
which is fully constrained by Gauss' Law in two dimensions). Since a Dirac fermion with anti-periodic boundary
conditions is equivalent to a single scalar with periodic boundary conditions in two dimensions, the
functional determinants satisfy \cite{Fujikawa:2003az}
\be
{\rm det}_f(\slashed{\pa})= [{\rm det}_B(-\sq)]^{-\frac{1}{2}}
\label{FermiBose}
\ee
so that for $N=1$ one obtains from (\ref{ZNf}) with (\ref{FermiBose})
\be
Z_{\rm Schw} = Z^{(N=1)} = \int [\cD\chi]\,\exp\left\{\frac{i}{2\pi}\!\int\!d^2x\left(\chi \sq \chi - \frac{e^2}{\pi}\, \chi^2 \right)\right\}
\label{ZSchw}
\ee
which is exactly the expression for a single real propagating pseudoscalar boson field $\chi$ with mass
$M^2_{\chi} = e^2/\pi$, recovering the well-known result for the Schwinger model~\cite{Schwinger:1962tp,Brown:1963}. 
Because of relation (\ref{FermiBose}), for $N>1$ (\ref{ZNf}) defines a theory of a single massive boson with mass 
$M^2_{\chi} = e^2/\pi$ and $N-1$ massless bosons \cite{Halpern:1975jc,Hetrick:1995wq,Hosotani:1998za}.

It is clear from the final form (\ref{ZSchw}) for $N=1$ that one has traded the original single fermion degree 
of freedom for a single boson degree of freedom $\chi$, which is a \textit{bona fide} propagating field in its own 
right, with its kinetic term $\chi\sq \chi$ generated by the axial anomaly. The number of overall local degrees of
freedom is conserved. Comparing the expressions for the currents (\ref{vecaxvec}) and (\ref{vecaxvecchi}), 
it is also clear that the boson field $\chi$ is bilinear in $\bpsi$ and $\psi$, and hence is made up of a 
fermion/anti-fermion pair. This is a relativistic version of the Cooper pairing phenomenon familiar in 
non-relativistic many-body theory, and the BCS theory of superconductivity \cite{Leggett:2006}.

We conclude this section with a few additional remarks. First, the anomaly may also be regarded in effect 
as giving rise to a gauge invariant mass term $A_{\m}^{\perp} A^{\perp\m}$ for the gauge field in (\ref{SeffA}), 
as the functional integral of the Schwinger model may also be written in the form
\be
Z_{\rm Schw} = \int [\cD A_{\m}^{\perp}]\,\exp\left\{\frac{iN}{2e^2}\!\int\!d^2x\left(A_{\m}^{\perp} \sq A^{\perp\,\m}
 - \frac{e^2}{\pi}\, A_{\m}^{\perp}A^{\perp\,\m} \right)\right\}
\label{ZSchwA}
\ee
when use is made of (\ref{SeffA}), (\ref{Ameas}) and (\ref{FermiBose}). However, this interpretation of a propagating
massive boson only makes sense because the anomaly through fermion pairing has rendered the gauge field 
into an effective propagating degree of freedom, whereas it was totally constrained by Gauss' Law in the classical
theory. Notice also that this interpretation does not require fixing a gauge, and as use of the gauge invariant 
measure (\ref{Ameas}) makes clear, the mass term is fully gauge invariant. This gauge invariant mass generation 
for a gauge field is basically the Stueckelberg mechanism \cite{Stueckelberg:1938zz,Aurilia:1980jz} for mass 
generation which would serve as a prototype for the Higgs mechanism in the Standard Model. In the limit of 
vanishing coupling $e \to 0$, the gauge field remains massless. The gauge field propagating mode with a 
finite screening length is a relativistic version of collective excitations familiar in many-body systems, 
induced \textit{e.g.}\ in superconductors (Meissner effect) and finite temperature plasmas (Debye screening).

Next we note that had we performed the functional integral in the opposite order, integrating first over the vector
potential, we would have obtained the gauge invariant result
\begin{align}
Z^{(N)}\! &= \int \prod_{i=1}^N [\cD \psi_i][\cD \bpsi_i]
\int \frac{[\cD A]}{{\rm Vol}\, [U(1)]}\, \exp \left\{iS_f[\psi, \bpsi; A]  + \frac{iN}{e^2}\, S_g [A]\right\}\nn
& = \int \prod_{i=1}^N [\cD \psi_i][\cD \bpsi_i] \,\exp\! \left\{i \sum_{j=1}^N\int\! d^2x\, \bpsi_j\,
\g^\m \! \!\stackrel{\leftrightarrow}{\ \pa_\m}\!\psi_j 
 -\frac{ie^2}{2N} \int\! d^2x\int\! d^2x'\, j^{\m}_x (\sq)^{-1}_{xx'}\, j_{\m\,x'}\!\right\}
 \label{intA}
\end{align}
which is a theory of $N$ massless fermions with a four-fermion current-current interaction between them.
Except for its non-locality this is again similar to the starting point for BCS theory \cite{Leggett:2006}.

Due to the conservation of the charge current $j^{\m}$, its space and time components are not 
independent, and
$j^1 = -(\pa_x)^{-1} \dot \r$, so that
\be
 \frac{1}{2}\int d^2x\int d^2x'\, j^{\m}_x \left(\frac{\eta_{\m\n}}{\sq}\right)_{xx'}\, j^{\n}_{x'} 
 = -  \frac{1}{2}\int dt \int dx\int dx'\, \r(t, x)\,  \frac{1}{\pa_x^2}\, \r(t,x')
\ee
is in fact a instantaneous Coulomb interaction between the two charge densities $\r = j^0$ at spatial 
positions $x$ and $x'$.  It is remarkable that this apparently non-local (but also apparently non-anomalous) 
theory of massless fermions interacting by their mutual long range Coulomb interaction becomes
the \textit{local} theory of a single \textit{non-interacting} but \textit{massive boson} $\chi$, together with
$N-1$ free fermions in (\ref{ZNf}) via the previous route of the axial anomaly. For strictly zero
coupling $e=0$, the free fermion and free boson representations are equivalent. However as soon 
as $e\neq 0$, no matter how small its magnitude, the attractive Coulomb interaction between the 
fermions and anti-fermions destabilizes the free massless fermion ground state, and leads to the 
ground state or vacuum of a massive bound state boson instead, again reminiscent of the
Cooper instability and pairing phenomenon \cite{Leggett:2006}. 

Finally we note that the effect of functionally integrating over the chiral boson in (\ref{Zchi})
is up to the boson determinant and use of the saddle point eq.\ (\ref{chieom}) equivalent to 
the previous form (\ref{intSeffA}), so that we may equally well write 
\be
Z_f^{(N)}[A]= [{\rm det}_f(\slashed{\pa})]^{N}\, \exp\{ iS_{\rm anom}[\chi; A]\}\Big\vert_{\sq \chi = \widetilde{F}}\ .
\label{intSeffchiA}
\ee
This shows that the one-loop generating functional of axial or electromagnetic current correlators in the original
free $N$-fermion quantum theory is mapped (up to a multiplicative constant independent of $A_{\m}$) to the tree diagrams 
of the classical bosonic action (\ref{Seffchi}), with the chiral boson field sourced by $\widetilde{F}$ according to 
(\ref{chieom}). The equivalence of the quantum one-loop fermion theory to tree level boson will be shown
explicitly in current correlation functions in the next section.

\subsection{Correlation Functions of Currents, Spectral Function and Sum Rule}
\label{sec:correlator-jj}

\begin{figure}[ht]
\centering
\begin{subfigure}{.35\textwidth}
\centering
\includegraphics{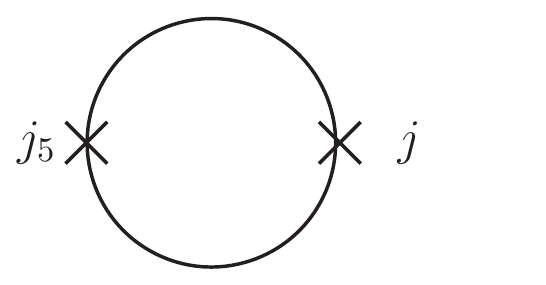}
\caption{The one-loop fermion polarization diagram.}
\label{fig:JJ-loop}
\end{subfigure}
\hspace*{1cm}
\begin{subfigure}{.35\textwidth}
\centering
\includegraphics{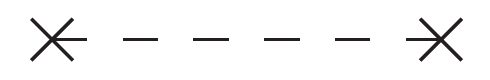}
\caption{The pseudoscalar tree diagram.}
\label{fig:JJ-tree}
\end{subfigure}
\caption{Equivalence of one-loop fermion polarization diagram with pseudoscalar tree.}
\label{fig:JJ}
\end{figure}

The exact quantum effective action (\ref{SeffA}) resulting from integrating out the fermions  
arises entirely from the one-loop diagram in Fig. \ref{fig:JJ-loop}, which in Fourier space is
\be
i\int d^2 x\, e^{ik\cdot x}\,\lag 0|{\cal T} j^\l_5 (x) j^\n(0)|0 \rag
= -\e^{\l\m} \,\Pi_\m^{\ \n} (k)\big\vert_{m=0, d=2} 
\label{J5J}
\ee
where ${\cal T}$ denotes time-ordering and
\be
\Pi^{\m\n}(k) = \frac{2N\,}{(2\pi)^{\frac{d}{2}}}\, \Gamma\left(2 - \tfrac{d}{2}\right) 
\big(k^\m k^\n - k^2 \eta^{\m\n}\big) \int_0^1 dx \, x(1-x)\,  [k^2x(1-x) + m^2]^{\frac{d}{2} - 2}
\label{PolEm}
\ee
is the vacuum polarization in general $d$ dimensions for $N$ fermions of mass $m$. For $d=2$ this is finite,
and for massless fermions $m=0$ it becomes simply
\be
\Pi^{\m\n}(k) \big\vert_{m=0, d=2} = \big(k^\m k^\n- k^2 \eta^{\m\n}\big)\,  \frac{N}{\pi k^2}\,.
\label{VacPol}
\ee
Multiplying (\ref{J5J}) by $ik_\l $ and substituting (\ref{VacPol}) gives the axial anomaly eq.\ (\ref{axanom}).
The anomaly for massless fermions is thus intimately connected to the pole at $k^2=0$ in (\ref{VacPol}).
 
The $1/k^2$ massless pole in (\ref{VacPol}) in the one-loop fermion polarization corresponds to the physical 
propagator of the single effective bosonic degree of freedom $\chi$ which is massless in the absence of 
electromagnetic interactions, becoming massive in the Schwinger model according to (\ref{ZSchw}). 
Comparing (\ref{vecaxvec}) and (\ref{vecaxvecchi}), it is clear that the pseudoscalar effective boson field 
$\chi$ is related to fermion bilinears. Indeed substituting (\ref{vecaxvecchi}) the one-loop fermion vacuum 
polarization $\Pi^{\m\n}(k)$ can be written in terms of a \textit{tree} amplitude for the boson $\chi$ according to
\begin{align}
\Pi^{\m\n}(k) &=  i\int d^2 x\, e^{ik\cdot x}\,\lag 0| {\cal T} j^\m(x) j^\n(0)| 0\rag\nn
&= \frac{N^2}{\pi^2} \e^{\m\l} \e^{\n\r} k_\l k_\r \int d^2 x\, e^{ik\cdot x}\,i\lag 0|{\cal T}\chi(x) \chi(0)|0\rag\nn
&=\frac{N^2}{\pi^2} \e^{\m\l} \e^{\n\r} k_\l k_\r \,\frac{\pi}{Nk^2} \nn
&=  \big(k^\m k^\n - k^2 \eta^{\m\n}\big)\,  \frac{N}{\pi k^2}
\label{Pichi}
\end{align}
where we have used the normalization of the $\chi$ propagator from (\ref{ZNf}). 
The $1/k^2$ pole is the signal of a propagating (pseudo)scalar degree of freedom in the quantum
theory not present in the classical action (\ref{Schwmod}). It is a quantum effect of the fermion
pairing in the two-particle fermion sector involving the correlation of two currents, hence four fermion
operators, which may be re-expressed in terms of an effective single particle bosonic theory and
tree amplitude (in which $\hbar$ is a parameter). It is clear from (\ref{J5J})-(\ref{Pichi}) that for
massless fermions $m=0$ and in the limit of vanishing coupling to the electric field, $e\to 0$, 
the pairing of fermions to form a massless boson and the massless pole in (\ref{VacPol}) 
is associated with the axial anomaly, quite apart from the classical gauge field action $S_{g}[A]$,
and in this limit the boson in (\ref{ZSchw}) or the propagating electric potential in (\ref{ZSchwA})
remains massless. Indeed considering the full gauge field inverse propagator function,
\textit{cf.}\ (\ref{ZSchwA}) with (\ref{Aperpdef}), 
\be
({\cal D}^{-1})^{\m\n} (k) = ({\cal D}_0^{-1})^{\m\n} (k) - \Pi^{\m\n}(k) = 
-N (k^{\m}k^{\n} - k^2\eta^{\m\n}) \left[\frac{1}{e^2} + \frac{1}{\pi k^2}\right]
\label{invprop}
\ee
we observe that when $e^2 > 0$, the massless $1/k^2$ pole combines with the classical contribution $1/e^2$ so that 
${\cal D}^{-1}$  vanishes at $-k^2 = M_{\chi}^2 =e^2/\pi$. This corresponds to a propagator $(e^2/N)(k^2 + M^2_{\chi})^{-1}$ 
with a pole at this value of $k^2$, which is that of a massive boson. The massless anomaly pole in the polarization 
(\ref{VacPol}) indicates a propagating bosonic excitation, but only when the classical gauge field action $S_g[A]$ is added 
to the fermion theory with a finite dimensionful coupling $e^2$ in (\ref{ZNf}) or (\ref{invprop}) does this boson couple
to the gauge field and is a finite mass for it generated.

The equality of the massless fermion loop (\ref{VacPol}) and massless boson tree amplitude (\ref{Pichi}) 
(with $e\to 0$) is illustrated diagrammatically in Fig. \ref{fig:JJ}. Notice that because of the linear dependence 
of the currents upon $\chi$ in (\ref{vecaxvecchi}), and the absence of any $A_{\m}$ dependence of the $\chi$ 
propagator itself, Fig. \ref{fig:JJ} is the only diagram generated by the tree effective action in
(\ref{intSeffchiA}). Correspondingly the correlation function of two currents is the only
connected correlation function in the theory, and the full equivalence of fermion loop to scalar tree
is contained in only the diagrams represented in Fig. \ref{fig:JJ}.
 
\begin{figure}[ht]
\centering
\def\svgwidth{0.7\columnwidth}
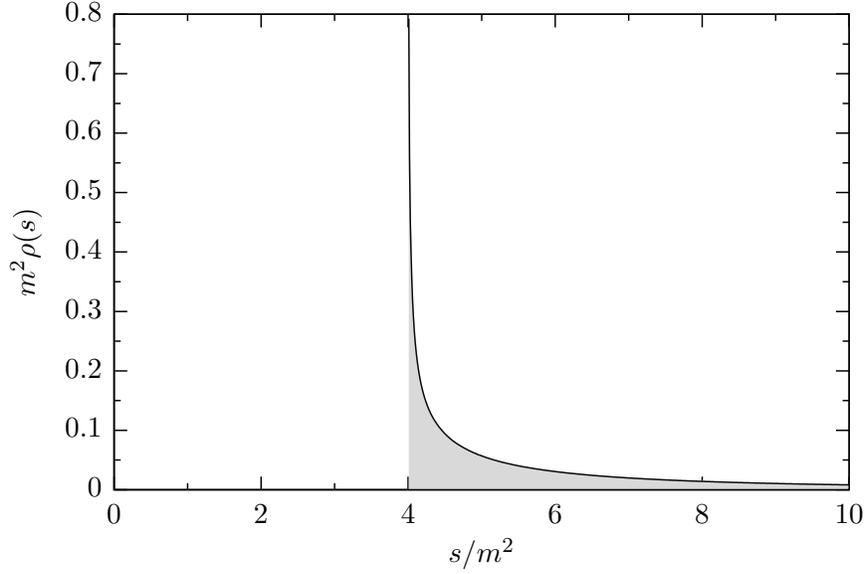
\caption{The spectral function (\ref{specJ}) for finite fermion mass $m$ as a function of $s/m^2$, shown here
for $N=1$. The area under the curve $\varrho_J(s)$ shown in gray obeys the finite sum rule (\ref{sumrule}).}
\label{fig:rho-jj}
\end{figure}

It is interesting to examine what happens to the massless $1/k^2$ pole if the theory is deformed
away from exactly zero fermion mass. Considering again the vacuum polarization with massive 
fermions in (\ref{PolEm}), we may introduce the spectral representation by inserting within
the Feynman parameter integral in (\ref{PolEm}) the identity
\be
1 = \int_0^{\infty} ds \, \d\left( s - \frac{m^2}{x(1-x)}\right)
\label{eq:spectralidentity1}
\ee
and interchanging the $s$ and $x$ integrals to obtain
\be
\Pi^{\m\n}(k)\big\vert_{d=2} =  (k^\m k^\n- k^2 \eta^{\m\n})\,\int_0^{\infty}ds\, \frac{\varrho_J(s)}{k^2 + s}\, 
\label{Pi2dm}
\ee
with
\be
\varrho_J(s) \equiv \frac{N}{\pi} \int_0^1 dx \,\d\left( s - \frac{m^2}{x(1-x)}\right) 
= \frac{2N}{\pi} \frac{m^2}{s^2} \frac{1}{\sqrt{1- \frac{4m^2}{s}}}\, \theta (s- 4m^2)\,.
\label{specJ}
\ee
This spectral function is illustrated in \figref{fig:rho-jj}.

As is clear from its definition (\ref{specJ}) or by direct integration, in $d=2$ dimensions the spectral function 
$\varrho_J(s)$ obeys the ultraviolet finite sum rule
\be
\int_0^{\infty} ds\,  \varrho_J(s) = \frac{N}{\pi}
\label{sumrule}
\ee
for any $m^2 \ge 0$. On the other hand from (\ref{specJ}), when $m \rightarrow 0$, $\varrho_J(s)$ vanishes 
for all $s> 0$. This is consistent with the sum rule only by $\varrho_J(s) $ becoming a $\d (s)$ function distribution 
in the limit of zero fermion mass. Indeed from the first expression in (\ref{specJ}) we see that
\be
\lim_{m \rightarrow 0} \varrho_J(s) = \frac{N}{\pi} \int_0^1 dx \,\d (s) = \frac{N}{\pi}\,\d (s) \,.
\label{rhom0}
\ee
Substituting this into (\ref{Pi2dm}) recovers (\ref{VacPol}) in the massless limit. Thus the effect
of the fermion mass perturbation is to spread the infinitely sharp $\d (s)$ peak in the spectral function
to a distribution in center of mass energy $s$ over a few times $m^2$ above the threshold $s > 4m^2$.
The ultraviolet sum rule expresses the fact that the pseudoscalar boson degree of freedom
remains in the two-particle fermion sector for any positive fermion mass, becoming however
a resonance rather than an isolated pole if $m> 0$. The fermion pairing due to the axial
anomaly does not disappear even for finite fermion mass, and because of the sum rule
(\ref{sumrule}) the boson field $\chi$ becomes the appropriate description again when $s \gg 4m^2$.

The Schwinger term in the commutator of currents is also directly related to the axial anomaly
and sum rule. The expectation value of the commutator of two currents is given by the discontinuity of the polarization
tensor from $k^0 + i \e$ to $k^0 -i\e$ , \textit{i.e.}
\begin{align}
&i\Big\lag \big[  j^\m (t,x), j^\n (t',x')\big]\Big\rag = i\int \frac{d^2k}{(2\pi)^2}\, e^{-ik^0(t-t') + i k^1(x-x')}\ 
2\, {\rm Im}\left[\Pi^{\m\n} (k^0 + i\e,k^1)\right]\nn
& = i\int \frac{d^2k}{2\pi}\, e^{-ik^0(t-t') + i k^1(x-x')}\,(k^\m k^\n - k^2 \eta^{\m\n})\int_0^{\infty} ds\, \varrho_J(s) 
\ {\rm sgn}(k^0) \ \delta\big(-(k^0)^2 + (k^1)^2 + s\big) \nn
& = (\eta^{\m\n} \sq - \pa^\m \pa^\n)\, \int_0^{\infty} ds\, \varrho_J(s)\, D(t-t', x-x'; s)
\label{jjcomm}
\end{align}
where
\begin{align}
D(t, x; s) &= i\int \frac{d^2k}{2\pi}\, e^{-ik^0t + i k^1x}\ {\rm sgn}(k^0) \,\d\big(-(k^0)^2 + (k^1)^2 + s\big) \nn
& =\int_{-\infty}^{\infty} \frac{dk^1}{2\pi}\, e^{ik^1x}\ \frac{\sin \big(t \sqrt{(k^1)^2 + s}\big)}{\sqrt{(k^1)^2 + s}} 
= \frac{1}{2}\, {\rm sgn}(t)\, \theta(t^2-x^2) \, J_0\Big(\sqrt{s(t^2-x^2)}\Big)
\label{Ddef}
\end{align}
is the Pauli-Jordan commutator function for a scalar field of mass $\sqrt s$ in two dimensions. 
Since $D(t,x; s)$ is an odd function of $t$ and satisfies
\be
\pa_t\,D(t, x; s)\big\vert_{t=0} = \delta(x)\qquad {\rm but} \qquad \pa_t^{2\ell}\, D(t, x; s)\big\vert_{t=0} = 0
\label{Dproperties}
\ee
for any even number of time derivatives and any $s$, only the $\m \neq \n$ term with one time derivative 
in (\ref{jjcomm}) survives when evaluated at $t=t'$, and we find the equal time commutator
\be
\Big\lag\big[ j^0 (t,x), j^1 (t,x')\big]\Big\rag = -i\, \pa_x\, \d (x-x')\int_0^{\infty} ds\, \varrho_J(s) 
= -\frac{iN}{\pi}\, \pa_x\,\delta(x-x')
\label{j0j1comm}
\ee
as a consequence of the sum rule (\ref{sumrule}). This is the expectation value of the anomalous Schwinger 
equal time commutator~\cite{Adam:1993fy} for $N$ identical fermions of any mass. 

In fact the Schwinger anomalous commutator (\ref{j0j1comm}) is exact at the operator level as may be seen  
from the boson field representation of the currents in (\ref{vecaxvecchi}), since
\be
\big[ j^0 (t,x), j^1 (t,x')\big] = -\frac{N^2}{\pi^2}\, [\pa_x \chi(t,x), \pa_t \chi (t,x')]  = 
 - \frac{iN}{\pi}\, \pa_x\, \d (x-x')
\label{Schwcomm}
\ee
as a consequence of the equal time canonical commutation relation of the $\chi$ field, 
\be
i\co{\dot\chi(t,x')}{\chi(t,x)} = \frac{\pi}{N}\,\d (x-x')
\label{chicomm}
\ee
normalized  as in (\ref{ZNf}). For massless fermions the unequal time
commutation function of currents becomes simply
\be
\big [ j^\m (t,x), j^\n (t',x')\big]\Big\vert_{m=0} =-\frac{iN}{\pi}(\eta^{\m\n} \sq - \pa^\m \pa^\n)
\, D_0(t-t', x-x')
\label{jjD0}
\ee
where
\be
D_0(t, x) \equiv D(t, x; s=0) = \frac{1}{2}\, {\rm sgn}(t)\, \theta(t^2-x^2)
\label{D0def}
\ee
is the commutator function for a massless scalar in two dimensions.

Thus the $1/k^2$ pole in the correlation function of massless fermion currents (\ref{VacPol}), the ultraviolet sum rule 
for the spectral function (\ref{sumrule}), the Schwinger term in the equal time commutator of currents (\ref{j0j1comm}), 
and the fermion pairing and bosonization formulae (\ref{vecaxvec}) and (\ref{vecaxvecchi}) are all related to and derivable 
from the axial anomaly (\ref{axanom}). The effects of the axial anomaly persist in the sum rule and Schwinger term even 
if the fermions are massive, although only if they are massless does the chiral boson pairing field $\chi$ describe a 
pseudoscalar state with a mass sharply defined by $\d$-function support only at $k^2=0$, rather than a broader 
resonance as in Fig. \ref{fig:rho-jj}, and only in the case of massless fermions is the theory exactly soluble.

\subsection{Boson Operators and the Schwinger Term}
\label{sec:FockspaceChiral}

The previous treatment of the Schwinger model by functional integral and covariant methods readily shows all
of its essential features. Fermion pairing in the Schwinger model may be realized explicitly also by canonical boson
operators in Fock space \cite{Wolf:1983,vonDelft:1998pk,Senechal:1999us}. We review this standard operator 
bosonization related to the axial anomaly in order to compare and contrast it with the corresponding pairing and 
effective scalar related to the 2D conformal anomaly in the next section.

Let us first consider the case of a single fermion ($N=1$). The Dirac eq.\ in the chiral representation gives
\be
(\pa_t \pm \pa_x) \psi_{\pm} = 0
\label{eq:DiracFock}
\ee
for the single component right and left moving massless chiral fields. Thus each may be expanded  in Fourier modes, 
\be
\psi_\pm(t,x)=\inv{\sqrt{L}}\sum_{q\geq\inv2}\left(b^{(\pm)}_{q}e^{-i\tilde k_q(t\mp x)}
+d^{(\pm)\dag}_{q}e^{i\tilde k_q(t\mp x)}\right)
\label{psiop}
\ee
where
\be
\tilde k_q=\frac{2\pi q}{L}
\ee
and $q$ is a half-integer for anti-periodic boundary conditions on the interval $x  \in [0, L]$. The fermion Fock
space operators obey the anti-commutation relations
\be
\aco{b^{(\pm)}_{q}}{b^{(\pm)\dag}_{q'}}_+=\d_{q,q'}=\aco{d^{(\pm)}_{q}}{d^{(\pm)\dag}_{q'}}_+
\label{eq:fermion-operator-commutators}
\ee
and the free fermion vacuum is defined by
\be
b^{(\pm)}_q | 0\rag = d^{(\pm)}_q | 0\rag = 0\,.
\label{bdvac}
\ee
To simplify notation somewhat one can define
\be
 c^{(\pm)}_{q\geq\inv2} \equiv b^{(\pm)}_q\,, \qquad\qquad  c^{(\pm)}_{q\leq-\inv2} \equiv d^{(\pm)\,\dag}_{-q}
\label{cqdef}
\ee
so that 
\be
\psi_\pm(t,x)= \inv{\sqrt{L}}\sum_{q\in\Z + \inv2 }c^{(\pm)}_q e^{- i\tilde k_q t}\,e^{\pm i \tilde k_q x}
\label{psiop-simple}
\ee
and the anti-commutation relations
\be
\aco{c^{(\pm)}_{q}}{c^{(\pm)\dag}_{q'}}_+=\d_{q,q'}\,,\qquad \aco{c^{(\pm)}_{q}}{c^{(\pm)}_{q'}}_+
= \aco{c^{(\pm)\dag}_{q}}{c^{(\pm)\dag}_{q'}}_+ = 0
\ee
hold, for all (both positive and negative) half-integers $q \in \Z + \inv 2$. 

Bearing in mind that normal ordering of the fermion operators is defined with respect to the fermion 
vacuum (\ref{bdvac}), and $c^{(\pm)\,\dag}_{q\leq-\inv2}$ as defined by the Hermitian conjugate of
(\ref{cqdef}) is an annihilation operator, the normal ordered fermion charge density operator for a single fermion 
\be
j^0 =\ :\!\bpsi(t,x)\g^0\psi(t,x)\!:\ =\ :\!\bpsi(t,x)\g^1\g_5\psi(t,x)\!:\  =\   :\!\psi^\dag_+\psi_+\!:\  +\  :\!\psi^\dag_-\psi_-\!:
\label{j0def}
\ee
and the current density operator
\be
j^1=\ :\!\bpsi(t,x)\g^1\psi(t,x)\!:\ =\  :\!\bpsi(t,x)\g^0\g_5\psi(t,x)\!:\ =\ :\!\psi^\dag_+\psi_+\!:\  -\  :\!\psi^\dag_-\psi_-\!:
\label{j1def}
\ee
can be expressed in terms of the fermion bilinears via
\be
:\!\psi^\dag_\pm \psi_\pm\!: \ = \frac{1}{L} \sum_{n \in \Z} \rho_n^{(\pm)}\, e^{-ik_nt}\, e^{\pm ik_n x}
\label{jpsipm}
\ee
with
\be
\rho^{(\pm)}_n \equiv  \sum_{q\in\Z + \inv2}:\! c^{(\pm)\dag}_{q-n}c_q^{(\pm)}\!\!: \ \equiv
\sum_{q\geq\inv2}c^{(\pm)\, \dag}_{q-n}c^{(\pm)}_q -\sum_{q\leq -\inv2} c^{(\pm)}_qc^{(\pm)\,\dag}_{q-n}\,.
\label{rhodef}
\ee
Note that
\be
k_n = \frac{2 \pi n}{L}
\label{kndef}
\ee
so that $\rho^{(\pm)}_n$ is defined by (\ref{rhodef}) for all integers $n$, and is periodic on the interval $x  \in [0, L]$.

From the hermiticity of (\ref{jpsipm}) or from (\ref{rhodef}) it follows (by shifting $q \rightarrow q+n$ and 
regrouping terms) that
\be
\rho^{(\pm)\,\dag}_n = \rho^{(\pm)}_{-n} \qquad \forall\, n\in\Z\,.
\label{rhodag}
\ee
Note that for $n> 0$ $\rho_n^{(\pm)}$ may also be written in terms of the physical fermion
creation and annihilation operators in the form
\be
\rho^{(\pm)}_{n>0} = \sum_{q = \inv2}^{n -\inv2} d^{(\pm)}_{n-q}b^{(\pm)}_q 
+ \sum_{q = n+ \inv2}^{\infty} b^{(\pm)\dag}_{q-n}b^{(\pm)}_q 
- \sum_{q  \le - \inv2} d^{(\pm)\dag}_{-q}d^{(\pm)}_{n-q}\,.
\label{rhoposn}
\ee
The $n=0$ densities 
\be
\rho^{(\pm)}_0 = Q_{\pm} = \int_0^L dx \ :\!\psi^\dag_{\pm}\psi_{\pm}\!: \ 
= \sum_{q\in\Z + \inv2}:\!c^{(\pm)\dag}_q c^{(\pm)}_q\!\!:\  =
\sum_{q\geq\inv2}c^{(\pm)\dag}_q\, c^{(\pm)}_q -\sum_{q\leq-\inv2} c^{(\pm)}_q\,c^{(\pm)\dag}_q
\label{chargedensities}
\ee
are total charge operators for right and left moving fermions respectively. 

It is clear that the mixed commutator of left and right movers $[\rho^{(\mp)}_n,\rho^{(\pm)}_{n'}]= 0$,
while a short calculation, \textit{cf.}\ \appref{sec:commutators}, shows that
\be
[\rho^{(\pm)}_n,\rho^{(\pm)}_{n'}] = n\, \d_{n,-n'} \qquad {\rm so\ that}\qquad 
[\rho^{(\pm)}_n,\rho^{(\pm)\,\dag}_{n'}] = n\,\d_{n,n'}\,.
\label{rhocomm}
\ee
This finite non-zero commutator for the Fourier moments of the charge densities is anomalous, 
since a naive computation ignoring the normal ordering in (\ref{rhodef}) and freely shifting the
$q$ indices in the unregulated sums gives zero. With proper normal ordering with respect
to the fermion vacuum, the expectation value of the equal time commutator currents is instead
from (\ref{j0def})-(\ref{jpsipm}) and (\ref{rhocomm})
\begin{align}
[j^0(t,x), j^1(t, x')]\big\vert_{N=1} &= \frac{1}{L^2} \sum_{n\in \Z} n e^{ik_n (x - x')} 
- \frac{1}{L^2} \sum_{n\in \Z} n e^{-ik_n (x - x')}\nn
& = \frac{1}{\pi L} \sum_{n \in \Z} k_n e^{ik_n (x - x')} = -\frac{i}{\pi}\, \pa_x \delta(x-x')
\label{currcomm}
\end{align}
for a single fermion ($N=1$). Thus (\ref{j0j1comm}) for the expectation value is verified to be an exact result, 
valid at the operator level, as is also expected from (\ref{Schwcomm}). 

Since the anomalous commutator (\ref{rhocomm}) is a $c$-number, the current algebra for $n >0$
is isomorphic to the canonical algebra of a bosonic field operator, which we now construct as follows.
Let us define
\be
a^{(\pm)}_n\equiv -\frac{i}{\sqrt{|n|}}\,\rho^{(\pm)}_n\,, \qquad n \neq 0\,, \qquad N=1
\label{andef}
\ee
which, for $n$ strictly positive, obey the canonical commutation relations, 
\be
[a^{(\pm)}_n, a^{(\pm) \dag}_{n'}] = \frac{1}{\sqrt{nn'}}\, [\r_n^{(\pm)}, \r_{n^\prime}^{(\pm)\,\dag}] =  \d_{n,n'}
\label{cancomm}
\ee
and construct the boson field operators
\be
\phi_{\pm}(t,x)=\sum_{n=1}^{\infty}\inv{\sqrt{4\pi n}}\left(a^{(\pm)}_{n}e^{-ik_n(t\mp x)}
+a^{(\pm)\dagger}_{n}e^{ik_n(t\mp x)}\right)
+ \phi^0_\pm(t, x)
\label{phiposn}
\ee
where $\phi^0_\pm$ is the contribution of the $n=0$ mode, which must be treated separately.
Since from (\ref{andef}) $a^{(\pm) \dag}_n = - a^{(\pm)}_{-n}$, the mode sum
in (\ref{phiposn}) may also be expressed as a sum over all the non-zero integers, but
as we wish to keep track of positive and negative energies in what follows, we keep $n\ge 0$.
The interpretation of the bosonic operators $a_n$ and $a_n^\dag$ for $n > 0$ in (\ref{andef}) 
is that they either move a fermion from an occupied state to an unoccupied state, or 
they create (or destroy) a particle-hole state. Since these operations do not change the fermion number, 
$a_n$ and $a_n^\dag$ commute with $Q_{\pm}$, and the Fock space they span has \emph{fixed charges} 
$Q_{\pm}$ or total numbers of left and right movers 
\cite{Wolf:1985,Manton:1985jm,Hetrick:1988yg,Senechal:1999us,vonDelft:1998pk}.
Hence the bosonization of the full fermion Fock space is incomplete without inclusion of the $n=0$ 
modes, which as we now show involves the raising and lowering operators of right and left moving 
fermion number.

The form of the zero mode completion of $\phi_{\pm}$ is determined by the following considerations. First, the $\phi_{\pm}^0$
are linearly independent and each must be a function only of $t\mp x$. Second, a limiting process $k_n \rightarrow 0$ 
of the mode functions in (\ref{phiposn}) shows that they can be at most only \textit{linear} functions of the variable $t\mp x$. 
Third, relations (\ref{j0def}) and (\ref{j1def}) determine the coefficients of the linear dependence of $\phi_{\pm}^0$ on 
$t\mp x$ in terms of $Q_{\pm}$. Finally, the normalization of the constant terms, to be called $R_{\pm}$, can be chosen 
so that
\be
\phi^0_\pm= \inv{2\sqrt{\pi}}R_\pm  + \frac{\sqrt{\pi}}{L}Q_{\pm}(t\mp x)
\label{zeromodes}
\ee
with the canonical commutation relations
\be
[R_{\pm}, Q_{\pm}] = i\,, \qquad [R_{\pm}, Q_{\mp}] = 0
\ee
between the Hermitian `coordinates' $R_{\pm}$ and the corresponding `momenta' $Q_{\pm}$ for the right and left
moving fields respectively. Thus the $R_{\pm}$ are the fermion number changing operators needed to span the 
original full fermion Hilbert space. Moreover since
\be
[\phi^0_\pm,\dot\phi^0_\pm]=\frac{i}{2L}
\label{commzero}
\ee
it is now easily verified that with the inclusion of the zero modes the equal time canonical commutation relations
\be
[(\phi_+ \pm \phi_-)_{t,x}\ , (\dot\phi_+ \pm \dot\phi_-)_{t,x'}] = \frac{i}{L}\sum_{n\in\Z}e^{ik_n(x-x')}=i\,\d (x-x')
\label{PhiETC}
\ee
are fulfilled on the finite interval $[0,L]$.  Note that the $n=0$ term in this sum comes from the zero mode commutators
(\ref{commzero}), which is necessary to complete the delta function coming from the non-zero mode Fock space.
Thus, with the zero modes (\ref{zeromodes}) included, the sum or difference of the right and left moving field operators
(\ref{phiposn}) each define complete canonical local boson fields, the sum a scalar and the difference
a pseudoscalar field. 

Taking the derivatives of the boson field operators (\ref{phiposn}) with respect to $x$ and $t$,
and using the definition (\ref{andef}), as well as (\ref{jpsipm}), (\ref{kndef}), (\ref{rhodag}),
and (\ref{zeromodes}), (\ref{j0def}) and (\ref{j1def}) become
\bes
\begin{align}
j^0 &= \frac{\,1}{\!\sqrt{\pi}\,}\,\pa_x\big(\phi_+ - \phi_-\big)\\
j^1 &=  -\frac{\,1}{\!\sqrt{\pi}\,}\, \pa_t\big( \phi_+ - \phi_-\big)\,,\qquad (N=1)
\end{align}
\label{j0j1phi}\ees
in terms of the full quantum pseudoscalar field including its zero mode contributions.
Thus the fermion current components can be expressed in terms of the pseudoscalar
boson, as expected by our previous consideration of the axial anomaly. The 
description of the current components in terms of derivatives of the scalar sum 
$\phi_+ + \phi_-$ is suited instead to the inequivalent dual theory in which the 
vector current is anomalous and the chiral symmetry is maintained at the quantum level.
The commutator of current components (\ref{jjD0}) again follows as does the Schwinger 
anomalous equal time commutator (\ref{Schwcomm}) directly from the canonical commutation 
relation (\ref{PhiETC}) and (\ref{j0j1phi}) at the operator level. 

When $N >1$, the anomalous commutator (\ref{currcomm}) acquires a factor of $N$.
Hence the currents themselves in (\ref{j0j1phi}) must acquire a factor of $\sqrt N$ in order for
the pseudoscalar boson $\phi_+ - \phi_-$ to remain canonically normalized by (\ref{PhiETC}), \textit{i.e.}
\bes
\begin{align}
j^0 &= \sqrt{\frac{N}{\!\pi}}\, \pa_x\big(\phi_+ - \phi_-\big)\\
j^1 &=  -\sqrt{\frac{N}{\!\pi}}\,\, \pa_t\big( \phi_+ - \phi_-\big)\,.
\end{align}
\label{j0j1phiN}\ees
Dividing by $\sqrt{N}$ this is equivalent to defining the canonical Fock space operators in (\ref{andef}) by
\be
a^{(\pm)}_n\equiv -\frac{i}{\sqrt{|n| N}}\,\rho^{(\pm)}_{n, N}, \qquad n \neq 0
\label{Nandef}
\ee
with $\rho^{(\pm)}_{n, N}$ the Fourier moments of the currents $j^0 \pm j^1$ for $N$ fermions (recall Eqs. \eqref{j0def}-\eqref{jpsipm}).

Comparing the rescaled currents with those in the functional integral representation (\ref{ZNf}) 
in (\ref{vecaxvecchi}), we see that the field $\chi$ of the previous section is related
to the canonically normalized quantum pseudoscalar operator field of this section by
\be
\chi = \sqrt{\frac{\pi}{N}}\ \big(\phi_+ - \phi_-\big)
\label{rescalePhi}
\ee
for $N$ identical fermions, consistent with the normalization of the commutation relation (\ref{chicomm}).
The scaling of the currents (\ref{j0j1phiN}) or (\ref{Nandef}) with $\sqrt N$ as opposed to linearly in $N$ 
is due to the fact that whereas \textit{classical} currents in (\ref{vecaxvecchi}) scale linearly with the total number
of particle species, the \textit{quantum fluctuations} in these currents encoded in (\ref{currcomm})
and the commutation relations (\ref{PhiETC}) are suppressed with respect to these by $\sqrt{\hbar/N}$.
The effective loop expansion parameter is therefore $\hbar/N$. A classical condensate of $\phi_{\pm}$ 
would be larger than the quantum fields in (\ref{phiposn}) by a factor of $\sqrt{N/\hbar}$,
and give a classical $\chi$ of order $N^0$ and currents in (\ref{vecaxvecchi}) of order $N$.

This establishes the complete equivalence of the covariant functional integral approach of
\secref{sec:chiral-review}, and the Fock space operator description of this section. In the latter approach 
the normal ordering prescription which takes proper account of the filled Dirac sea is critical
to obtaining correct finite results consistent with the covariant anomaly encoded in the
amplitude (\ref{J5J}) and Schwinger commutator (\ref{currcomm}). Indeed the anomaly itself
may be regarded as a consequence of the Dirac sea filled to an infinite depth \cite{Adam:1993fy}.

An interesting point to notice about the construction of the boson operators (\ref{phiposn})
is that whereas the non-zero modes are strictly periodic on the interval, the zero modes (\ref{zeromodes}) 
are not. Instead only the exponential operators $\exp\, ( 2i \sqrt{\pi} \phi_{\pm})$ are periodic in the 
representation where $Q_{\pm}$ are diagonal and take on integer values.  The operators 
$U_{\pm} = \exp (i R_{\pm})$ and $U^\dag = U^{-1} = \exp (-i R_{\pm})$ are the raising and lowering 
operators needed to change the right and left moving fermion numbers $Q_{\pm}$ by one unit. 
Together with the $n > 0$ bosonic Fock space operators in (\ref{andef}) they span the entire original 
fermionic Fock space, thereby  completing the bosonization and making it fully invertible in terms of 
$\exp\, ( 2i \sqrt{\pi} \phi_{\pm})$  (`re-fermionization'). In the condensed matter literature the exponentials 
of the zero modes (\ref{zeromodes}) are referred to as \textit{Klein factors} \cite{vonDelft:1998pk}. 

In the field theory context these zero modes are called \textit{winding modes} \cite{Manton:1985jm,Burgess:1993np}, 
because the fixed $Q_{\pm}$ sectors of the Hilbert space are sectors of fixed Chern-Simons winding number 
\be
N_{\rm CS} \equiv \int K^{\m}\, d\Sigma_{\m} = -\frac{1}{2\pi} \int \e^{\m\n}A_{\n} d\Sigma_{\m} =  \frac{1}{2\pi} \int_0^L A_1 dx
\label{NCS}
\ee
given by the time component of the topological current $K^{\m} = -\frac{1}{2\pi} \e^{\m\n}A_{\n}$ integrated 
over the spatial interval $[0,L]$. The Chern-Simons number (\ref{NCS}) is invariant under gauge transformations 
periodic on the interval, but changes by an integer $N_{\rm CS} \rightarrow N_{\rm CS} + q$ under `large' gauge 
transformations $A_1 \rightarrow A_1 -i U_q^{-1}\pa_x U_q$ where $U_q(x) = \exp (2 \pi iqx/L)$ is the holonomic 
winding of the $U(1)$ phase $q$ times as $x$ varies over $[0,L]$. The electric charge $Q_{+} + Q_{-}$ is
conserved, but the axial charge $Q_5 = Q_{+} - Q_{-}$ changes by $2 N q$ units under such a transformation.
The winding sectors are in one-to-one correspondence with the integers characterizing the topologically 
distinct mappings of the $U(1)$ Wilson loop phase $\exp (i \int_0^L dx A_1) = \exp (i \oint dx^\m A_\m)$ 
winding around the non-contractible loop of spatially periodic interval $[0,L]$ with its endpoints identified,
which is the mapping ${\mathbb S}^1 \rightarrow {\mathbb S}^1$ with fundamental group $\Pi_1 ({\mathbb S}^1) = {\mathbb Z}$. 
The phase of the Wilson loop may also be recognized as the Aharonov-Bohm phase of the non-contractible loop 
around the spacetime tube ${\mathbb R} \times {\mathbb S^1}$ thought of as a `solenoid'  threaded by a 
quantized magnetic flux in the three-dimensional flat spacetime in which the two-dimensional cylinder can be embedded. 

\subsection{Intermediate Pair States of \texorpdfstring{$\lag j j\rag$}{<jj>}}
\label{sec:JJ}

In order to see explicitly how the fermion pairing into an effective bosonic degree of freedom
works in detail, and how the fermion loop can be represented as a boson tree as in Fig.
\ref{fig:JJ}, we consider next the intermediate Fock states that contribute to the current-current
correlation function (\ref{VacPol}) in both the massless fermionic and bosonic representations.

In the original fermionic representation the vacuum polarization (\ref{VacPol}), given by the
one-loop diagram of \figref{fig:JJ-loop}  implies that the cut intermediate states are two-particle fermion states.
Considering first the case of $N=1$, the general on-shell normalized two-fermion state is
\be
| q, s; q', s'\rag \equiv b^{(s)\dag}_{q} d^{(s')\dag}_{q'} \,| 0 \rag
\label{twofermion}
\ee
in the representation (\ref{psiop}), where the indices $s,s'= \pm$ distinguish left and right moving fermions.
The half-integer indices $q, q' \ge \inv2$ are to be summed over all allowed positive values
and $s, s' = \pm$ in the intermediate state sum.

On the other hand, in the bosonic representation the on-shell normalized single boson state is
\be
| n,s\rag = a^{(s )\dag}_{n}|0 \rag = -\frac{i}{\sqrt{n}} \, \sum_{q = \inv2}^{n-\inv2} b^{(s)\dag}_q d^{(s)\, \dag}_{n-q}|0\rag
\label{oneboson}
\ee
which is a particular \textit{coherent superposition} of two-fermion states (\ref{twofermion}). Thus it is not
obvious \textit{a priori} that the sum over this very different restricted set of intermediate states will yield
the same result for $\lag j j\rag$ as the that of summing over all two-fermion states  (\ref{twofermion}) 
with no restrictions. We will now show that nevertheless the sum over the coherent boson states
(\ref{oneboson}) coincides with the sum over unrestricted two-fermion states (\ref{twofermion}).

For the sum over general two-fermion intermediate states we first note that
\be
\lag 0 | j^\m(t,x) | q,{s}; q', {s'}\rag = \exp \left[-i\, t\,E(q , q' ) + i\,x\, p_{s,s'}(q , q' )\right]\, 
\lag 0 | j^\m (0,0)  | q,{s} ; q',{s'}\rag
\label{jmatrix}
\ee
where 
\be
E(q , q') = \frac{2\pi}{L} (q + q' ) \,,
\qquad p_{s,s'}(q ,q') =  \frac{2\pi}{L} \left(s q + s' q'\right)
\label{Eandp}
\ee
is the energy and momentum respectively of the two-particle state (\ref{twofermion}). Because the two
chiralities do not mix in the massless fermion limit, \textit{cf.}\ (\ref{j0def})-(\ref{j1def}), the only non-zero matrix
elements of (\ref{jmatrix}) involve 
\be
\lag 0 | \psi^{\dag}_{\pm} \psi_{\pm}(0) | q, {s}; q', {s'}\rag = \frac{1}{L} \,\d_{s, \pm}\, \d_{s', \pm}
\ee
and therefore only states with the same helicity $s = s'$ will contribute to the intermediate state sum. 
In that case from (\ref{Eandp}) the matrix element (\ref{jmatrix}) depends only upon $n = q + q' $. Then 
at fixed $n\ge 1$, the sum over $q = n- q'$ ranges from $\inv2$ to $n - 1/2$. 
The sum over the complete set of two-fermion intermediate states (\ref{twofermion}) gives
\begin{align}
\lag 0 | j^0(t,x)\,  j^0(t',x') | 0\rag &= \lag 0 | j^1(t,x)\, j^1(t',x') | 0\rag \nn
& =\sum_{s, s' = \pm}\hspace{-3mm} \sum_{\quad q, q' \ge \inv 2} 
\lag 0 | j^0(t,x) | q, {s}; q', {s'}\rag \lag  q, {s}; q',{s'}\ | j^0(t',x') | 0\rag\nn
& = \frac{1}{L^2} \sum_{s = \pm} \sum_{n =1}^{\infty} \sum_{q= \inv2}^{n - \inv2}\, \exp[-ik_n (t-t')] \, \exp[is k_n (x-x')]\nn
& = \frac{1}{L^2} \sum_{n=1}^{\infty}\,n\,e^{-ik_n (t-t')} \left[ e^{ik_n (x-x')}  + e^{-ik_n (x-x')}\right]
\label{j0j0fermion}
\end{align}
since
\be
\sum_{q =\inv2}^{n - \inv2} 1 = n\,.
\label{sumq}
\ee
Likewise we obtain
\be
\lag 0 | j^0(t,x)\,  j^1(t',x') | 0\rag = \frac{1}{L^2} \sum_{n_ =1}^{\infty}n\,e^{-ik_n (t-t')} \left[ e^{ik_n (x-x')}  - e^{-ik_n (x-x')}\right]
\label{j0j1fermion}
\ee
for the $\lag j^0 j^1\rag$ matrix element.

On the other hand, beginning with the single boson intermediate state (\ref{oneboson}) we find the matrix element
\bes
\begin{align}
 \lag 0 | j^0(t,x) | n, s \rag &= -\frac{i}{2\pi} \, \frac{k_n}{\sqrt{n}}\,\exp \left[-i\, k_n (t -s x) \right]\, \lag 0 | a_n^{(s)}a_n^{(s)\,\dag} | 0\rag\nn
&= -\frac{i \sqrt{n}}{L} \, \exp \left[-i\, k_n (t -s x) \right]\\
 \lag 0 | j^1(t,x) | n, s \rag &= -\frac{is}{2\pi} \, \frac{k_n}{\sqrt{n}}\,\exp \left[-i\, k_n (t -s x) \right]\, \lag 0 | a_n^{(s)}a_n^{(s)\,\dag} | 0\rag\nn
&= -\frac{is \sqrt{n}}{L} \, \exp \left[-i\, k_n (t -s x) \right]
\end{align}
\ees
by (\ref{cancomm}), (\ref{phiposn}), and (\ref{j0j1phi}). Hence the sum over single boson intermediate states is
\begin{align}
\lag 0 | j^0(t,x)  j^0(t',x') | 0\rag &= \lag 0 | j^1(t,x)  j^1(t',x') | 0\rag
=\sum_{s = \pm} \sum_{n=1}^{\infty} \lag 0 | j^0(t,x) | n, s\rag \lag  n, s | j^0(t',x') | 0\rag \nn
& = \frac{1}{L^2} \sum_{s = \pm} \sum_{n =1}^{\infty}\, n\, \exp[-ik_n (t-t')] \, \exp[is k_n (x-x')]\nn
& = \frac{1}{L^2} \sum_{n=1}^{\infty}\,n\,e^{-ik_n (t-t')} \left[ e^{ik_n (x-x')}  + e^{-ik_n (x-x')}\right]
\label{j0j0boson}
\end{align}
and likewise 
\be
\lag 0 | j^0(t,x)  j^1(t',x') | 0\rag = \frac{1}{L^2} \sum_{n =1}^{\infty}n\,e^{-ik_n (t-t')} \left[ e^{ik_n (x-x')}  - e^{-ik_n (x-x')}\right]
\label{j0j1boson} 
\ee
which coincide with the results obtained with the arbitrary two-fermion states inserted, (\ref{j0j0fermion})
and (\ref{j0j1fermion}) respectively. Naturally, the same result is obtained if we use the explicit fermion pair
representation of the single boson state (\ref{oneboson}) defined by (\ref{rhoposn}) and (\ref{andef}), if
again the sum (\ref{sumq}) is used.

By either method the results for the current-current correlator may be summarized as
\be
\lag 0 | j^{\m}(t,x)  j^{\n}(t',x') | 0\rag = -\frac{iN}{\pi} \, \left(\eta^{\m\n}\sq - \pa^{\m}\pa^{\n}\right)
G_0^{\ >}(t-t', x-x')
\label{jjG0}
\ee
where
\begin{align}
G_0^{\ >}(t-t', x-x') &= i\, \Big\lag \big(\phi_+ - \phi_-\big)_{t,x}\big(\phi_+ - \phi_-\big)_{t',x'}\Big\rag_{\rm nonzero} \nn
& = \frac{i}{4\pi} \sum_{n=1}^{\infty} \frac{1}{n} \, e^{-ik_n(t-t')} \left[e^{i k_n(x-x')} + e^{-ik_n(x-x')}\right]
\label{G0def}
\end{align}
is the Wightman function of the canonically normalized massless pseudoscalar field $\phi_+ - \phi_-$
defined in the previous section in the periodic interval $[0,L]$, with the zero mode removed, since 
it does not contribute to (\ref{jjG0}). For $N$ identical species of fermions we have multiplied by $N$
to obtain (\ref{jjG0}). The commutator and Schwinger term in the continuum limit is as before, \textit{cf.}\ (\ref{j0j1comm}).
If the time-ordered product of currents is considered instead, the Feynman Green's function
\be
G_0(t-t', x-x') = \theta(t-t') G_0^>(t-t', x-x') + \theta(t'-t) G_0^>(t'-t, x-x')
\label{FeynG}
\ee
for the free massless boson is obtained, whose Fourier transform is $1/k^2$ in the continuum limit.
This shows the complete equivalence between the two-fermion massless intermediate states contributing 
to the vacuum polarization $\Pi^{\m\n} = i\lag {\cal T} j^{\m}(x) j^{\n}(x')\rag$ loop diagram (\ref{VacPol}),
and the fermion paired single boson intermediate states contributing to the corresponding tree diagram 
in Fig. \ref{fig:JJ}. 

\section{Fermion Pairing and Scalar Boson of the Conformal Anomaly}
\label{sec:gravity}
\subsection{Covariant Path Integral and Effective Action in Curved Spacetime}
\label{subsec:GravFunctInt}

In addition to chiral symmetry the action $S_f$ of (\ref{Schwmod}) for massless fermions also has an apparent 
conformal symmetry. To make this explicit it is useful to generalize the fermionic action to curved spacetime
with general spacetime metric $g_{\m\n}(x)$. This is the external field taking the place of the gauge
field in this case. With the usual minimal coupling to the local zweibein frame field $e^a_{\ \m}(x)$ the 
fermion action in curved spacetime reads
\be
S_f[\psi,\bpsi; g, A] = -\sum_{j=1}^N \int d^2x\, [{\rm det}\,e_{\ \m}^a] \ 
\bpsi_j ( -i\gamma^a\,E^{\m}_{\ a} \!\!\stackrel{\leftrightarrow}{\ \nabla_\m} + m)\psi_j
\label{Sfcurved}
\ee
where $E^{\m}_{\ a}(x) \equiv  \eta_{ab} \,g^{\m\n}(x) e^b_{\ \n}(x)$ is the inverse of $e^a_{\ \m}(x)$, and
\be
\stackrel{\leftrightarrow}{\ \nabla_\m} \equiv  \! \stackrel{\leftrightarrow}{\ \pa_\m}
+ \omega_\m- i A_\m = \frac{1}{2} \big(\stackrel{\rightarrow}{\pa_\m}
-  \stackrel{\leftarrow}{\pa_\m}\big) + \omega_\m- i A_\m 
\label{covderiv}
\ee
is the double edged covariant derivative. This is defined in terms of the curved spacetime spin connection
\be
\omega_{ab\,\m} = E^{\n}_{\ [a}\eta_{b]c}\, \nabla_{\m}e^c_{\ \n} \qquad {\rm by}\qquad
\omega_{\m} =  \tfrac{1}{2} \Sigma^{ab} \omega_{ab\, \m} = \tfrac{1}{2} \Sigma^{ab} E^{\n}_{\ a}\eta_{bc}\, \nabla_{\m}e^c_{\ \n} 
\label{spincon}
\ee
in the absence of torsion, where $\Sigma^{ab} = \frac{1}{4}[\g^a, \g^b]$ and anti-symmetrization of any
two tensor indices is defined by $t_{[ab]} \equiv (t_{ab} - t_{ba})/2$. We have included a fermion mass $m$
in (\ref{Sfcurved}), although we are primarily interested in massless fermions $m=0$. The zweibein and its 
inverse satisfy
\be
e_{\ \m}^a(x)\, e_{\ \n}^b(x)\, \eta_{ab} = g_{\m\n}(x)\,,\qquad e_{\ \m}^a(x)\, E^{\m}_{\ b}(x) = \d^a_{\ b}\,,\qquad
{\rm det}\,e_{\ \m}^a = \sqrt{-g} = \sqrt{-{\rm det}\,g_{\m\n}}
\label{zweibein}
\ee
where the Greek curved spacetime indices must now be distinguished from the tangent space
Latin indices, and the tangent space Dirac matrices $\g^a$ may be taken to be the same as those 
in flat spacetime of the previous section, \eqnref{gamdef}, with this replacement of spacetime indices $\m,\n,\dots$ by tangent
space indices $a,b,\dots$

The variation of (\ref{Sfcurved}) with respect to the zweibein produces a stress tensor with both symmetric
and anti-symmetric terms. The anti-symmetric term is proportional to the divergence of spin density of the fermions
which couples to torsion in the Cartan approach to gravity. Since in this paper we do not consider torsion, 
we restrict ourselves to the symmetrized $T^{\m\n}_f$. This symmetrized fermion stress-energy tensor is given by 
\be
T^{\m\n}_f =  \frac{\eta^{ab}E_{\ b}^{\m}\!\!}{[{\rm det}\,e_{\ \rho}^c]}  \frac{\delta S_f}{\delta e_{\ \n}^a}\bigg|_{\rm sym.}
= \sum_{j=1}^N\left(- i \bpsi_j \gamma^{(\mu} {\stackrel{\leftrightarrow}{\nabla}}\,^{\nu )}\psi_j 
- g^{\mu\nu} \bpsi_j (-i \gamma^{\lambda} \!\!\stackrel{\leftrightarrow}{\ \nabla_\lambda} +\, m )\psi_j\right)
\label{SETF}
\ee
which is classically both covariantly conserved
\be
\nabla_\m T^{\m\n}_f=0
\label{Tconserv}
\ee
and traceless $g_{\m\n}T^{\m\n}_f = T^\m_{\ \m\, f} =0$ for $m=0$, by use of the eqs. of motion. These express the 
invariance of $S_f$ under both general coordinate transformations and conformal transformations 
\begin{align}
 &\psi \rightarrow e^{-\s/2} \psi\,, &&
E_{a}^{\ \m} \to e^{-\s} E_{a}^{\ \m}\,, &&
g_{\m\n} \to e^{2\s} g_{\m\n} \,,&&
{\rm det}\,e_{\ \m}^a \to e^{2\s}\, {\rm det}\,e_{\ \m}^a
\end{align}
in the massless case \cite{Bertlmann-book}.

Just as in the case of vector and chiral invariance for $m=0$ in flat spacetime, both of these classical invariances 
cannot be maintained at the quantum level and at least one must be abandoned. The Equivalence Principle requires 
(\ref{Tconserv}) for consistent coupling to gravity, with the result that once enforced at the quantum level, one finds that 
conformal invariance must be violated. In a general curved spacetime background the 
conformal anomaly is \cite{Davies:1976ei,Brown:1976wc,Birrell:1982ix}
\be
\big\lag T^\m_{\ \m}\big\rag_f =\frac{N}{24\pi}R
\label{anom2D}
\ee
in terms of the Ricci curvature scalar $R$. Since
\be
\d S_f = \frac{1}{2} \sqrt{-g}\ T^{\m\n}_{f}\, \d g_{\m\n} =  \sqrt{-g} \,T^\m_{\ \m\, f}\, \d \s
\ee
under a conformal transformation, and 
\be
\sqrt{-g}\, R = \sqrt{-\bar g}\, \bar R - 2 \,\sqrt{-\bar g}\, \sqb \s
\label{Rsig}
\ee
is linear in $\s$ when the metric is parameterized in the form
\be
g_{\m\n} = e^{2 \s} \bar g_{\m\n}
\label{metconf}
\ee
with $\bar g_{\m\n}$ is a fixed fiducial metric, the anomaly eq.\ (\ref{anom2D}) gives
\be
\frac{\d \G_{\rm eff}}{\d \s} = \sqrt{-g} \,T^\m_{\ \m\, f} = \frac{N}{24\pi} \left(\sqrt{-\bar g}\, \bar R - 2 \,\sqrt{-\bar g}\, \sqb \s\right)
\label{sigvar}
\ee
which is linear in $\s$, and thus may be integrated directly in a manner analogous to (\ref{axanom})-(\ref{intSeffA}),
to obtain the quantum effective action quadratic in $\s$,
\be
\G_{\rm eff}[\s; \bar g] = \frac{N}{24\pi} \int d^2x  \sqrt{-\bar g} \left(-\s \sqb \s + \s \bar R\right)\,.
\label{Gameffsig}
\ee
Then, solving (\ref{Rsig}) for $\s$ and using the conformal invariant property of the wave operator
in two dimensions, $ \sqrt{-\bar g} \sqb =  \sqrt{-g} \sq$, we obtain 
\be
\G_{\rm eff}[\s; \bar g] = S_{\rm anom}[g] - S_{\rm anom}[\bar g]
\label{Gamdiff}
\ee
in terms of the non-local but fully covariant effective action \cite{Polyakov:1981rd,Mottola:2006ew}
\be
S_{\textrm{anom}}[g]=-\frac{N}{96\pi}\int\!\text{d}^2x\sqrt{-g}\int\!\text{d}^2x'\sqrt{-g'}R(x)(\sq^{-1})_{xx'}R(x')
\,. \label{Sanom}
\ee
Since all two dimensional metrics are locally conformally flat, the fiducial metric $\bar g_{\m\n}= \eta_{\m\n}$ may
be taken to be flat, $\bar R =0$, and its action $S_{\rm anom}[\bar g]$,  viewed as an integration constant of the variation 
(\ref{sigvar}), may be set to zero (up to possible contributions from the non-trivial topology of $\bar g_{\m\n}$).
Thus the effect of integrating out the massless fermions is \cite{Dettki:1993cr}
\be
Z_f^{(N)}[g]= \int \prod_{i=1}^N [\cD \psi_i][\cD \bpsi_i] \exp\{iS_f[\psi, \bpsi; g, A=0]\} 
= [{\rm det}_f(-i\slashed{\pa})]^{N}\,\exp\{ iS_{\rm anom}[g]\}
\label{Zg}
\ee
in a general background metric $g_{\m\n}(x)$, with the background gauge potential $A_\m$ set to zero. 

The similarity between the axial and conformal anomalies and their effective actions, (\ref{SeffA}) and
(\ref{Sanom}) is striking. The appearance of the massless scalar propagator $(\sq^{-1})_{xx'}$ again
suggests an effective massless boson field is associated with this anomaly. Indeed in a similar manner 
to the introduction of the pseudoscalar boson field $\chi$ in the axial anomaly case, one may introduce 
a scalar boson field $\varphi$, with the local effective action
\be
S_{\rm anom}[\varphi; g] = \frac{N}{48\pi}\int\!\text{d}^2x\sqrt{-g}\left(\tfrac{1}{2}\varphi\sq\varphi+ R\varphi\right)
\label{Sanomloc}
\ee
so that $\varphi$ satisfies the eq.\ of motion
\be
-\sq \varphi = - \frac{1}{\sqrt{-g}}\pa_{\m} \left(\sqrt{-g}\, g^{\m\n}\,\pa_{\n}\varphi\right) = R
\label{phieom}
\ee
and has the massless propagator
\be
\int d^2 x\, e^{ik\cdot x}\, i\lag {\cal T} \varphi (x) \varphi (0)\rag = \frac{48\pi}{N}\, \frac{1}{k^2}\,.
\label{phiprop}
\ee
Integrating out the $\varphi$ boson field returns the action (\ref{Sanom}), allowing us to write
\be
Z_f^{(N)}[g] =[{\rm det}_f(-i\slashed{\pa})]^{N} [{\rm det}_B(-\sq)]^{\frac{1}{2}}\!
\int [\cD\varphi]\,\exp\left\{ i S_{\rm anom}[\varphi; g]\right\}\,.
\label{Zphi}
\ee
This form of the generating functional with (\ref{Sanomloc})  is quite analogous to the local form 
(\ref{Zchi}) with (\ref{Seffchi}). Note also that as the chiral anomaly term $\widetilde F$ is
a topological density, so too is the conformal anomaly term $R$, the spacetime integral of which
is a topological invariant, proportional to the Euler characteristic of a manifold with Euclidean signature.
Thus the Einstein action in two dimensions is a topological invariant, analogous to the $\theta \widetilde F$
term in the Schwinger model, and the $\varphi$ field remains massless unless a term proportional
to $R^2$, analogous to the $\widetilde F^2/2 = -F_{\m\n}F^{\m\n}/4$ Maxwell term, is added to the 
gravitational action. As in the chiral case the conformal anomaly $\sqrt{-g} R$ is a 
topological density and a total derivative,
\be
\sqrt{-g}R =\pa_{\m}\big(\sqrt{-g}\, \Omega^{\m}\big)\qquad {\rm with} \qquad 
\Omega^{\m} = 2 E^{\m}_{\ a}E^{\n}_{\ b}\, \omega^{ab}_{\ \ \n} = 2 g^{\l [ \m}E^{\n]}_{\ c}\,  \nabla_{\n} e^c_{\ \l} 
\label{Omdef}
\ee
in $d=2$. The topological nature of $\sqrt{-g} R$ as the Euler density leads to invariance of the 
scalar action (\ref{Sanomloc}) under constant shifts $\varphi \to \varphi + \varphi_0$ with \cite{Mottola:2006ew}
\be
J^{\m} = \frac{1}{4\pi}\, \big(\nabla^{\m} \varphi + \Omega^{\m}\big) = \frac{1}{4\pi}\, \big(g^{\m\n}\pa_{\n} \varphi + \Omega^{\m}\big) 
\label{Jtildef}
\ee
the corresponding gauge (\textit{i.e.}\ frame) dependent Noether current, which is analogous to (\ref{j5tilde}).
It is covariantly conserved $\nabla_{\m}J^{\m} = 0$ by virtue of (\ref{phieom}) and (\ref{Omdef}).

We note next an important difference between the cases of the axial and conformal anomalies.
Whereas in (\ref{Zchi}) all the dependence on the background $A_{\m}$ is contained in 
$S_{\rm anom}[\chi; A]$ given by the local action (\ref{Seffchi}), in (\ref{Zphi})
the boson determinant ${\rm det}_B(-\sq)$ still contains dependence upon the
metric $g_{\m\n}$ through the Laplace-Beltrami wave operator in (\ref{phieom}).
This is a reflection of the fact that whereas in the axial case, the effective boson field $\chi$ 
is neutral, and does not itself carry either a vector or axial charge, the $\varphi$ field 
has an energy-momentum in the gravitational field, given explicitly by
\be
\hspace{-3mm} T^{\m\n}_B[\varphi;g] = \frac{2}{\sqrt{-g}}\frac{\d S_{\rm anom}[\varphi;g]}{\d g_{\m\n}}
= \frac{N}{24\pi}\left(\nabla^\m\nabla^\n\varphi - g^{\m\n}\sq\varphi
+ \tfrac{1}{2} \nabla^\m\varphi\nabla^\n\varphi - \tfrac{1}{4} g^{\m\n}\nabla_\l\varphi\nabla^\l\varphi\right)\
\label{Tphi}
\ee
with a non-linear coupling to the metric. Note also that unlike the currents (\ref{vecaxvecchi}) which depend
only linearly on the boson field $\chi$, the stress-energy current (\ref{Tphi}) contains quadratic
terms in $\varphi$. As a result, if $\varphi$ is treated as a \textit{bone fide} quantum field in its own right,  
and functionally integrated over as in (\ref{Zphi}), it has its own conformal anomaly, which would 
effectively shift  $N$ in (\ref{anom2D}) to $N+1$. This is taken account of in (\ref{Zphi}) by the fact 
that the boson determinant  det$_B(-\sq)$ also depends on the metric,  and cancels the shift of $N$ 
to $N+1$, thereby restoring equality to the original fermion functional integral (\ref{Zg}).

If $N > 1$ the correction of the bosonic determinants dependence upon $g_{\m\n}$ could be handled by 
replacing the full det$_B(-\sq)$ by its flat spacetime counterpart, det$_B(-\sqb)$, while simultaneously 
replacing $N$ by $N-1$ in $S_{\rm anon}$. The conformal anomaly of (\ref{Tphi}) then shifts $N-1$ back 
to $N$. Notice however, that the fermion and boson flat spacetime determinants only cancel
when $N=1$ due to (\ref{FermiBose}), and if $N=1$ the action for the $\varphi$ field multiplied 
by $N-1$ would vanish identically. 

This difficulty can be avoided if we consider a scalar field $\bPhi$ which is defined by
\be
\bPhi = \sqrt{\frac{N}{48\pi}}\, \varphi
\label{canPhi}
\ee
and hence is canonically normalized independently of $N$. In terms of $\bPhi$ the anomaly effective action
(\ref{Sanomloc}) becomes
\be
S_{\rm eff}[\bPhi; g] = \int\!\text{d}^2x\sqrt{-g}\left(\tfrac{1}{2}\bPhi\sq\bPhi+  \sqrt{\frac{N}{48\pi}} \,R\,\bPhi\right)\,.
\label{SanombPhi}
\ee
If $N$ is now shifted to $N-1$ to compensate for the anomaly of the $\bPhi$ field itself, we obtain
\be
Z_f^{(N)}[g] =[{\rm det}_f(-i\slashed{\pa})]^{N-1} 
\int [\cD\bPhi]\,\exp\left\{i \int\!\text{d}^2x\sqrt{-g}\left(\tfrac{1}{2}\bPhi\sq\bPhi+  \sqrt{\frac{N-1}{48\pi}} \,R\,\bPhi\right)\right\}
\label{ZNphi}
\ee
where we have used (\ref{FermiBose}) for the flat spacetime determinants. Varying the exponential in (\ref{ZNphi}) 
with respect to the metric gives the canonically normalized bosonic stress-energy tensor 
\be
T_{\bPhi}^{\m\n} =  \nabla^\m\bPhi\nabla^\n\bPhi - \tfrac{1}{2} g^{\m\n}\nabla_\l\bPhi\nabla^\l\bPhi
+ \sqrt{\frac{N-1}{12\pi}}\, \Big(\nabla^\m\nabla^\n\bPhi - g^{\m\n}\sq\bPhi\Big)
\label{TbPhi}
\ee
which should be identical to the fermion expression (\ref{SETF}) at the operator level, after appropriate
regularization.

Note that if $N=1$, the linear $R\bPhi$ coupling in (\ref{ZNphi}) vanishes entirely and we obtain simply
\be
Z_f^{(N=1)}[g] = \int [\cD\bPhi]\,\exp\left\{\frac{i}{2} \int\!\text{d}^2x\sqrt{-g}\, \bPhi\sq\bPhi\,\right\}
\label{Zgone}
\ee
completely re-expressing the original free fermionic functional integral (\ref{Zg}) in terms of a free bosonic one
for $N=1$. The stress-energy tensor of the boson (\ref{TbPhi}) now lacks a term linear in $\bPhi$ and is simply
given by the stress-energy tensor of a single canonically normalized boson, which is nothing else
than the stress-energy tensor of the massless boson of the usual $N=1$ Schwinger model (with coupling $e=0$). 
In the absence of any coupling to the gauge field the chiral anomaly vanishes and the same result for
the stress-energy tensor is obtained if either the scalar or pseudoscalar linear combination of left and right
movers in (\ref{PhiETC}) is used. 

The equality of the fermion and boson stress-energy tensors for $N=1$
is somewhat non-trivial, since $\bPhi$ is composed of fermion bilinears and $T^{\m\n}_{\bPhi}$ apparently
contains four-fermion operators. However, after normal ordering, these four-fermion terms can be shown 
to vanish identically due to Fermi-Dirac statistics, and equality of the remaining terms in the stress-energy tensors
of one fermion and one boson was shown explicitly in \cite{Coleman:1969zi}. The two stress-energy tensors
clearly have the same conformal anomaly in curved spacetime because a single boson has the same 
anomaly as a single fermion in two dimensions. Thus in this approach nothing has been either 
gained or lost in replacing a single quantum fermion by a single quantum boson. If $N>1$ treating the
boson $\bPhi$ as a \textit{bona fide} full canonical quantum field with a linear $\sqrt{(N-1)/48\pi}$ coupling
to the scalar curvature should continue to work for all correlation functions involving the
stress-energy tensor. However when $N>1$ there are terms both linear and quadratic in $\bPhi$ in the stress
tensor and it is no longer obvious how to identify the scalar boson with fermion pairs.

There is an alternate approach to handling the bosonization and stress-energy tensor correspondences
which is in some ways closer in spirit to the chiral bosonization and purely linear relation of the currents
(\ref{vecaxvecchi}) to the boson field. Since in that case the chiral boson $\chi$ carries no
$U(1)$ vector or $U_A(1)$ axial charge, it does not contribute its own term to the axial anomaly
when quantized, as $\varphi$ does, accounted for by the dependence of the determinant 
$[{\rm det}_B(-\sq)]^{\frac{1}{2}}$ upon the metric $g_{\m\n}$. However, if we simply drop
the functional integral over the quantum field $\varphi$, then there is no need for the compensating 
boson determinant in (\ref{Zphi}), and no need to shift $N$. This amounts to neglecting all scalar boson 
loop diagrams, and treating $\varphi$ instead as an effectively classical field satisfying (\ref{phieom})
and contributing only to tree diagrams, in which case the axial and conformal cases are more similar. 
Then
\begin{align}
Z_f^{(N)}[g] &=[{\rm det}_f(-i\slashed{\pa})]^{N} \exp\left\{i S_{\rm anom}[\varphi; g]\right\}\!\Big\vert_{\sq \varphi = -R}\nn
&= \text{const.} \times \exp\left\{i S_{\rm eff}[\bPhi; g]\right\}\! \Big\vert_{\sq \bPhi = -\sqrt{\frac{N}{48\pi}}R}
\label{Ztree}
\end{align}
with $\bPhi$ treated as an independent field, the variation of $S_{\rm eff}$ with respect to which leads to its 
eq.\ of motion $\sq \bPhi = -\sqrt{\frac{N}{48\pi}}R$. This approach may also be obtained from (\ref{ZNphi}) 
in the large $N$ approximation, in which the quantum effects of a single $\varphi$ boson, and the prefactor 
$[{\rm det}_B(-\sq)]^{\frac{1}{2}}$ are of order one, and hence suppressed by $1/N$ in comparison 
to $N\gg 1$ fermions. However, (\ref{Ztree}) is valid for \textit{any} $N$, including $N=1$, since the functional
integral over $\bPhi$ and the $[{\rm det}_B(-\sq)]^{\frac{1}{2}}$ prefactor in (\ref{ZNphi}) precisely cancel 
each other. The analogous equivalence can be seen in the chiral case from Eqs. (\ref{SeffA})-(\ref{intSeffA}) 
by substituting the eq.\ of motion (\ref{chieom}) for the chiral boson field $\chi$ in (\ref{Seffchi}).
Since the effective tree action in the chiral case is exactly quadratic in $A_{\m}$, and the currents
are purely linear in $\chi$, this generates only the single tree graph of Fig. \ref{fig:JJ-tree}.
As we shall see, in the gravitational case (\ref{Ztree}) is closest to the chiral bosonization in the Schwinger 
model in that only the terms linear in the quantum $\bPhi$ field in the stress-energy tensor (\ref{Tphi}) 
need to be considered to identify the boson field (\ref{canPhi}) as composed of fermion pairs, analogous 
to the currents (\ref{vecaxvecchi}) linear in $\chi$, but in addition, it leads to the remarkable result that the arbitrary 
variations of (\ref{Ztree}) with respect to the metric $g_{\m\n}$ generate all one-loop correlation functions of 
quantum stress-energy tensors of the fermions by tree graphs of the scalar boson field (\ref{canPhi}).

\subsection{Correlation Functions, Spectral Function and Sum Rule}
\label{sec:correlation-gravity}

\begin{figure}[ht]
\centering
\includegraphics[width=0.6\columnwidth]{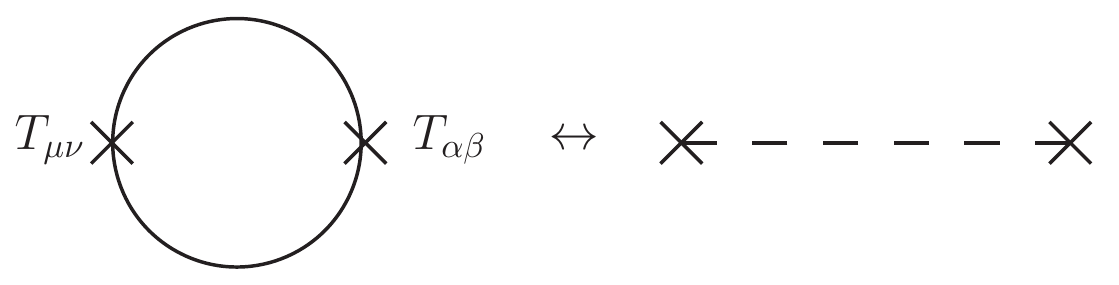}
\caption{The one-loop fermion $\lag TT\rag$ polarization.}
\label{fig:TT}
\end{figure}

We illustrate the second approach in terms of scalar tree graphs first with the simplest non-trivial correlation 
function of the one-loop stress-tensor polarization function of the fermions
\be
\Pi^{\m\n\a\b}(k)\equiv \frac{i }{N}\int d^2 x\, e^{ik\cdot x}\,\lag {\cal T}\, T^{\m\n}_f (x) T^{\a\b}_f(0)\rag
\label{TPolf}
\ee
in flat spacetime illustrated in Fig. \ref{fig:TT}.

\begin{figure}[ht]
\centering
\def\svgwidth{0.7\columnwidth}
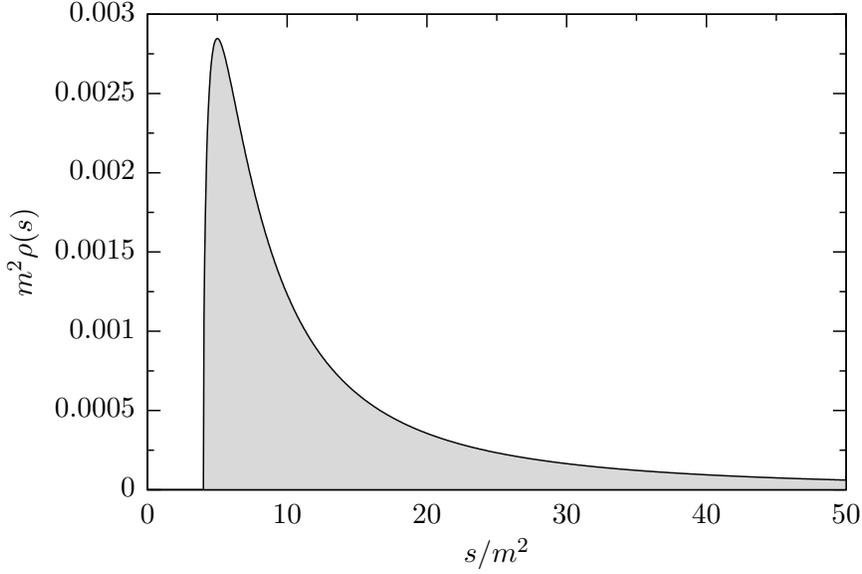
\caption{The area under the curve $\varrho_T(s)$ (shown here for $N=1$) obeys the sum rule (\ref{Tsumrule}),
and in the limit $m \rightarrow 0$ becomes a $\d$-function according to (\ref{zeromassrhoT}).}
\label{fig:rho-TT}
\end{figure}

In $d=2$ dimensions the tensor structure of this polarization is completely determined by the 
conservation law (\ref{Tconserv}) and Ward identities following from it. The result for massive fermions is
\be
\Pi^{\m\n\a\b}(k)\big\vert_{d=2} = \frac{N}{4\pi}\, (\eta^{\m\n}k^2 - k^{\m}k^{\n})(\eta^{\a\b}k^2 - k^{\a}k^{\b})
 \int_0^1 dx \, \frac{x(1-x)(1-2x)^2} {[k^2x(1-x) + m^2]}\,.
\label{VacPolFm}
\ee
Using once more the identity \eqref{eq:spectralidentity1} and interchanging the $s$ and $x$ integrals we obtain
\be
\Pi^{\m\n\a\b}(k)\big\vert_{d=2} = (\eta^{\m\n}k^2 - k^{\m}k^{\n})(\eta^{\a\b}k^2 - k^{\a}k^{\b})
\int_0^\infty ds\,\frac{\varrho_T(s)}{k^2+s}
\label{Pim}
\ee
where
\be
\varrho_T(s)\equiv \frac{N}{4\pi}\int_0^1 dx \, (1-2x)^2\,\d\left(s-\frac{m^2}{x(1-x)}\right)
=\frac{N}{2\pi}\frac{m^2}{s^2}\sqrt{1-\frac{4m^2}{s}}\, \th(s-4m^2)\,.
 \label{specT}
\ee
As in the previous case of the current spectral function (\ref{specJ})-(\ref{sumrule}),
the stress-tensor spectral function (\ref{specT}) obeys a UV finite sum rule, in this case
\be
\int_0^\infty ds\,\varrho_T(s)=\frac{N}{4\pi} \int_0^1dx\, (1-2x)^2 = \frac{N}{12\pi}
\label{Tsumrule}
\ee
which is illustrated in \figref{fig:rho-TT}.

In the massless case $m=0$, the representation (\ref{specT}) shows that $\varrho_T(s)$
becomes a $\d$-function concentrated at $s=0$,
\be
\lim_{m\to0}\varrho_T(s)=\frac{N}{4\pi}\int_0^1 dx \, (1-2x)^2\,\d(s)=\frac{N}{12\pi}\,\d(s)
\label{zeromassrhoT}
\ee
corresponding to 
\be
\Pi^{\m\n\a\b}(k)\big\vert_{m=0, d=2} =  (\eta^{\m\n}k^2 - k^{\m}k^{\n})(\eta^{\a\b}k^2 - k^{\a}k^{\b})\, \frac{N}{12\pi k^2} 
\label{VacPolF0}
\ee
which like (\ref{VacPol}) exhibits a massless scalar pole.

Hence as in the case of the axial anomaly, a correlated boson state appears in the correlation function of two fermions 
associated with the anomalous current. Both the anomaly and the bosonic excitation associated with it survive when the 
theory is deformed away from its conformal limit of $m=0$. The boson is broadened into a resonance as in Fig. \ref{fig:rho-TT}
obeying an ultraviolet finite sum rule (\ref{Tsumrule}) for any mass. The anomalous contribution of the
correlation function is also easily separated from the non-anomalous contribution by writing the trace of (\ref{Pim}) 
in the form
\begin{align}
\Pi^{\m\ \a\b}_{\ \m}(k)\big\vert_{d=2} &= k^2 \,(\eta^{\a\b}k^2 - k^{\a}k^{\b})
\int_0^\infty ds\,\frac{\varrho_T(s)}{k^2+s} \nn
&= (\eta^{\a\b}k^2 - k^{\a}k^{\b}) \,\left[ \int_0^\infty ds\,\varrho_T(s) -  \int_0^\infty ds\,\varrho_T(s) \frac{s}{k^2+s}\right]\nn
&=  \frac{N}{12\pi}\,(\eta^{\a\b}k^2 - k^{\a}k^{\b}) \left[ 1 - 6m^2  \int_{4m^2}^\infty  \frac{ds}{s\,(k^2+s)}\, \sqrt{1-\frac{4m^2}{s}}\right]\,.
\label{Pimtrace}
\end{align}
The first term in this last expression, in which numerator and denominator cancel and the sum rule (\ref{Tsumrule}) is used,
is the anomalous contribution independent of $m$, while the second term is the finite non-anomalous contribution, expected 
even classically for the trace when $m >0$, but which vanishes for $m=0$.

The expectation value of the commutator of two fermion stress-energy tensors can be written as a weighted spectral integral
\be
i\Big\lag \big[T^{\m\n}_f (t,x), T^{\a\b}_f(t',x')\big]\Big\rag 
= \frac{N}{12\pi} \big(\eta^{\m\n}\sq - \pa^\m \pa^\n\big) \big(\eta^{\a\b} \sq- \pa^\a \pa^\b\big) 
\, \int_0^{\infty} \!ds\, \varrho_T(s)\, D(t-t', x-x';s)
\label{TTcomm}
\ee
analogous to (\ref{jjcomm}). Here and for the remainder of \secref{sec:gravity}, $\sq$ will denote the
flat Minkowski space wave operator. Using the properties of the commutator function $D$ (\ref{Ddef})-(\ref{Dproperties}),
only the terms with an odd number of time derivatives survive at equal times $t=t'$, so that from
(\ref{Pim}) and the sum rule (\ref{Tsumrule}) we obtain the anomalous commutator expectation values 
\be
\Big\lag\big[T^{00}_f (t,x), T^{01}_f(t,x')\big]\Big\rag 
= \Big\lag\big[T^{11}_f (t,x), T^{01}_f(t,x')\big]\Big\rag =  \frac{iN}{12\pi}\,\pa_x^3\,\delta(x-x')
\label{TTanomcom}
\ee
for any $m$, all other equal time commutators vanishing. This is the expectation value of
the Schwinger term for the stress-energy tensor commutators in two dimensions \cite{Tomiya:1985br,Ebner:1987pg,Bertlmann:2000da},
which is independent of fermion mass.

When the fermion mass vanishes, we now compare the result (\ref{VacPolF0}) and (\ref{TTcomm}) with 
their counterparts in the bosonic theory. Because the boson stress-energy tensor (\ref{Tphi}) has
terms both quadratic and linear in $\varphi$, there will be both a one-loop and tree level contribution
to the $\lag TT\rag$ bosonic correlation function. The loop contribution which is of order $N^0$
gives rise to the boson quantum anomaly contribution which has the effect of shifting $N$ by one,
as discussed in the previous \secref{subsec:GravFunctInt}. In order to avoid this shift and
match the fermion loop to a boson tree diagram with a single fermion pair intermediate state,
we neglect the $\varphi$ loop, and compute only the tree level contribution from the terms linear 
in the $\varphi$ stress-energy tensor, consistent with (\ref{Ztree}).
For the correlation function (\ref{TPolf})
this amounts to considering only the term linear in the boson field, \textit{i.e.}
\be
T^{\m\n}_{\bPhi\,\rm lin} = \frac{N}{24 \pi}\, \big( \pa^{\m} \pa^{\n} - \eta^{\m\n} \sq \big)\, \varphi
= \sqrt{\frac{N}{12 \pi}}\ \Big( \pa^{\m} \pa^{\n} - \eta^{\m\n} \sq \Big)\, \bPhi
\label{linT}
\ee
which gives
\begin{align}
\Pi^{\m\n\a\b}_{\bPhi\, {\rm lin}}(k) &= i\int d^2 x\, e^{ik\cdot x}\,\lag {\cal T} T^{\m\n}_{B\ \rm lin} (x) T^{\a\b}_{B\ \rm lin}(0)\rag \nn
&= \left( \frac{N}{12\pi}\right) \big(\eta^{\m\n}k^2 - k^{\m}k^{\n}\big)\big(\eta^{\a\b}k^2 - k^{\a}k^{\b}\big)
\int d^2 x\, e^{ik\cdot x}\,i\lag {\cal T}\bPhi (x) \bPhi (0)\rag\nn
&= (\eta^{\m\n}k^2 - k^{\m}k^{\n})(\eta^{\a\b}k^2 - k^{\a}k^{\b})\, \frac{N}{12\pi k^2} 
\label{TPolB}
\end{align}
after use is made of the canonical normalization of the $\bPhi$ field in (\ref{canPhi}).
The result (\ref{TPolB}) coincides with (\ref{VacPolF0}). Likewise, if we compute the 
commutators appearing in (\ref{TTanomcom}) we obtain agreement at the operator level from the
equal time commutator function of the massless scalar $\varphi$ field, normalized according to
(\ref{phiprop}), neglecting any anomalous commutator of the quadratic $\varphi$ stress-energy tensor
itself, according to (\ref{Ztree}).

\subsection{Stress-Energy Tensor, Virasoro Algebra and Schwinger Term of Fermions}
\label{TFermion}

In order to determine the precise form of the fermion pairing into a boson related to the conformal anomaly in the operator
representation,  we return to the Fock space representation for the fermions introduced in \secref{sec:FockspaceChiral}, 
starting with a single fermion: $N=1$. Using the Dirac equation \eqref{eq:DiracFock}, we find 
\bes
\begin{align}
T^{00}_f = T^{11}_f&=\frac{i}{2}\left(\psi^\dag_+\pa_t\psi_+-\pa_t\psi^\dag_+\psi_++\psi^\dag_-\pa_t\psi_--\pa_t\psi^\dag_-\psi_-\right)
\label{T00F}\\
T^{01}_f = T^{10}_f &= \frac{i}{2}\left(\psi^\dag_+\pa_t\psi_+-\pa_t\psi^\dag_+\psi_+-\psi^\dag_-\pa_t\psi_-+\pa_t\psi^\dag_-\psi_-\right)
\end{align}
\ees
for the (unregularized) fermion stress-energy tensor. Upon inserting the Fock space expansion \eqref{psiop-simple},
normal ordering, and using the regularization described in \secref{sec:casimir} to subtract the zero-point energy in
the infinite domain $1/L \to0$, leading to the finite Casimir energy density \eqref{CasEnerDen}, we obtain (for $N=1$)
\bes
\begin{align}
T^{00}_f=T^{11}_f &=\frac{2\pi}{L^2}\sum_{n\in\Z}\left(\bL^{(+)}_n e^{-ik_nt}e^{ik_nx} 
+\bL^{(-)}_n e^{-ik_nt}e^{-ik_nx}\right) -\frac{\pi}{6L^2} \label{T00} \\
T^{01}_f=T^{10}_f &=\frac{2\pi}{L^2}\sum_{n\in\Z}\left(\bL^{(+)}_n e^{-ik_nt}e^{ik_nx}
-\bL^{(-)}_n e^{-ik_nt}e^{-ik_nx}\right)
\end{align}\label{Tfermion}
\ees
where the fermionic Virasoro generators are defined by \cite{Goddard:1986bp}
\begin{align}
\bL^{(\pm)}_n&=\sum_{q\in\Z_{\inv2}}\left(q-\frac{n}{2}\right):\!\!c^{(\pm)\dagger}_{q-n}c^{(\pm)}_q\!\!:
\ = \sum_{q\geq\inv2}\left(q-\frac{n}{2}\right)c^{(\pm)\dagger}_{q-n}c^{(\pm)}_q
-\sum_{q\leq-\inv2}\left(q-\frac{n}{2}\right)c^{(\pm)}_qc^{(\pm)\dagger}_{q-n}
 \label{ViragenF}
\end{align}
for the left and right moving fermions respectively. Note that the Virasoro generators satisfy $\bL^{(\pm)\dag}_n=\bL^{(\pm)}_{-n}$
and have zero vacuum expectation value $\lag0|\bL^{(\pm)}_n|0\rag=0$ due to normal ordering, which affects only $\bL^{(\pm)}_0$.

The commutator of two Virasoro generators yields the Virasoro algebra, \textit{cf.}\ \appref{sec:virasoro},
\begin{align}
\co{\bL^{(\pm)}_n}{\bL^{(\pm)}_{n'}}&=(n-n')\bL^{(\pm)}_{n+n'}+\frac{n(n^2-1)}{12}\,\d_{n,-n'}
\label{Virasoro}
\end{align}
for the right and left moving fermions separately. As in the case of the current moments $\rho_n^{(\pm)}$, 
normal ordering with respect to the fermion vacuum and its filled Dirac sea leads to an anomalous commutator,
the central term in the Virasoro algebra above. For $N$ fermions we have
\be
\bL^{(\pm)}_{n,N}\equiv \sum_{j=1}^N\bL^{(\pm),j}_{n}
 =\sum_{j=1}^N\sum_{q\in\Z_{\inv2}}\left(q-\frac{n}{2}\right):\!\!c^{(\pm),j\dagger}_{q-n}c^{(\pm),j}_q\!\!: 
 \label{LlargeN}
\ee
and hence
\be
\co{\bL^{(\pm)}_{n,N}}{\bL^{(\pm)}_{n',N}} = (n-n')\bL^{(\pm)}_{n+n',N}+\frac{N}{12} \,n(n^2-1)\d_{n,-n'}
\label{VirasoroN}
\ee
which now features a factor $N$ in the central extension. Converting this relation to position space
and taking account of the finite shift of the Casimir term in (\ref{T00}) gives the equal time commutator
\begin{align}
\co{T^{00}_f(t,x)}{T^{01}_f(t,x')}
&=-i\left(T^{00}_f(t, x)+T^{00}_f(t, x')+\frac{\pi}{3L^2}\right)\,\pa_x\,\d(x-x') \nn
&\quad + \frac i{12\pi}\,\pa_x\left(\pa_x^2+\frac{4\pi^2}{L^2}\right)\d(x-x') \nn
&=-i\Big(T^{00}_f(t,x)+T^{00}_f(t,x')\Big)\pa_x\,\d(x-x')+ \frac{iN}{12\pi}\,\pa_x^3\,\d(x-x')\,, 
\label{SchwT}
\end{align}
showing the relation to the Schwinger contact term. The sign of the Schwinger term here agrees with earlier 
work \cite{Tomiya:1985br,Ebner:1987pg} but apparently disagrees with \cite{Bertlmann:2000da}.

\subsection{Classical Scalar Condensate and Quantum Casimir Energy}
\label{sec:casimir}

In the axial case considered previously we remarked on the possibility of a classical condensate current
scaling with $N$. However, with no sources for this current we were free to set it and the expectation value
of $\chi$ to zero in the interval $[0,L]$. On the other hand in the gravitational case, in the finite interval $[0,L]$, 
$N$ massless fermions with anti-periodic boundary conditions have a finite Casimir energy density~\cite{Birrell:1982ix}
\be
\lag 0|T_{00\, f}|0\rag_R=-\frac{\pi N}{\,6L^2}
\label{CasEnerDen}
\ee
proportional to $N$. In the fermion representation this can be computed from the quantum stress-energy tensor 
of the fermions (\ref{SETF}) by introducing a cutoff in the sum over modes and subtracting the cutoff dependent 
contribution in the infinite $L$ limit, effectively setting the quantum zero point energy to be zero in that limit. For 
finite $L$ this subtraction leaves behind the finite energy density (\ref{CasEnerDen}) as the cutoff is 
removed \cite{Birrell:1982ix}. It may also be computed by $\z$-function methods, as follows  by substituting the
mode expansion (\ref{psiop-simple}) in (\ref{T00F}). We obtain the unrenormalized, infinite sum
\be
\lag 0|T_{00\, f}|0\rag = \lag 0|T_{11\, f}|0\rag = -\frac{2N}{L} \sum_{q \ge \inv2} \tilde k_q = -\frac{4\pi N}{L^2} \sum_{q \ge \inv2}\, q\,.
\label{infsum}
\ee
The generalized Riemann $\z$ function is defined by
\be
\z_R(s, a) = \sum_{n=0}^{\infty}\, (n + a)^{-s}\,, \qquad {\rm Re}(s) > 1
\label{Riemzeta}
\ee
which defines a function which can be analytically continued to $s=-1$, so the sum over half-integers in (\ref{infsum}) has
a finite part which can be defined by
\be
\sum_{n=0}^{\infty}\, \left(n + \tfrac{1}{2}\right)^{-s}\Big\vert_{s=-1}= 
\z_R\left(-1 , \tfrac{1}{2}\right) = -\frac{1}{2}\,B_2\left(\tfrac{1}{2}\right) = 
-\frac{1}{2}\left(\frac{1}{4} -\frac{1}{2} +\frac{1}{6}\right) = \frac{1}{24}
\label{Bern}
\ee
where $B_2(a) = a^2 -a + \inv6$ is the second Bernoulli polynomial.
Substituting (\ref{Bern}) into (\ref{infsum}) gives (\ref{CasEnerDen}).

In the boson representation the Casimir energy (\ref{CasEnerDen}) is a leading order in $N$ effect,
corresponding to a condensate $\bar \varphi$ and may be calculated from the stress-energy tensor of the boson field 
(\ref{Tphi}) by purely \textit{classical} means. To find the correct classical condensate field $\bar \varphi$ for
the periodically identified space on the finite interval $[0,L]$, we recognize first from (\ref{metconf}), (\ref{Gameffsig}) 
and (\ref{Sanomloc}) that $e^\varphi = e^{2 \sigma}$ may be thought of as the conformal factor that transforms a 
fixed fiducial metric $\bar g_{\m\n}$ to the metric $g_{\m\n}$ of interest. Then we note that the infinite
${\mathbb R}^2$ plane may be mapped to the cylinder by the following conformal transformation. Introducing polar 
coordinates $(r,\theta)$ in infinite Euclidean ${\mathbb R}^2$ gives
\be
d\bar\t^2 + d\bar x^2 = dr^2 + r^2 d\theta^2 = r_0^2 \,e^{2\eta} ( d\eta^2 + d\theta^2 )
\label{cylinder}
\ee
where $\eta = \ln (r/r_0)$ ranges from $-\infty$ to $\infty$, and $\theta \in [0, 2 \pi ]$, which describes a cylinder. 
Relabelling $\theta = 2 \pi x/L$ and analytically  continuing (\ref{cylinder}) to $\eta = 2 \pi i t/L$, $\bar \tau = i \bar t$ 
allows us to write the line element for the real Lorentzian  time cylinder as a conformal transformation of Lorentzian 
infinite flat spacetime by
\be
-dt^2 + dx^2 = e^{\bar \varphi} \,( - d\bar t^2  + d \bar x^2)
\label{confcyl}
\ee
with 
\be
\bar \varphi =-\frac{4i\pi t}{L}
\label{phibar}
\ee
after identifying $r_0 = L/2 \pi$. Therefore, taking the fixed fiducial metric to be that of infinite Lorentzian flat spacetime
to have vanishing energy density as before, the stress-energy tensor in the periodically identified domain $[0,L]$
may be computed by substituting the classical condensate $\bar\varphi$ of (\ref{phibar})
into $T^{\m\n}_B[\bar \varphi]$ of (\ref{Tphi}) to obtain
\begin{align}
\overline T_{\m\n} &\equiv T_{\m\n\, B}[\bar \varphi] = \frac{N}{96\pi}
\left(2\,\pa_\m\bar\varphi\,\pa_\n\bar\varphi - \eta_{\m\n}\eta^{\a\b}\pa_\a \bar\varphi\,\pa_\b \bar\varphi\right)\nn
& = - \frac{\pi N}{\,6L^2} \,\big(2\, \d^0_{\,\m}\,\d^0_{\,\n} + \eta_{\m\n}\big)
\label{Tcond}
\end{align}
which is a traceless stress-energy tensor, $\overline T^{\m}_{\ \m} = 0$ with $T_{00}[\bar \varphi] = \lag 0|T_{00\, f}|0\rag$ 
of (\ref{CasEnerDen}). 

Thus, the quantum Casimir energy of the fermions may be computed from the stress-energy tensor of the scalar boson 
$\varphi$, viewed as a \textit{classical condensate} with value (\ref{phibar}) obtained by a conformal transformation 
from infinite flat spacetime. Being a linear function of $t$, the condensate (\ref{phibar}) satisfies
\be
\sq \bar\varphi = 0
\label{sqphibar}
\ee
consistent with (\ref{phieom}) in a spacetime with zero curvature, $R=0$. If $L \rightarrow \infty$ for fixed $t$,
$\bar\varphi \to0$. For finite $L$ periodicity, the linear growth of (\ref{phibar}) without 
bound as $t \rightarrow \pm \infty$ is analogous to the linear time dependence of the winding modes 
(\ref{zeromodes}) in the usual chiral bosonization of the Schwinger model in states for which the 
background charges $Q_{\pm} \neq 0$. Indeed, (\ref{phibar}) should be viewed as a solution 
of (\ref{sqphibar}) only in the distributional sense, with source `charges' at $t= \pm \infty$. 

That the non-trivial conformal transformation (\ref{confcyl}) does indeed give rise to the equivalent of
the Chern-Simons number (\ref{NCS}) but for the Euler characteristic, \textit{viz.}
\be
Q_{\rm CS} = \frac{1}{4 \pi} \int \Omega^{\m} d\Sigma_{\m}
\label{QCS}
\ee
follows by direct calculation from (\ref{Omdef}) in the general conformal frame 
$e^a_{\ \m} = \exp(\s) \,\d^a_{\ \m}$ and $g^{\m\n} = \exp(-2\s)\, \eta^{\m\n}$, 
related to the flat fiducial metric of ${\mathbb R}^2$. Then (\ref{Omdef}) gives
\be
\Omega^{\m} = -2\, g^{\m\n}\, \pa_{\n} \s 
\ee
and 
\be
Q_{\rm CS}\big\vert_{{\mathbb R} \times {\mathbb S}^1}= -\frac{1}{2 \pi} \int g^{\m\n} \pa_{\n} \bar\s \, d \Sigma_\m 
= -\frac{i}{L} \int_0^L dx = -i 
\label{topcharge}
\ee
when evaluated for the conformal transformation $\bar \s = \bar\varphi/2 = -2\pi i t/L$ in the $(t, x)$ coordinates 
of the cylinder (\ref{confcyl}) on the periodic interval $x \in [0,L]$, for which $g^{00} = -1$ and $d\Sigma_0 = dx$. 

In this case the Chern-Simons charge corresponding to the conformal anomaly is one-to-one with the holonomic windings 
of the zweibein frame field $e^a_{\ \m}$ in the $SO(2) \simeq U(1)$ tangent space as $x$ varies over the interval $[0,L]$.
This is again the mapping ${\mathbb S}^1 \to {\mathbb S}^1$ with homotopy group $\Pi_1({\mathbb S}^1) = {\mathbb Z}$. 
The imaginary unit charge (\ref{topcharge}) is the result of analytically continuing the definition of the conserved 
topological Euler number and its corresponding secondary Chern-Simons form defined on Riemannian manifolds
to the pseudo-Riemannian signature metric (\ref{confcyl}) of Lorentzian spacetime. The physical interpretation
of this background conformal charge is that the fermion vacuum defined on the anti-periodically identified finite 
interval $[0,L]$ may be regarded as being filled uniformly with a classical scalar condensate (\ref{phibar}) of 
fermion pairs, whose density is determined by topology or global boundary conditions on ${\mathbb R} \times {\mathbb S}^1$, 
with a finite negative energy density and pressure $\r = p = - \pi N/6L^2$, equal to its quantum Casimir energy.

\subsection{The Scalar Boson of the Conformal Anomaly: Canonical Field}
\label{sec:canonical}

In the case of the electric current algebra and axial anomaly, the anomalous Schwinger term is the only 
non-trivial commutator, and being a c-number, the algebra of the Fourier components of the current 
density (\ref{rhocomm}) is easily mapped to the canonical algebra of a bosonic field by a simple rescaling 
with $\sqrt{n}$ through (\ref{andef})-(\ref{cancomm}), which becomes $\sqrt{nN}$ for $N$ fermions.
In the case of the stress-energy tensor and conformal anomaly, the Virasoro algebra is non-abelian 
already at the classical level, and the stress-energy tensor (\ref{Tphi}) or (\ref{TbPhi}) also contains 
terms quadratic in the scalar boson field $\varphi$ or $\bPhi$. As already remarked, these quadratic terms 
in $T^{\m\n}_{\ B}$ or $T^{\m\n}_{\bPhi}$ lead to the scalar boson field having its own conformal anomaly and 
shifting the coefficient of the central term by one unit. Note that the canonical bosonic representation (\ref{TbPhi})
makes it clear that the fermion stress-energy tensor does not have a simple homogeneous scaling with $N$
but rather is a sum of terms with different scalings. Finally the boson field has a classical imaginary 
expectation value or condensate $\bar \varphi$ of (\ref{phibar}) to account for the Casimir energy in the 
finite interval $[0,L]$, so that a simple Hermitian boson field construction of the kind found in the charge 
current case also cannot be appropriate here.

These apparent roadblocks to a construction of a boson-fermion operator correspondence
are related, and are all removed by identification of the boson field with that part of the fermion
stress-energy tensor that scales with $\sqrt{N}$. We note that although from (\ref{LlargeN}) 
one might expect the Virasoro generators $\bL^{(\pm)}_{n,N}$ to scale linearly with $N$, and 
indeed the c-number contribution in (\ref{T00}) corresponding to the Casimir energy (\ref{CasEnerDen})  
and the condensate $\bar\varphi$ does scale linearly with $N$, the commutator in (\ref{VirasoroN}) 
quadratic in $\bL^{(\pm)}_{n,N}$ itself scales at most linearly with $N$. Thus the quantum 
operator part of the Virasoro generators that give rise to the central term in  (\ref{VirasoroN}) 
scale only as $\sqrt{N}$, just as the currents do in (\ref{j0j1phiN}), and it is this part that
we can identify with the part of the boson stress-energy tensor (\ref{TbPhi}) linear in the canonically
normalized $\bPhi$ field that scales in the same way as $\sqrt{N}$.

In order to make this identification of terms in the fermion and boson stress-energy tensor that
scale as $\sqrt{N}$, we need first to subtract the condensate part which scales linearly
with $N$. Therefore let us first define the quantum field $\hat \varphi$ by the shift
\be
\varphi = \bar\varphi + \hat\varphi
\ee
with $\bar\varphi$ the condensate (\ref{phibar}), and substitute this 
into the stress-energy tensor $T_{\m\n\, B}$ for the boson field (\ref{Tphi}) obtaining
\begin{align}
T_{\m\n\, B}[\bar\varphi + \hat\varphi]  &= \overline T_{\m\n\, B}
+ \frac{N}{48\pi}\left( 2\, \pa_\m\pa_\n\hat\varphi  + 2\, \dot{\bar\varphi}\,\d^0_{\,(\m}\pa_{\n)} \hat\varphi 
+ \eta_{\m\n}\,\dot{\bar\varphi}\,\dot{\hat\varphi} \right) \nn
& \quad +\, \frac{N}{96\pi}\left(2\, \pa_\m\hat\varphi\,\pa_\n\hat\varphi
- \eta_{\m\n}\,\pa_\l\hat\varphi\,\pa^\l\hat\varphi \right) 
\label{scalarT}
\end{align}
where we have used the fact that $\sq \bar\varphi = \sq \hat \varphi = 0$, and $\overline T_{\m\n\, B}$ 
is the condensate contribution (\ref{Tcond}). Thus at finite $L$ in the presence of the scalar condensate 
$\bar\varphi$ the term linear in the quantum field $\hat \varphi$ becomes
\be
T_{\m\n\, B\,\rm lin} = \frac{N}{48\pi}\left( 2\, \pa_\m\pa_\n\hat\varphi  + 2\, \dot{\bar\varphi}\,\d^0_{\,(\m}\pa_{\n)} \hat\varphi 
+ \eta_{\m\n}\,\dot{\bar\varphi}\,\dot{\hat\varphi} \right)
\label{TBlin}
\ee
instead of (\ref{linT}). 

If we introduce boson operators
\begin{align}
\Phi_{\pm}(t,x)&=\sum_{n=1}^{\infty}\inv{\sqrt{4\pi n}}\left(\ba^{(\pm)}_{n}e^{-ik_n(t\mp x)}
+{\ba}^{(\pm)}_{-n}e^{ik_n(t\mp x)}\right)
+ \Phi^0_\pm(t, x) \,,\nn
\Phi^0_\pm(t, x)&=\inv{2\sqrt{\pi}}\bR_{\pm}+\frac{\sqrt{\pi}}{L}(t\mp x)\,\bQ_{\pm}
\label{Phican}
\end{align}
analogous to those introduced previously in the chiral bosonization (\ref{phiposn}),
and require the commutation relations
\be
[\ba^{(\pm)}_n,\ba^{(\pm)}_{-n'}]=\sgn{n}\, \d_{n,n'}\,, \qquad [\bR_{\pm},\bQ_{\pm}]=i 
\label{canorm}
\ee
then $\Phi_+ + \Phi_-$ is a canonically normalized scalar field independent of $N$. 
As we shall see shortly, it is important that we do not assume that $\Phi_{\pm}$ is Hermitian,  
so that unlike in the chiral case, $\ba_{-n} \neq \ba^\dag_n$. Because of the normalization
in the anomaly action (\ref{Sanomloc}), the quantum scalar field $\hat \varphi$ is
related to the canonically normalized scalar field defined through (\ref{Phican}) by
\begin{align}
\hat \varphi &= \sqrt{\frac{48\pi}{N}} \, \Big( \Phi_+ + \Phi_-\Big)\nn
&= \sqrt{\frac{12}{N}}\,\sum_{\pm}\left\{\sum_{n\neq 0} \inv{\sqrt{|n|}}\, \ba^{(\pm)}_{n} e^{-ik_n(t\mp x)}
+ \bR_{\pm}+\frac{2\pi(t\mp x)}{L}\,\bQ_{\pm}\,\right\}         \,.
\label{twophis}
\end{align}
Thus it is clear that the terms linear in $\hat \varphi$ in the anomaly boson energy-momentum
tensor (\ref{scalarT}) are proportional to $N/\sqrt N = \sqrt N$ and dominate in the large $N$
limit over the terms quadratic in $\hat \varphi$ in (\ref{scalarT}) which are of order $N/(\sqrt N)^2=1$.
The $\sqrt N$  leading order terms linear in $\hat\varphi$ in  (\ref{scalarT}) give
\begin{align}
T^{00}_{\ B,\rm lin} \pm T^{01}_{\ B,\rm lin} &= \frac{N}{24\pi} \left(\pa_t - \frac{2\pi i}{L}\right) 
\big(\pa_t \mp \pa_x\big) \,\hat \varphi\nn
&= -\frac{4\pi}{L^2} \sqrt{\frac{N}{12}} \left\{ \sum_{n\neq 0} \frac{n(n+1)}{\!\!\sqrt{|n|}}  \ba^{(\pm)}_{n} e^{-ik_n(t\mp x)}
+ i \bQ_{\pm}\right\}
\label{Tlin}
\end{align}
after substituting for $\bar\varphi$ and $\hat \varphi$ from (\ref{phibar}) and (\ref{twophis}) respectively.
Comparing (\ref{Tlin}) to the corresponding fermion terms from (\ref{Tfermion}) we find that 
\be
\ba^{(\pm)}_n = -\frac{\!\!\sqrt{12\,|n|}}{n(n+1)}\ \frac{\bL^{(\pm)}_{n,N}}{\sqrt N\ }
 \qquad \forall\  n\neq 0, -1
\label{aLrel}
\ee
to leading order in $N\gg 1$.  In this limit the shift of $N$ by one unit and the quantum contributions
of the quadratic terms in the $\bPhi$ stress-energy tensor (\ref{TbPhi}) can be neglected, corresponding
at the operator level to the neglect of the non-abelian terms in the Virasoro algebra. Indeed
since $\ba^{(\pm)}_n$ are canonically normalized independent of $N$, under the identification 
(\ref{aLrel}), $\bL^{(\pm)}_{n,N}$ scales as $\sqrt{N}$ and the classical non-abelian terms in 
(\ref{VirasoroN}) are suppressed relative to the central term scaling linearly in $N$. Thus in this 
limit the canonically normalized boson field Fock space operators are given by the
Virasoro generator moments of the fermion stress-energy tensor, in analogy with
the corresponding identification of the chiral boson with the moments of the electromagnetic 
current densities for $N$ fermions in (\ref{Nandef}). 

Following the discussion at the end of \secref{subsec:GravFunctInt}, the relation (\ref{aLrel})
can also be regarded as an exact relation identifying the canonical boson field operator
for any $N$, provided that all Wick contractions of the $\bPhi$ field leading to loop diagrams 
are neglected. This follows from the fact that the effective loop expansion parameter is $\hbar/N$
and neglecting loops in the boson tree effective action $S_{\rm eff}$ of (\ref{Ztree})
corresponds precisely to neglecting the non-anomalous commutator of the Virasoro generators
(\ref{VirasoroN}).

The $n=0, -1$ modes are not defined by (\ref{aLrel}). Together with the $n=+1$ Virasoro generator, the
$\bL_{n, N}^{(\pm)}$ for $n=0, \pm 1$ form an $SL(2,\mathds{C})$ global sub-algebra with vanishing
central term in (\ref{VirasoroN}). Hence it clearly is not possible to identify these generators with
boson particle modes. They are instead analogous to the collective coordinate or coherent $n=0$
total charge mode in the chiral bosonization of \secref{sec:FockspaceChiral}, and must hence be treated 
separately. Indeed these modes are related to the condensate $\bar \varphi$, since changing its sign
$\bar\varphi\to \bar\varphi^* =-\bar\varphi$, which gives the same condensate energy density
has the effect of shifting the subtlety from the $n=-1$ mode to the $n=+1$ mode.

An $SL(2,\mathds{C})$ transformation on the $n=\pm 1$ Fock space bosonic operators takes the form
\be
\left(\begin{array}{c}\ba_1\\ \ba_{-1}\end{array}\right) = 
\left(\begin{array}{cc} A & B\\ C& D\end{array}\right)
\left(\begin{array}{c}\ba_1'\\ \ba_{-1}'\end{array}\right) 
 \,,\qquad AD-BC =1
\ee
and preserves the canonical commutation relation (\ref{canorm}). The parameters $A, B, C, D$ 
satisfying $AD-BC = 1$ can always be chosen to depend on $n$ in such a way that the limit
\be
\lim_{n\to -1}\left\{(1+ n) \ba_n\right\} =\lim_{n\to 1}\left\{(1-n)\ba_{-n}\right\} = \lim_{n\to 1}\left\{(1-n)(C  \ba'_n + D \ba'_{-n})\right\}
\ee
is finite, \textit{e.g.}
\be
A=B + \frac{ (1-n)}{\a} = \frac{\b}{2} + \frac{ (1-n)}{\a} \,,\qquad\qquad C=D = \frac{\a}{1-n}
\ee
for any $\a, \b$ remaining finite as $n \to 1$. Then (\ref{aLrel}) gives
\bes
\begin{align}
\lim_{n\to -1}\left\{(1+ n)\, \ba_n^{(\pm)}\right\} &= \a^{(\pm)} \, (\ba^{(\pm)\,\prime}_1 + \ba^{(\pm)\,\prime}_{-1})
= -\sqrt{\frac{12}{N}}\,  \bL_{-1, N}^{(\pm)}   \\
\lim_{n \to 1} \left\{(1+n)\,\ba_n^{(\pm)}\right\} &= \b^{(\pm)}\,(\ba^{(\pm)\,\prime}_1 + \ba^{(\pm)\,\prime}_{-1})
= -\sqrt{\frac{12}{N}}\,  \bL_{1, N}^{(\pm)} 
\end{align}
\label{aone}\ees
to leading order in large $N$. These expressions remain finite in the $n \to \mp 1$ limits respectively provided 
the $\ba'_{\pm 1}$ are finite. Since by this (singular) $SL(2,\mathds{C})$ transformation $\bL_{-1, N}^{(\pm)}$ 
becomes proportional to $\bL_{1, N}^{(\pm)}$, they commute with each other. 

The $n=0$ constant mode of the boson is identified with $\bL_{0, N}^{(\pm)}$ according to (\ref{Tfermion}) and (\ref{Tlin})
\be
\bQ_{\pm} = i \,\sqrt{\frac{12}{N}}\,\bL_0^{(\pm)}
\label{Qzero}
\ee
to leading order in large $N$. Since the Virasoro generators are Hermitian this implies that the 
$\bQ_{\pm}$ are anti-Hermitian. Note also that with the identification (\ref{aLrel})
\be
\ba^{(\pm)\dag}_n=\frac{n-1}{n+1}\,\ba^{(\pm)}_{-n}
\label{ba-n}
\ee
the non-zero modes of scalar boson field $\hat\varphi$ also are not Hermitian, due to the
finite imaginary shift of the condensate $\bar\varphi$. In the continuum limit $L\rightarrow \infty$
where the condensate $\bar\varphi \to 0$, equivalent to  $|n|  \gg 1$ in (\ref{ba-n}), the quantum
field $\hat\varphi$ becomes Hermitian.  

As with the $n \neq 0, \pm 1$ modes, the identifications (\ref{aone}) and (\ref{Qzero}) may be regarded 
as exact for any finite $N$, provided the loops of the quantum $\bPhi$ field are neglected with respect 
to tree diagrams, corresponding to neglect of all the mutual commutators of $\bL_0^{(\pm)}, \bL_{\pm1, N}^{(\pm)}$.

\subsection{Intermediate Pair States of \texorpdfstring{$\lag TT\rag$}{<TT>}}
\label{sec:IntermedTT}

We have seen in \secref{sec:correlation-gravity} that the one-loop $\lag TT\rag$-correlator of fermionic 
quantum fields is given exactly by the tree graph of stress-energy tensors in terms of the scalar $\varphi$ in
the local form of the Polyakov action corresponding to the conformal anomaly \eqref{Sanomloc}.
Thus we expect that the intermediate states of fermion bilinear pairs described by $\varphi$ will reproduce
the fermion intermediate states of the $\lag TT\rag$-correlator, similarly to the situation in \secref{sec:JJ}.

Considering first the case of $N=1$, and inserting once more a complete set of the general on-shell 
two-fermion states \eqref{twofermion}, this time into the polarization \eqref{VacPolF0}, requires the evaluation of
\be
\lag 0 | T^{\m\n}_f(t,x) | q,s; q',s'\rag = \exp \left[-i\, t\,E(q , q' ) + i\,x\, p_{s,s'}(q, q' )\right]\, \lag 0 | T^{\m\n}_f (0,0)  | q,s; q', s'\rag
\label{Tmatrix}
\ee
where energy and momentum are given in \eqref{Eandp}. Since the two chiralities do not mix for massless fermions, 
we only have contributions for $s=s'$, \textit{i.e.}
\bes
\begin{align}
\lag 0 | T^{00}_f (0,0)  | q, s ;q', s'\rag &= \lag 0 | T^{11}_f (0,0)  | q, s ;q', s'\rag =\frac{1}{2L} (\tilde k_{q} - \tilde k_{q'})\, \d_{ss'}\\
\lag 0 | T^{01}_f (0,0)  | q, s ;q', s'\rag &= \lag 0 | T^{10}_f (0,0)  | q, s ;q', s'\rag =\frac{s}{2L} (\tilde k_{q} - \tilde k_{q'})\, \d_{ss'}\,.
\end{align}
\ees
Thus, defining as before $n = q + q' \ge 1$, so that
\be
\tilde k_{q} - \tilde k_{q'} = \frac{4\pi}{L} \left(q - \frac{n}{2}\right)
\ee
we find
\begin{align}
&\lag 0 | T^{00}_f(t,x)\,  T^{00}_f(t',x') | 0\rag
=\sum_{s, s' = \pm}\hspace{-3mm} 
\sum_{\quad q_{s}, q'_{s'} \ge \inv 2} \lag 0 | T^{00}_f(t,x) | q, s ;q', s'\rag \lag  q, s; q',s'\ | T^{00}_f(t',x') | 0\rag\nn
& = \left(\frac{2\pi}{L^2}\right)^2\sum_{s = \pm} \sum_{n =1}^{\infty} \sum_{q= \inv2}^{n - \inv2}\left(q-\frac{n}{2}\right)^2 \exp[-ik_n (t-t')] \, 
\exp[is k_n (x-x')]+\left(\frac{\pi}{6L^2}\right)^2\nn
& = \frac{4\pi^2}{L^4} \sum_{n=1}^{\infty}\frac{n(n^2-1)}{12}e^{-ik_n (t-t')} \left[ e^{ik_n (x-x')}  
+ e^{-ik_n (x-x')}\right]+\left(\frac{\pi}{6L^2}\right)^2\nn
& =  \lag 0 | T^{11}_f(t,x)\,  T^{11}_f(t',x') | 0\rag = \lag 0 | T^{01}_f(t,x)\,  T^{01}_f(t',x') | 0\rag 
\label{T00T00F}
\end{align}
where we have used
\be
\sum_{q =\inv2}^{n - \inv2} \left(q-\frac{n}{2}\right)^2 = \frac{n(n^2-1)}{12}\,.
\ee
The constant shift is due to the (square of the) Casimir energy \eqref{CasEnerDen}.
Likewise we obtain
\begin{align}
 \lag 0 | T^{00}_f(t,x)  T^{01}_f(t',x') | 0\rag
&= \left(\frac{2\pi}{L^2}\right)^2\sum_{s = \pm} s \sum_{n =1}^{\infty} 
 \sum_{q= \inv2}^{n - \inv2}\left(q-\frac{n}{2}\right)^2 \exp[-ik_n (t-t')] \, \exp[is k_n (x-x')]\nn
&=\frac{4\pi^2}{L^4} \sum_{n=1}^{\infty}\frac{n(n^2-1)}{12}e^{-ik_n (t-t')} \left[ e^{ik_n (x-x')}  - e^{-ik_n (x-x')}\right]\,
\label{T00T01F}
\end{align}
for the mixed $\lag T^{00}T^{01}\rag = \lag T^{11}T^{01}\rag$ matrix element with a complete set of
arbitrary two-fermion states inserted.

On the other hand from the bosonic viewpoint the only intermediate states which should contribute to these 
matrix elements are the single boson states 
\be
| n, \pm\rag = \sqrt{\frac{n+1}{n-1}}{\ba}^{(\pm )\dag}_{n}|0 \rag = -\frac{\!\!\sqrt{12}}{n(n^2-1)}\ \frac{1}{\sqrt N}
\sum_{j=1}^N\sum_{q\in\Z_{\inv2}}\left(q+\frac{n}{2}\right):\!\!c^{(s),j\dagger}_{q+n}c^{(s),j}_q\!\!: | 0\rag
\qquad \forall \,n>1
\label{bosonint}
\ee
consisting of the coherent fermion pair described by (\ref{ViragenF}) and (\ref{aLrel}).
The normalization factor is required due to \eqref{ba-n} so that $\lag n, \pm | n, \pm\rag =1$
is properly normalized. The only term needed in the tree diagram is the term
linear in the quantum field $\hat \varphi$ given by (\ref{TBlin}), for which
\bes
\begin{align}
T^{00}_{B\,,\rm lin} &= T^{11}_{B\,,\rm lin} = \frac{N}{48 \pi} (2\, \ddot{\hat\varphi} + \dot{\bar\varphi}\, \dot{\hat\varphi})
= \frac{N}{24 \pi} \left(\pa_t - \frac{2\pi i}{L} \right)\pa_t \hat \varphi \,,\\
T^{01}_{B\,,\rm lin} &= -\frac{N}{48 \pi} (2\, \pa_t\pa_x{\hat\varphi} + \dot{\bar\varphi}\, \pa_x\hat\varphi)
= -\frac{N}{24 \pi} \left(\pa_t - \frac{2\pi i}{L} \right)\pa_x\hat \varphi
\,.
\end{align}
\ees
Substituting then (\ref{twophis}) we obtain
\bes 
\begin{align}
\lag 0 | T^{00}_{B\,,\rm lin} (t,x)  | n, s\rag &= \frac{1}{2\pi} \sqrt{\frac{N}{12n}} \left(-ik_n - \frac{2\pi i}{L}\right) (-ik_n)\, 
e^{-ik_n(t-s x)} \,\sqrt{\frac{n-1}{n+1}}\nn
& = -\frac{2\pi}{L^2}\sqrt{\frac{N}{12}}\sqrt{n(n^2-1)}\, e^{-ik_n(t-s x)} \,,\\
\lag 0 | T^{01}_{B\,,\rm lin} (t,x)  | n, s\rag &= -\frac{1}{2\pi} \sqrt{\frac{N}{12n}} 
\left(-ik_n - \frac{2\pi i}{L}\right) (is k_n) e^{-ik_n(t-s x)} \,\sqrt{\frac{n-1}{n+1}}\nn
& = -\frac{2\pi s}{L^2}\sqrt{\frac{N}{12}}\sqrt{n(n^2-1)}\, e^{-ik_n(t-s x)}
\,.
\end{align}
\ees
Thus from the linear terms in the boson stress-energy tensor we find the matrix elements and  intermediate state sum
\begin{align}
&\lag 0 | T^{00}_{B\,,\rm lin}(t,x)\,  T^{00}_{B\,,\rm lin} (t',x') | 0\rag 
=\sum_{s=\pm}\sum_{n=1}^{\infty}\, \lag 0 | T^{00}_{B\,,\rm lin} (t,x)  | n, s\rag \lag n,s | T^{00}_{B\,,\rm lin} (t',x')  | 0\rag\nn
&= \frac{4\pi^2N}{L^4} \sum_{n=1}^{\infty}\frac{n(n^2-1)}{12}e^{-ik_n (t-t')} \left[ e^{ik_n (x-x')}  + e^{-ik_n (x-x')}\right]
+\left(\frac{\pi N}{6L^2}\right)^2\nn
& =  \lag 0 | T^{11}_{B\,,\rm lin}(t,x)\,  T^{11}_{B\,,\rm lin} (t',x') | 0\rag = \lag 0 | T^{01}_{B\,,\rm lin} (t,x)\,  T^{01}_{B\,,\rm lin} (t',x') | 0\rag 
\end{align}
while
\begin{align}
\lag 0 | T^{00}_{B\,,\rm lin}(t,x)  T^{01}_{B\,,\rm lin}(t',x') | 0\rag
&=\sum_{s=\pm}\sum_{n=1}^{\infty} \lag 0 | T^{00}_{B\,,\rm lin} (t,x)  | n, s\rag \lag n,s | T^{01}_{B\,,\rm lin} (t',x')  | 0\rag\nn
&=\frac{4\pi^2N}{L^4} \sum_{n=1}^{\infty}\frac{n(n^2-1)}{12}e^{-ik_n (t-t')}\! \left[ e^{ik_n (x-x')}  - e^{-ik_n (x-x')}\right]
\end{align}
which coincide with the two-fermion intermediate state results of (\ref{T00T00F}) and (\ref{T00T01F}) respectively.

Thus the results for the $\lag TT\rag$ matrix elements of summing over a complete set of two-fermion states 
(\ref{twofermion}) in the fermion loop is identical to that obtained by saturating the intermediate state sum by the particular
fermion pair state represented by the coherent state boson (\ref{bosonint}) and (\ref{aLrel}) with only
the linear term in the boson stress-energy tensor (\ref{TBlin}) in a tree amplitude. This is exactly analogous to the chiral 
boson coherent intermediate state in the $\lag jj\rag$ correlator of Sec. \ref{sec:JJ}, computed in the tree
amplitude of Fig. \ref{fig:JJ}.

By either method the results for the stress-energy tensor correlator may be written in terms of the free scalar function 
$G_0^>$ of (\ref{G0def}) in the form
\bes
\begin{align}
\lag 0 | T^{00}(t,x)\,  T^{00}(t',x') | 0\rag &= \frac{-iN}{12\pi} \left(\pa_x^4 + \frac{4\pi^2}{L^2}\pa_x^2\right)G_0^>(t-t', x-x')\\
\lag 0 | T^{00}(t,x)\,  T^{01}(t',x') | 0\rag &= \frac{iN}{12\pi} \left(\pa_t\pa^3_x + \frac{4\pi^2}{L^2}\pa_t\pa_x\right)G_0^>(t-t', x-x')
\end{align} \label{TTall}
\ees
so that the only equal time commutator which is non-zero is
\begin{align}
\lag 0 | \left[T^{00}(t,x)\,  T^{01}(t,x')\right] | 0\rag &= \frac{iN}{12\pi} \left(\pa_x^3 + \frac{4\pi^2}{L^2}\pa_x\right)
\left(\frac{1}{L} \sum_{n \in {\mathbb Z}} e^{ik_n(x-x')} \right)\nn
& =\frac{iN}{12\pi} \left(\pa_x^3 + \frac{4\pi^2}{L^2}\pa_x\right)\d(x-x')\,,
\end{align}
the $n=0$ constant mode vanishing under the derivatives. This is in agreement with the Schwinger term
computed in (\ref{SchwT}).

In the limit of infinite $L$ we may also write the results (\ref{TTall}) in the covariant form
\be
i\lag 0 | T_{\m\n}(t,x)\,  T_{\a\b}(t',x') | 0\rag \to \frac{N}{12\pi} \Big(\eta_{\m\n} \sq - \pa_{\m}\pa_{\n}\Big)\left(\eta_{\a\b} \sq - \pa_{\a}\pa_{\b}\right)
G_0^>(t-t', x-x')
\ee
in agreement with (\ref{VacPolF0}), when time-ordering is taken into account and the Wightman function $G_0^>$ is
replaced by the Feynman massless boson propagator (\ref{FeynG}) whose Fourier transform is $1/k^2$.

Since the time-ordered Feynman propagator is the analytic continuation of the Euclidean propagator,
we can express the Euclidean stress-energy tensor correlation function in the form
\be
\big\lag T^{\m\n}(x)\,  T^{\a\b}(y) \big\rag_E = \frac{N}{12\pi} \Big(\d^{\m\n} \pa^2 - \pa^{\m}\pa^{\n}\Big)\left(\d^{\a\b} \pa^2 - \pa^{\a}\pa^{\b}\right)
G_{\bPhi}(|x-y|)\Big\vert_{M \to 0}
\label{EuclTTG}
\ee
where
\be
G_{\bPhi}(|x|) = \int \frac{d^2P}{(2\pi)^2} \, \frac{e^{iP\cdot x}}{P^2 + M^2} = \frac{1}{2\pi} \, K_0\big(M|x|\big) 
\to -\frac{1}{2\pi} \, \ln \big(M |x|\big) + \text{const.}
\label{Euclprop}
\ee
is the Euclidean propagator for a massive scalar field in $d=2$ dimensions, satisfying
\be
(-\pa^2 + M^2) \, G_{\bPhi}(|x|) \to - \pa^2 \, G_{\bPhi}(|x|) =  \d^{(2)}(x)\,.
\label{propeq}
\ee
Here $x$ and $y$ denote Euclidean $2$-vectors with norm $|x| =   \sqrt{x^{\m}x^{\m}}$, and $\pa^2 \equiv \d^{\m\n} \pa_{\m}\pa_{\n}$ denotes
the Laplacian. The mass $M$ is inserted as an infrared regulator. In the limit $M\to 0$ the propagator (\ref{Euclprop}) becomes a logarithm 
and the mass dependence drops out under the derivatives in (\ref{EuclTTG}) or (\ref{propeq}), so that the limit to zero mass can be safely taken. 
Thus in Euclidean space the correlation function of two stress-energy tensors of massless fermions (or in fact any conformal field in $d=2$) becomes
\be
\big\lag T_{\m\n}(x)\,  T_{\a\b}(y) \big\rag_E = -\frac{N}{24\pi^2\!}\ \Big(\d_{\m\n} \pa^2 - \pa_{\m}\pa_{\n}\Big)\Big(\d_{\a\b} \pa^2 - \pa_{\a}\pa_{\b}\Big)
 \, \ln \big(|x-y|\big)
\label{EuclTT}
\ee
These results show the complete equivalence between the fermion loop and the bosonic tree calculations of all
components of the $\lag TT\rag$ correlator, where only the linear term (\ref{TBlin}) of the boson stress-energy tensor is
retained, after the finite shift in the finite spatial interval $[0,L]$. This correspondence between fermion loops
and bosonic trees can be extended to arbitrary numbers of stress-energy tensor correlations in Euclidean space next.

\section{Stress Tensor Correlators: Fermion Loops and Scalar Trees}
\label{sec:correlation-functions}
In \secref{sec:correlation-gravity} we showed that the one-loop correlation function $\lag TT\rag$ of
$N$ fermions can be represented by the simple two-point linear tree diagram illustrated in \figref{fig:TT}.
The intermediate states of fermion pairs in the one-loop correlation function are exactly the
correlated pairs of the single boson (\ref{bosonint}). In the chiral bosonization of Sec. \ref{sec:FockspaceChiral}
there are no connected diagrams of correlation functions with more than two currents, so that the mapping
of fermion to boson intermediate states needs to be checked only in the case of $\lag jj\rag$. For the stress
tensor on the other hand, correlation functions of arbitrary numbers of stress-energy tensors appear, due to
the essentially non-abelian, non-linear coupling of the metric to matter. The purpose of this section is to 
demonstrate that the correlation functions of arbitrary numbers of stress-energy tensors $\lag TT\dots T\rag$
of the fermions at one-loop order are mapped to scalar linear tree diagrams with exactly the same
boson intermediate states as found for the two-point function in Sec. \ref{sec:IntermedTT}.

\subsection{\texorpdfstring{$\lag TT\dots T\rag$}{<TT...T>} Correlators: Ward Identities}
\label{sec:arbitraryTs}

\begin{figure}[ht]
\begin{subfigure}{.5\textwidth}
\centering
\includegraphics[width=0.4\columnwidth]{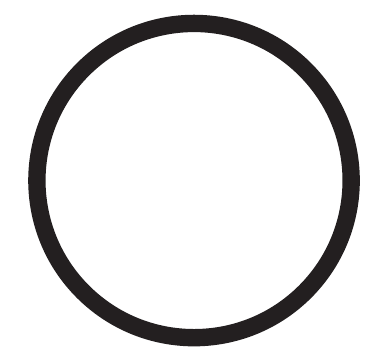}
\caption{The fermion one-loop effective action (\ref{GameffF}), in a background gravitational field.}
\label{fig:fermionSeff}
\end{subfigure}
\hspace*{1cm}
\begin{subfigure}{.4\textwidth}
\centering
\includegraphics{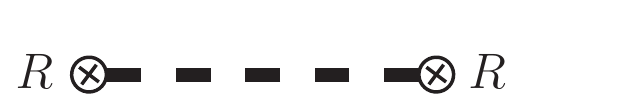}
\caption{The scalar tree effective action, with the propagator in a background gravitational field represented by a bold dashed line.}
\label{fig:scalarSeff}
\end{subfigure}
 \caption{Effective action for fermion loops and scalar trees.}
\end{figure}

To establish this equivalence between fermion loops and scalar trees for arbitrary $n$-point functions of
the stress-energy tensor, we shall make extensive use of the Ward identities these correlation functions obey.
Formally computing the functional integral over the $N$ massless fermion fields in a general
fixed background metric $g_{\m\n}$ as in \secref{subsec:GravFunctInt} gives the effective action
\be
S_{\rm E}[g]= \hbar N\,\text{Tr}\ln\!\big(\!-i\g^aE_a^{\ \m}\!\!\stackrel{\leftrightarrow}{\ \nabla_\m}\big)
 \label{GameffF}
\ee
illustrated in Fig. \ref{fig:fermionSeff} after continuation to Euclidean space. On the other hand, computing 
the effective action explicitly yields the non-local Polyakov action (\ref{Sanom}), or by (\ref{Ztree}), the
local form (\ref{SanombPhi}) in terms of the canonically normalized scalar $\bPhi$. The one-loop fermion 
and boson tree versions of the Euclidean effective action are illustrated in Figs. \ref{fig:fermionSeff} and \ref{fig:scalarSeff}
respectively. The $\hbar N$ loop factor in (\ref{GameffF}) and Fig.  \ref{fig:fermionSeff} is taken into account in (\ref{SanombPhi})
by $(\sqrt{\hbar N})^2$ from the two factors  of the quantum anomaly coefficient in the source of $\sq \bPhi$ from each end
of the tree diagram in Fig. \ref{fig:scalarSeff}. 

The one-point and general $n$-point gravitational vertex functions are defined in Euclidean space by
\bes
\begin{align}
\G^{\m\n}(x) &= \big\lag T^{\m\n}(x)\big\rag=\frac{2}{\sqrt{g(x)}}\ \var{S_E[g]}{g_{\m\n}(x)} \label{Tn1}\\
\G^{\m_1\n_1\dots\m_n\n_n}(x_1,\ldots,x_n) &= \frac{2^n}{\sqrt{g(x_1)}\dots\sqrt{g(x_n)}}\ 
\frac{\d^{n}S_E [g]}{\d g_{\m_1\n_1}(x_1)\dots \d g_{\m_n\n_n}(x_n)}\label{Tn}
\end{align}
\label{Gamdef}\ees
which are represented by connected one-loop diagrams in the fermion theory, because
$S_E$ is the generating functional of connected one-particle irreducible (1PI) diagrams. 
In (\ref{Gamdef}) and the remainder of this section we drop the subscript $_E$  on the correlation
functions, since we work entirely in Euclidean space ${\mathbb R}^2$.
The multiple variations with respect to the metric in (\ref{Tn}) produce connected 
$n$-point functions of fermion stress-energy tensors at different points, 
as well as local contact terms from varying the explicit dependence of the stress-energy tensor
itself upon the metric, 
\begin{align}
\G^{\m_1\n_1\dots\m_n\n_n}(x_1,\ldots,x_n) &= \big\lag T^{\m_1\n_1}(x_1)\dots T^{\m_n\n_n}(x_n)\big\rag_c\nn
&\quad +\  \frac{2}{\sqrt{g(x_1)}} \Big\lag\frac{\d T^{\m_2\n_2}(x_2)}{\d g_{\m_1\n_1}(x_1)}\ 
T^{\m_3\n_3}(x_3)\dots T^{\m_n\n_n}(x_n)\Big\rag_c + \dots 
\label{Gamall}
\end{align}
where the ellipsis contains all possible multiple local variations of the stress-energy tensor insertions.
If all the points are distinct, $x_i \neq x_j$ for all $i\neq j$, then these latter contact terms are absent, and
$\G^{\m_1\n_1\dots\m_n\n_n}$ is simply the connected Euclidean correlation function (denoted simply by 
the subscript $c$) of $n$ stress-energy tensors at $n$ different points in the first line of (\ref{Gamall}). 

Since $S_E[g]$ is a scalar invariant which incorporates the conformal anomaly, the product 
$\sqrt{g(x)}\lag T^{\m\n}(x)\rag$ satisfies the local conservation and trace identities
\bes
\begin{align}
\sqrt{g}\ \nabla_{\n}\big\lag T^{\m\n}\big\rag &= \pa_{\n} \big(\sqrt{g} \,\lag T^{\m\n}\rag\big) 
+ \G^{\m}_{\ \l\n}\big(\sqrt{g}\,\lag T^{\l\n}\rag\big) = 0\label{TIdents1}\\
g_{\m\n}\, \big(\sqrt{g}\, \big\lag T^{\m\n}\big\rag\big) &=\frac{N}{24\pi}\, \sqrt{g}\,R
\label{TIdents}
\end{align}
\ees
in an arbitrary background metric $g_{\m\n}(x)$, where  $\G^{\m}_{\ \l\n}$ is the Christoffel connection.
By varying these fundamental identities multiple times one arrives at the Ward identities satisfied 
by the general $n$-point correlation functions \eqref{Gamdef}, which when evaluated finally in the flat metric
$g_{\m\n} = \bar g_{\m\n} $ are
\bes
\begin{align}
\pa_{\n_1}\bar\G^{\m_1\n_1\dots\m_n\n_n}(x_1,\ldots,x_n)
&=-\left(\sum\limits_{i=2}^n\d^{(2)}(x_1-x_i)\,\pa^{\m_1}_{x_i}\!\right)\, \bar\G^{\m_2\n_2\dots\m_n\n_n}(x_2,\ldots,x_n)
\label{WI1n} \\
\d_{\m_1\n_1}\bar\G^{\m_1\n_1\dots\m_n\n_n}(x_1,\ldots,x_n)&= -2\left(\sum\limits_{i=2}^n\d^{(2)}(x_1-x_i)\right)
\bar\G^{\m_2\n_2\dots\m_n\n_n}(x_2,\ldots,x_n) \nn
&\quad +\ \frac{N}{24\pi}\ 
\frac{2^{n-1}\d^{n-1}\big(\sqrt{g(x_1)}R(x_1)\big)}{\d g_{\m_2\n_2}(x_2)\ldots\d g_{\m_n\n_n}(x_n)}\bigg\vert_{g_{\m\n} = \d_{\m\n}}
\label{WI2n}
\end{align}
\label{WardIds}\ees
where we have used Cartesian coordinates in which the flat space metric $\bar g_{\m\n} = \d_{\m\n}$.
Eqs.( \ref{WardIds}) are distributional identities that vanish as any point is removed to infinity (so that total derivatives of 
$\d$-function  terms may be dropped). It is important to note that the Ward identities we used for our derivations enforce covariance 
at the expense of the trace anomaly.

As an example let us derive the $n=2$ case of \eqref{WI2n}. Using \eqref{Tn1}, we can write \eqref{TIdents} as
\be
2g_{\mu\nu}(x)\frac{\delta S_E[g]}{\delta g_{\mu\nu}(x)}=\frac{N}{24\pi}\, \sqrt{g}\,R.
\ee
Now let us take the variation with respect to $g_{\alpha\beta}(x')$ to obtain
\be
2g_{\mu\nu}(x)\frac{\delta^2 S_E[g]}{\delta g_{\mu\nu}(x)\delta g_{\alpha\beta}(x')}
+2\delta^{(2)}(x-x')\frac{\delta S_E[g]}{\delta g_{\alpha\beta}(x')}=\frac{N}{24\pi}\frac{\delta\big( \sqrt{g(x)}\,R(x)\big)}{\delta g_{\alpha\beta}(x')}
\,.
\ee
One only needs to multiply this expression by a factor of $2/\sqrt{g}$, use the definitions \eqref{Gamdef}, and evalute 
the resulting expression in flat space to obtain the desired result. Applying this procedure iteratively, \eqref{WI2n} can 
be shown to hold for any value of $n$. Similar manipulations are needed to obtain \eqref{WI1n} from \eqref{TIdents1}, 
where now the terms in the right-hand side of \eqref{WI1n} come from variations of the Christoffel symbol in \eqref{TIdents1}.

The latter trace identity (\ref{WI2n}) simplifies due to the fact that the variation of $\sqrt{g}R$ vanishes for $n > 2$. To prove
this note first that in $d=2$ dimensions an arbitrary metric variation can be written as a conformal transformation plus a diffeomorphism
\be
\d g_{\m\n} = \d_{\s} g_{\m\n} + \d_{\xi}g_{\m\n} \equiv 2 \s g_{\m\n} + \nabla_{\m}\xi_{\n} +  \nabla_{\n}\xi_{\m}\,.
\ee
Since $\sqrt{g} R$ is a scalar density which is invariant under diffeomorphisms up to surface terms, its local variation with 
respect to $\d_{\xi}g_{\m\n}$ vanishes and it is sufficient to consider only conformal deformations of the metric $\d_{\s} g_{\m\n}$.  
With the conformal parameterization of (\ref{metconf}), the relation (\ref{Rsig}) evaluated for $\bar g_{\m\n}$ the Euclidean
flat space metric is
\be
\sqrt{g}\, R = -2\sqrt{\bar g}\sqb\s = -2\, \d^{\m\n}\pa_{\m}\pa_{\n} \s =  -2\, \pa^2 \s
\label{Rsigflat}
\ee
in flat Cartesian coordinates. Eq.\ (\ref{Rsigflat}) shows that $\sqrt{g}R$ depends only linearly on the conformal metric variation 
$\d_{\s} g_{\m\n} = 2 \s \d_{\m\n}$. Hence  all higher order variations of $\sqrt{g}R$ around flat space beyond the first must 
vanish identically, \textit{i.e.}
\be
\frac{\d^{n-1}\big(\sqrt{g(x)}R(x)\big)}{\d g_{\m_2\n_2}(x_2)\ldots\d g_{\m_n\n_n}(x_n)}\bigg\vert_{g_{\m\n} = \d_{\m\n}} \!\!= 0\,,
\qquad n> 2\,,\ \ d=2\,.
\label{delRn}
\ee
For $n=2$ the first variation is given by
\be
\frac{\d\big(\sqrt{g(x)}R(x)\big)}{\d g_{\a\b}(y)}\bigg\vert_{g_{\m\n} = \d_{\m\n}} \!\!=  \left( - \d^{\a\b} \pa^2 + \pa^{\a}\pa^{\b}\right) \d^{(2)}(x-y)
\ee
in Cartesian coordinates of flat space, so that for the two-point function, the Ward identities (\ref{WardIds}) 
are \cite{Osborn:1991gm,Osborn:1993cr,DiFrancesco:1997nk,Coriano:2012wp,Serino:2014ysa}
\bes
\begin{align}
\pa_\n^x\big\lag T^{\m\n}(x)T^{\a\b}(y)\big \rag &= 0 \,, \label{TTcons}\\
\d_{\m\n}\big\lag T^{\m\n}(x)T^{\a\b}(y)\big\rag &= \frac{N}{12\pi}\, \left( - \d^{\a\b} \pa^2 + \pa^{\a}\pa^{\b}\right) \d^{(2)}(x-y)
\,, \label{TTtrace}\end{align}
\ees
where we have used $\lag T^{\a\b}(y)\rag = 0$ in flat space.

We may use the identity (\ref{WI2n}) recursively together with (\ref{delRn}) to derive identities for correlation functions 
with multiple insertions of traces $T(x_j)\equiv \d_{\m\n}T^{\m\n}(x_j)$, obtaining
\begin{align}
&\big\lag T(x_1)\ldots T(x_\ell)T^{\m_{\ell+1}\n_{\ell+1}}(x_{\ell+1})\ldots T^{\m_n\n_n}(x_n)\big\rag\nn
&=(-2)^\ell\bigg[\prod_{k=1}^\ell\Big(\!\sum_{i_k = k +1}^n \d^{(2)}(x_k-x_{i_k})\Big)\bigg] 
\big\lag  T^{\m_{\ell+1}\n_{\ell+1}}(x_{\ell+1})\ldots T^{\m_n\n_n}(x_n)\big\rag\,, \quad n> \ell+1\ge 2
\label{traceWI-multiple}
\end{align}
while in the special case $n=\ell + 1$,
\begin{align}
&\big\lag T(x_1)\ldots T(x_{n-1})T^{\a\b}(x_n)\big\rag\nn
&=(-2)^{n-2}\bigg[\prod_{k=1}^{n-2}\Big(\!\sum_{i_k = k +1}^n \d^{(2)}(x_k-x_{i_k})\Big)\bigg] 
\frac{N}{12\pi}\, \left( - \d^{\a\b} \pa^2 + \pa^{\a}\pa^{\b}\right) \d^{(2)}(x_{n-1}-x_n)
\label{traceand1}
\end{align}
which reduces to (\ref{TTtrace}) for $n=2$ and $x_1=x$, $x_2=y$.

\subsection{The Holomorphic Representation in Coordinate Space}

To compare the fermion loops to scalar trees it will be particularly convenient to work in Euclidean
time $\tau =it$ and the complex coordinates \cite{Ginsparg:1988ui,DiFrancesco:1997nk}
\be
z=x + i \tau\,, \qquad \bar{z} =x - i\t \,, 
\ee
so that the Euclidean flat line element and non-zero components of the metric are given by
\be
ds^2=d\t^2+dx^2=dz\,d\bar{z} \,, \qquad \bar g_{z\bar z} = \tfrac{1}{2}\,,\qquad \bar g^{z\bar z} = 2\,.
\label{compmetric}
\ee
In these complex coordinates the fermion energy momentum tensor \eqref{Tfermion} has components%
\footnote{We drop the subscript $f$ in the fermion stress-energy $T_f$ in this section for notational simplicity.}
\bes
\begin{align}
T_{zz}&=- \frac{i}{2} \left(:\!\psi^\dagger_+(z){\pa_z}\psi_+(z)\!:-:\!{\pa_z}\psi^\dagger_+(z)\psi_+(z)\!:\right) \\
T_{\bar z\bar z}&=\frac{i}{2}\left(:\!\psi^\dagger_-(\bar z){\pa_{\bar z}}\psi_-(\bar z)\!:-:\!{\pa_{\bar z}}\psi^\dagger_-(\bar z)\psi_-(\bar z)\!:\right)
\label{Tzzfermion}
\end{align}
\ees
while the mixed component $T_{z\bar z}$ vanishes on shell upon using the Dirac equations  of motion 
$\pa_z\psi_-=0=\pa_{\bar z}\psi_+$. These components satisfy the conservation law 
\begin{align}
 \pa_z T_{z\bar{z}}+\pa_{\bar{z}}T_{zz}&=0\,, &
 \pa_{\bar{z}} T_{z\bar{z}}+\pa_{z}T_{\bar{z}\bar{z}}&=0
 \,,
\end{align}
since $T_{zz}$ depends only on $z$ while $T_{\bar{z}\bar{z}}$ depends only on $\bar z$.

The fermion propagators are
\bes
\begin{align}
\lag\psi_+(z_1)\psi^{\dag}_+(z_2)\rag &=\frac{i}{2\pi}\inv{z_1-z_2} \equiv S_+(z_1-z_2)\\
\lag\psi_-(z_1)\psi^{\dag}_-(z_2)\rag &=-\frac{i}{2\pi}\inv{\bar z_1-\bar z_2} \equiv S_-(\bar z_1-\bar z_2)
\label{psizzprop}
\end{align}
\ees
in this notation (for one fermion). These are Green's functions of the chiral Dirac operator since, \textit{e.g.}
\be
-i\g^0\g^{\m}\frac{\pa}{\pa x_1^{\m}}S_+  = -2i\pa_{\bar z_1} S_+ = 
\frac{1}{\pi} \,\pa_{\bar z_1}\frac{1}{z_1-z_2} = -4 \pa_{\bar z_1}\pa_{z_1} G_{\bPhi} = \d^{(2)} (z_1-z_2)
\label{Diracprop}
\ee
where
\be
G_{\bPhi} (z_1-z_2) = -\frac{1}{2 \pi} \ln \left(M |z_1-z_2|\right) = - \frac{1}{4\pi} \big[ \ln (z_1-z_2) +  \ln (\bar z_1-\bar z_2)\big] + \text{const.}
\label{bPhiprop}
\ee
is the Euclidean scalar propagator (\ref{Euclprop}) and we have used $\sq = 4 \pa_{\bar z}\pa_z$  in complex coordinates.

Using (\ref{Tzzfermion}) and (\ref{psizzprop}), and keeping track of signs for the anti-commutation of
fermion fields, we find for the one-loop two-$T$ correlator of $N$ fermions
\begin{align}
\big\lag\, T_{z_1z_1}T_{z_2z_2}\big\rag &=-\frac{N}{4}\,\Big[2\pa_{z_1}S_+(z_1-z_2)\ \pa_{z_2}S_+(z_1-z_2)
-2S_+(z_1-z_2)\, \pa_{z_1}\pa_{z_2}S_+(z_1-z_2)\Big]\nn
& =\frac{N}{2(2\pi)^2}\,\inv{(z_1-z_2)^4}
\,. \label{TTcomplex}
\end{align}

For the canonically normalized boson field $\bPhi$ associated with the effective action of the conformal
anomaly \eqref{SanombPhi}, the energy momentum tensor components in complex coordinates of flat space are
\bes
\begin{align}
T_{zz}[\bPhi] &= \sqrt{\frac{N}{12\pi}}\, \pa_z\pa_z\bPhi\ + :\!\pa_z\bPhi\pa_z\bPhi\!:  \label{TzzbPhi}\\
T_{z\bar z}[\bPhi] &= -  \sqrt{\frac{N}{12\pi}}\, \pa_{\bar z}\pa_z\bPhi \label{TzzbarPhi}
\end{align}
\label{TPhi}\ees
with the anti-holomorphic $T_{\bar z \bar z}$ obtained from $T_{zz}$ by replacing $z$ by $\bar z$.
Its Euclidean propagator is given by (\ref{Euclprop}) or (\ref{bPhiprop}). Note that the scaling of the linear term in $\bPhi$ 
with $\sqrt{N}$ implies that exactly two insertions of the linear vertex is necessary to match the
$N$ scaling of the fermion loop correlations functions. For the two-point function this is just
the simple tree diagram of Fig. \ref{fig:TT}, already computed in \secref{sec:gravity}, \textit{cf.}\ \eqref{EuclTT},
which in complex coordinates is
\be
\big\lag T_{z_1z_1} T_{z_2z_2}\big\rag_c =  \frac{N}{12\pi}\,\big\lag\pa_{z_1}^2\bPhi(z_1)\pa_{z_2}^2\bPhi(z_2)\big\rag
=\frac{N}{12\pi}\,\pa_{z_1}^2\pa_{z_2}^2 G_{\bPhi}(z_1-z_2)
 =\frac{N}{2(2\pi)^2}\,\inv{(z_1-z_2)^4}
\label{TTscalar}
\ee
which coincides with (\ref{TTcomplex}), and corresponds to \eqref{VacPolF0} Fourier transformed to Euclidean position
space in complex coordinates, where all four indices are $z$. The massless pole of \eqref{VacPolF0} is nothing else 
but the scalar propagator $G_{\bPhi}$, and the projection operators $(\eta^{\m\n}k^2-k^\m k^\n)$ in coordinates
(\ref{compmetric}) become four derivatives $\pa_z^4$ of $G_{\bPhi}$.

Taking one trace in the two-point function yields
\be
\big\lag 4T_{z_1\bar{z}_1} T_{z_2z_2}\big\rag_c =\frac{N}{12\pi} \pa_{z_2}^2\d^{(2)}(z_1-z_2)
\label{2Ttrace}
\ee
no matter whether it is computed via a fermion loop or from a tree using (\ref{Tphi}).
However this contribution arises in each representation in quite a different way,
which merits some comment. In the fermion loop computation in position space 
the result (\ref{TTcomplex}) is well-defined if the points $z_1 \neq z_2$ are distinct.
The possible local contribution at $z_1 = z_2$ is undefined without further information.
That information comes from the requirement of covariance, or equivalently, the
enforcement of the Ward identity (\ref{TTtrace}). In covariant language the conservation
eq.\ (\ref{TTcons}) forces the tensor structure of the two-point function to be proportional
to the projector in (\ref{VacPolF0}) or $(\eta_{\m\n} \pa^2 - \pa_{\m}\pa_{\n}) (\eta_{\a\b} \pa^2 - \pa_{\a}\pa_{\b})$
in position space. This tensor structure determines the trace in terms of the non-trace
components so that the trace is obtained by replacing $\pa_{z_1}^2$ by $-\pa^2 = -4 \pa_{z}\pa_{\bar z}$,
which because of  (\ref{propeq}) or (\ref{Diracprop}) gives a well-defined contact term at $z_1=z_2$.
If one had chosen the trace anomaly to be absent, at the price of the loss of the covariant 
conservation eq.\ (\ref{TTcons}), this contact term would be absent. In contrast, in the
tree effective action of the boson field $\bPhi$, covariant conservation at the non-zero
trace anomaly has already been imposed, and the non-zero trace (\ref{2Ttrace}) follows
from the explicit non-zero trace contribution of the term linear in $\bPhi$ (\ref{TzzbarPhi})
inserted at the endpoints of the simple tree diagram in Fig. \ref{fig:TT}, canceling
the scalar propagator by (\ref{propeq}) and leading to a well-defined contact term
consistent with the anomalous trace Ward identity (\ref{2Ttrace}). The scaling with $N$
of loops and trees also arises in different ways, and helps to understand how a quantum
loop proportional to $\hbar$ can be equivalent to scalar tree diagrams. This is possible
because the anomaly effective action for $\bPhi$ has a linear $\bPhi R$ vertex 
which is itself proportional to $\hbar$ from the quantum anomaly it encodes. Two
insertions of this vertex and one $\bPhi$ propagator canonically normalized yield 
a factor of $N \hbar^2/\hbar = N\hbar$, exactly the same factor in the quantum one-loop
diagram of $N$ fermions. Additional insertions of the stress-energy tensor vertex always
bring with them the same number of additional propagators in either the loop
or tree representation, so that this factor of $N\hbar$ is unchanged for the arbitrary
$n$-point correlation function, once the two-point function is fixed by the anomaly.

\begin{figure}[ht]
\vspace{-5mm}
\centering
\includegraphics[width=1\columnwidth]{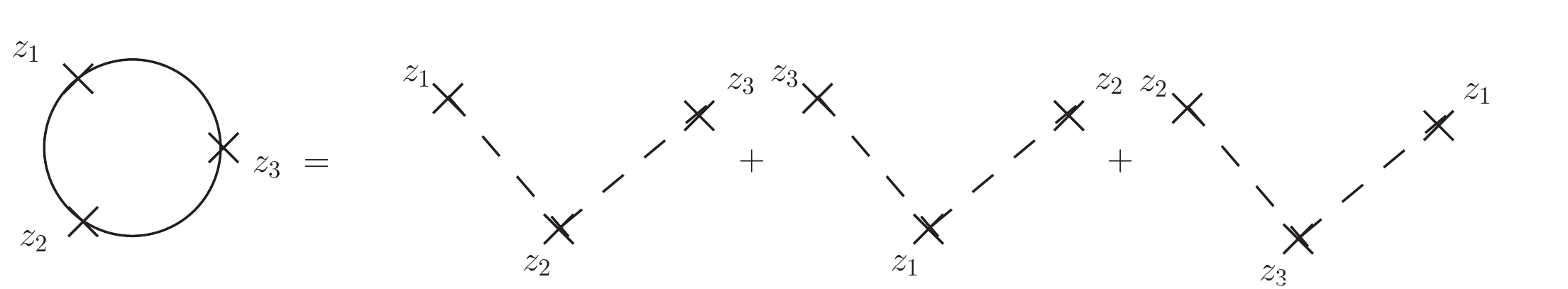}
\caption{Equivalence of fermion loop to scalar $\bPhi$ trees for $n=3$.}
\label{fig:TTT-illustr}
\end{figure}

For the higher $n$-point functions, we start with the simplest case of purely holomorphic $T_{zz}$ components.
For three insertions of $T_{zz}$, whether the $N$ fundamental quantum fields are fermions or bosons we find 
\cite{Osborn:1993cr,Erdmenger:1996yc,Erdmenger:1997gy,DiFrancesco:1997nk}:
\begin{align}
&\big\lag T_{z_1z_1}T_{z_2z_2}T_{z_3z_3}\big\rag_c= 
N \big[\pa_{z_1}\pa_{z_2}G_{\bPhi}(z_1-z_2)\big]\big[\pa_{z_2}\pa_{z_3}G_{\bPhi}(z_2-z_3)\big]
\big[\pa_{z_3}\pa_{z_1}G_{\bPhi}(z_3-z_1)\big]\nn
&=- \frac{N}{(2\pi)^3}\inv{\left(z_1-z_2\right)^2\left(z_2-z_3\right)^2\left(z_3-z_1\right)^2} \nn
& =-\frac{N}{3(2\pi)^3}\left[\inv{\left(z_2-z_1\right)^3\left(z_1-z_3\right)^3}  + \inv{\left(z_1-z_2\right)^3\left(z_2-z_3\right)^3}
+ \inv{\left(z_2-z_3\right)^3\left(z_3-z_1\right)^3}\right]
\label{3Tcomplex}
\end{align}
where the last line follows from the algebraic identity 
\be
3\left(z_1-z_2\right)\left(z_2-z_3\right)\left(z_3-z_1\right)=\left(z_1-z_2\right)^3+\left(z_2-z_3\right)^3+\left(z_3-z_1\right)^3.
\label{3Tident}
\ee
This identity is just what is needed for the identification with three tree graphs in $T_{zz}[\bPhi]$ as depicted in \figref{fig:TTT-illustr},
where the middle point containing the one $T_{zz} [\bPhi]$ vertex insertion that is quadratic in $\bPhi$ is $z_1, z_2$ and $z_3$ 
respectively. Thus, for example the first term corresponds to the tree graph
\begin{align}
\frac{N}{12\pi}\big\lag\pa_{z_1}^2\bPhi(z_1)\!:\!\pa_{z_2}\bPhi(z_2)\pa_{z_2}\bPhi(z_2)\!:\!\pa_{z_3}^2\bPhi(z_3)\big\rag_c
&=\frac{N}{6\pi}\big[\pa_{z_1}^2\pa_{z_2}G_{\bPhi} (z_1-z_2) \big]\big[\pa_{z_2}\pa_{z_3}^2G_{\bPhi}(z_2-z_3)\big]\nn
&=-\frac{N}{3(2\pi)^3}\inv{\left(z_1-z_2\right)^3\left(z_2-z_3\right)^3}
\end{align}
after taking account of the two allowed Wick contractions and substituting (\ref{bPhiprop}). Summing the three contributions
represented in Fig. \ref{fig:TTT-illustr} and using (\ref{3Tident}) gives (\ref{3Tcomplex}).

The case of $n=3$ shows that equality of loops and trees can be achieved upon summing over permutations of all allowed
graphs in each case with the proper symmetry factor(s), and only after a decomposition expressed by the algebraic identity
(\ref{3Tident}) is utilized. For the general case of $n$ holomorphic stress-energy tensor correlators at distinct points, $(z_1, \ldots, z_n)$ 
there are $n!$ permutations of the positions of the $z_i$ on the fermion loop. Due to cyclic symmetry of the one-loop graphs,
$n$ of them lead to the same expression, and furthermore the mirror reflection of any loop gives again the same
expression. Thus for the $n$-point correlator there are $(n-1)!/2$ distinct one-loop graphs that one must sum over.
For $n=3$ this is the single expression (\ref{3Tcomplex}). However on the scalar tree side there is mirror reflection
symmetry but no cyclic symmetry of $n$ points, so that the number of distinct tree diagrams is $n!/2$, giving the $3$
expressions in the last line of (\ref{3Tcomplex}).  
Before we turn to a general proof, it is illustrative to examine one more explicit example of $n=4$.
The one-loop graph with the four holomorphic stress-energy tensors of $N$ fundamental fermions or bosons
is
\begin{align}
&\big\lag T_{z_1z_1}T_{z_2z_2}T_{z_3z_3}T_{z_4z_4}\big\rag_c  \nn
&= N\big[\pa_{z_1}\pa_{z_2}G_{\bPhi} (z_1-z_2)\big]\big[\pa_{z_2}\pa_{z_3}G_{\bPhi}(z_2-z_3)\big]
\big[\pa_{z_3}\pa_{z_4}G_{\bPhi}(z_3-z_4)\big]\big[\pa_{z_4}\pa_{z_1}G_{\bPhi}(z_4-z_1)\big]\nn
&\quad + (1\leftrightarrow 2) +  (1\leftrightarrow 4)\nn
& =\frac{N}{(2\pi)^4}\inv{\left(z_1-z_2\right)^2\left(z_2-z_3\right)^2\left(z_3-z_4\right)^2\left(z_4-z_1\right)^2}
+ (1\leftrightarrow 2) +  (1\leftrightarrow 4)\,.
 \label{4Tcomplex}
\end{align}
Since $(4-1)!/2=3$ there are two additional distinct expressions with $1\leftrightarrow 2$ and $1\leftrightarrow 4$ interchanged.
For the scalar boson $\bPhi$ tree diagrams there are $4!/2=12$ distinct permutations with $2$ of the points at the ends
with the term linear in $\bPhi$ in (\ref{TzzbPhi}) inserted, and the other $2$ points in the interior with the term quadratic in 
$\bPhi$ in (\ref{TzzbPhi}) inserted. A typical term of this kind from the scalar tree with fixed sequence $(1234)$ is
\begin{align}
&\frac{N}{12\pi}\big\lag\pa_{z_1}^2\bPhi(z_1)\!:\!\pa_{z_2}\bPhi(z_2)\pa_{z_2}\bPhi(z_2)\!:\,  
:\!\pa_{z_3}\bPhi(z_3)\pa_{z_3}\bPhi(z_3)\!:\! \pa_{z_4}^2\bPhi(z_4)\big\rag_c\nn
&=\frac{N}{3\pi}\big[\pa_{z_1}^2\pa_{z_2}G_{\bPhi} (z_1-z_2) \big]\big[\pa_{z_2}\pa_{z_3}G_{\bPhi}(z_2-z_3)\big]
\big[\pa_{z_3}\pa_{z_4}^2G_{\bPhi}(z_3-z_4)\big]\nn
&=\frac{N}{3(2\pi)^3}\inv{\left(z_1-z_2\right)^3\left(z_2-z_3\right)^2\left(z_2-z_3\right)^3}\,.
\end{align}
Summing over the $12$ distinct terms of this kind from the scalar trees yields (\ref{4Tcomplex}) when the algebraic identity
\begin{align}
&\inv3\Bigg\{\inv{\left(z_1-z_2\right)^3\left(z_2-z_3\right)^2\left(z_3-z_4\right)^3}
 +\inv{\left(z_2-z_3\right)^3\left(z_3-z_4\right)^2\left(z_4-z_1\right)^3} \nn
 &\quad +\inv{\left(z_3-z_4\right)^3\left(z_4-z_1\right)^2\left(z_1-z_2\right)^3}
 +\inv{\left(z_4-z_1\right)^3\left(z_1-z_2\right)^2\left(z_2-z_3\right)^3} +(1\leftrightarrow2)+(1\leftrightarrow4)\Bigg\}\nn
&= \Bigg\{\inv{\left(z_1-z_2\right)^2\left(z_2-z_3\right)^2\left(z_3-z_4\right)^2\left(z_4-z_1\right)^2}
  + \inv{\left(z_1-z_3\right)^2\left(z_3-z_2\right)^2\left(z_2-z_4\right)^2\left(z_4-z_1\right)^2} \nn
 & \quad  +\inv{\left(z_1-z_2\right)^2\left(z_2-z_4\right)^2\left(z_4-z_3\right)^2\left(z_3-z_1\right)^2}\Bigg\}
\label{4Tident}
\end{align}
is used. This relation \eqref{4Tident} can be checked directly. However the algebraic identity needed 
at each order $n$ grows rapidly more complicated with $n$, making direct verification of the equivalence of loop 
and tree graphs for general $n$ impractical.

In order to prove the equivalence between fermion loops and scalar $\bPhi$ trees for an arbitrary number $n$ of holomorphic
(or anti-holomorphic) stress-energy tensor insertions we make use of the following Ward identity
\cite{Belavin:1984vu,Teschner:2001rv,DiFrancesco:1997nk} 
\begin{align}
\big\lag T_{z_1z_1}T_{z_2z_2}\ldots T_{z_{n+1}z_{n+1}}\big\rag &=-\frac{1}{2\pi}\sum_{i=1}^n
\left\{\frac2{(z_{n+1}-z_i)^2}+\inv{(z_{n+1}-z_i)}\diff{\ }{z_i}\right\}
\big\lag T_{z_1z_1}\ldots T_{z_nz_n}\big\rag \nn
&\quad + \ \frac{N}{2 (2\pi)^2} \sum_{i=1}^n\frac{1}{(z_{n+1}-z_i)^4} 
\,\big\lag T_{z_1z_1}\ldots T_{z_{i-1}z_{i-1}}T_{z_{i+1}z_{i+1}}\ldots T_{z_nz_n}\big\rag\,.
\label{WIBPZ}
\end{align}
This Ward identity is derived by combining the two Ward identities for conservation and trace of the stress-energy
tensor (\ref{WardIds}), together with an additional identity following from invariance under Lorentz boosts \cite{DiFrancesco:1997nk}, or 
alternately from the conformal transformation properties of the stress-energy tensor under arbitrary holomorphic conformal transformations
\cite{Belavin:1984vu}, which includes the previous three as special cases. The Ward identity \eqref{WIBPZ} holds for any
correlation function, connected or not. For $n=1$ the entire contribution to the connected two-point correlator comes from the 
last anomaly term in (\ref{WIBPZ}), as verified by (\ref{TTcomplex}). However for $n\ge 2$ this latter term corresponds 
to the sum of disconnected graphs in which $T_{z_1z_1}$ is Wick contracted with any of the other $T_{z_i z_i}$. Thus, 
restricting to the connected diagrams generated by the 1PI variations of $\G_{\rm eff}$ defined in (\ref{Gamdef}), we have simply
\be
\hspace{-3mm}\big\lag T_{z_1z_1}T_{z_2z_2}\ldots T_{z_{n+1}z_{n+1}}\big\rag_c =-\frac{1}{2\pi}\sum_{i=1}^n
\left\{\frac2{(z_{n+1}-z_i)^2}+\inv{(z_{n+1}-z_i)}\diff{\ }{z_i}\right\}
\big\lag T_{z_1z_1}\ldots T_{z_nz_n}\big\rag_c
\label{WIBPZconn}
\ee
for $n \ge 2$. Hence all holomorphic or anti-holomorphic stress-energy tensor connected correlation functions for $n \ge 2$
are determined by the fundamental two-point function recursively. A similar identity applies for the anti-holomorphic sector 
by replacing all $z$ by $\bar z$. 

The Ward identity (\ref{WIBPZconn}) generates $n$-point connected correlation functions from $n-1$ point correlation functions, 
and must hold for either the loops and the trees separately. The relation between the two which we shall prove by
means of this identity is 
\be
 \sum_{{\cal P}[i_1,\ldots i_n]}\G_n^{\rm loop}(z_{i_1},\ldots,z_{i_n})=\frac{n}{3}\sum_{{\cal P}[i_1,\ldots i_n]}
 \G_n^{\rm tree}(z_{i_1}, z_{i_2}; z_3\ldots,z_{n-2}; z_{i_{n-1}}, z_{i_n})
\label{Gamlooptree}
\ee
where the sum is over all $n!$ permutations of the arguments $z_i$ and we define the fundamental loop and tree expressions by
\be
\G_n^{\rm loop}(z_1,\ldots,z_n) \equiv (-2\pi)^n\big\lag T_{z_1z_1}\ldots T_{z_nz_n}\big\rag_c=\frac1{(z_1-z_2)^2\ldots(z_k-z_{k+1})^2\ldots(z_n-z_1)^2}
\label{Gamloop}
\ee
and
\begin{align}
&\G_n^{\rm tree}(z_1, z_2; z_3 \ldots, z_{n-2}; z_{n-1}, z_n) \equiv (-2\pi)^n\big\lag T_{z_1z_1}[{\bPhi}]\ldots T_{z_nz_n}[{\bPhi}]\big\rag_c\nn
&=\inv3\frac1{(z_1-z_2)^3(z_2-z_3)^2\ldots(z_k-z_{k+1})^2\ldots(z_{n-2}-z_{n-1})^2(z_{n-1}-z_n)^3}
\label{Gamtree}
\end{align}
respectively, with the endpoints of the linear tree $z_1$ and $z_n$ distinguished in the latter case, the corresponding
factors being cubic in the differences rather than quadratic as all the other terms. Since $2n$ of the $n!$ permutations 
are equal on the loop side, and pairs of terms are always equal on the tree side, the relative factor becomes just $1/3$ 
if the sums are taken only over those permutations which yield strictly different expressions.

If we look first at (\ref{Gamloop}) we notice that the loop with $n+1$ points can be generated by inserting an additional 
point $y$ somewhere in the loop, \textit{i.e.} 
\begin{align}
\frac1{(z_1-z_2)^2\ldots(z_k-z_{k+1})^2\ldots(z_n-z_1)^2}\to
 \frac1{(z_1-z_2)^2\ldots(z_k-y)^2(y-z_{k+1})^2\ldots(z_n-z_1)^2}
 \label{np1loop}
\end{align}
and this procedure can be performed in $n$ different ways (since the graph has $n$ internal lines). Summing over all such 
possibilities takes into account that instead of $(n-1)!/2$ one has now $n!/2$ distinct graphs for $\G_{n+1}^{\text{loop}}$.

The same expression can be derived using the Ward identity (\ref{WIBPZconn}) as follows. Since $z_i$ appears twice in 
$\G_n^{\rm loop}$, every derivative in the curly bracket of (\ref{WIBPZconn}) generates two terms. Furthermore, 
there are two different derivative terms in the sum of the curly bracket acting on any given $(z_k-z_{k+1})^{-2}$ factor in 
$\G_{n}^{\text{loop}}$, \textit{i.e.}\ that with $i=k$ and $i=k+1$. It is therefore convenient to combine subexpressions in the following way.
In the sum over $i$ in \eqref{WIBPZconn} we select two summands $i=k$ and $i=k+1$ corresponding to a line
connecting the two adjacent points $z_k$ and $z_{k+1}$ in $\G_n^{\rm loop}$. For these particular terms the operator in 
\eqref{WIBPZconn} gives
\bea
&&\hspace{-8mm}\left\{\!\frac2{(y-z_k)^2\!}+\frac1{y-z_k}\diff{\ }{z_k}+\frac2{(y-z_{k+1})^2\!}+\frac1{y-z_{k+1}}\diff{\ }{z_{k+1}}\!\right\}\!\!
\frac{1}{(z_{k-1} - z_k)^2 (z_k -z_{k+1})^2(z_{k+1}-z_{k+2})^2\!} \nn
&&\hspace{-8mm}=\!\left\{\!\frac2{(y-z_k)^2\!}+\frac2{(y-z_k)(z_{k-1}-z_k)\!} -\frac2{(y-z_k)(z_k-z_{k+1})\!} +\frac2{(y-z_{k+1})^2\!}
+\frac2{(y-z_{k+1})(z_k-z_{k+1})} \right. \nn
&&\hspace{5mm} \left. -\frac2{(y-z_{k+1})(z_{k+1}-z_{k+2})}\right\}\frac{1}{(z_{k-1}- z_k)^2 (z_k-z_{k+1})^2(z_{k+1}-z_{k+2})^2}\,.
\eea
Leaving aside the terms involving the outer points $z_{k-1}$ and $z_{k+2}$, and focusing only on those depending
on $z_k$ or $z_{k+1}$, in the last set of curly brackets we collect $1/2$ of the terms involving either one of them
in $(y-z_k)^2$ or $(y-z_{k+1})^2$, together with the full contribution of terms involving both $z_k$ and $z_{k+1}$ 
through $(z_k-z_{k+1})$ to obtain
\begin{align}
&\left\{\inv{(y-z_k)^2}-\frac2{(y-z_k)(z_k -z_{k+1})} +\inv{(y-z_{k+1})^2}+\frac2{(y-z_{k+1})(z_k-z_{k+1})}\right\}
\frac{1}{(z_k-z_{k+1})^2}\nn
&=\left\{\inv{(y-z_k)^2}+\inv{(y-z_{k+1})^2}-\frac2{(y-z_k)(y-z_{k+1})}\right\}\frac{1}{(z_k-z_{k+1})^2}\nn
&=\left[\inv{(y-z_k)}-\inv{(y-z_{k+1})}\right]^2 \frac{1}{(z_k-z_{k+1})^2}\nn
& =\frac{1}{(z_k-y)^2(y-z_{k+1})^2}
\label{middle}
\end{align}
which is exactly the result  (\ref{np1loop}) of inserting the new point $y$ into the loop graph between the adjacent
points $z_k$ and $z_{k+1}$. The other half of the $(y-z_k)^2$ or $(y-z_{k+1})^2$ terms together with the terms involving $z_{k-1}$ 
and $z_{k+2}$ are to be combined in the similar subexpressions corresponding to an insertion of the new vertex at $y$
between $z_{k-1}$ and $z_k$ and between $z_{k+1}$ and $z_{k+2}$ pairwise in the same way. Hence, the sum of all subexpressions 
generated by the Ward identity (\ref{WIBPZconn}) corresponds to inserting the new point $y$ on all $n$ possible lines in the loop, 
generating $n$ different one-loop graphs with $n+1$ insertions of $T_{zz}$ from every single one-loop graph 
with $n$ insertions of $T_{zz}$. Since there are $(n-1)!/2$ distinct terms in the loop diagrams with $n$ insertions
of $T_{zz}$, there are $n!/2$ distinct terms with $n+1$ insertions, as a check of our previous counting.

We will now show that the same result holds for the tree diagrams and $\G^{\rm tree}_n$. In order to generate $\G_{n+1}^{\text{tree}}$ 
from $\G_{n}^{\text{tree}}$ there are three distinct ways of inserting the extra point (which again we denote by $y$):
\begin{enumerate}
\itemsep=0pt
\item One can insert $y$ as a new end point changing the expression (\ref{Gamtree}) by the factor 
$(z_1-z_2)/(y-z_1)^3$, if the first endpoint on the left is chosen, or by the factor $(z_{n-1} -z_n)/(z_n-y)^3$
if the last endpoint on the right is chosen instead;
\item One can insert $y$ on the line immediately attached to one of the endpoints changing the expression (\ref{Gamtree})
by the factor $\frac{(z_1-z_2)^3}{(z_1-y)^3(y-z_2)^2}$ if on the left, and  $\frac{(z_{n-1}-z_n)^3}{(z_{n-1} - y)^2 (y-z_n)^3}$
if on the right;
\item Or one can insert $y$ somewhere in the middle of the tree graph, making a change precisely like in (\ref{np1loop}).
\end{enumerate}
The factors of possibilities (1) and (2) combine to 
\be
\frac{(z_1-z_2)}{(y-z_1)^3}\left[1-\frac{(z_1-z_2)^2}{(y-z_2)^2}\right]=\frac{(z_1-z_2)(y-2z_2+z_1)}{(y-z_1)^2(y-z_2)^2}
\label{yleft}
\ee
for the left side, with an analogous expression for possibilities (1) and (2) on the right side of the tree.

On the other hand the terms generated by the Ward identity \eqref{WIBPZ}, with $i=1,2$ 
\begin{align}
&\left\{\frac2{(y-z_1)^2}+\frac1{y-z_1}\diff{\ }{z_1}+\frac2{(y-z_2)^2}+\frac1{y-z_2}\diff{\ }{z_2}\right\}
\G^n_{\rm tree}(z_1,z_2; z_3,\ldots z_{n-2};z_{n-1},z_n) \nn
&=\left\{\frac2{(y-z_1)^2}-\frac{3}{(y-z_1)(z_1-z_2)} 
 +\frac2{(y-z_2)^2}+\frac{3}{(y-z_2)(z_1-z_2)}-\frac{2}{(z_2-z_3)}\right\}\G^n_{\rm tree}
\,. \label{lefttree}
\end{align}
Setting aside the term containing the interior point $z_3$, we combine the $3$ terms depending on the endpoint
$z_1$ with half of the term depending on $(y-z_2)^2$ to obtain the factor
\begin{align}
&\frac{2}{(y-z_1)^2}-\frac{3}{(y-z_1)(z_1 - z_2)} + \inv{(y-z_2)^2}+\frac{3}{(y-z_2)(z_1-z_2)}\nn
&=\frac{2}{(y-z_1)^2}+\inv{(y-z_2)^2}-\frac{3}{(y-z_1)(y-z_2)}\nn
&=\left[\frac{2}{(y-z_1)}-\inv{(y-z_2)}\right]\left[\inv{(y-z_1)}-\inv{(y-z_2)}\right]
 =\frac{(z_1-z_2)(y-2z_2+z_1)}{(y-z_1)^2(y-z_2)^2}
\end{align}
which coincides with  \eqref{yleft}. The same relation holds for the combination of possibilities (1) and (2) on the 
other end of the tree graph. The remaining terms of \eqref{lefttree} and its right side counter part, together with all other terms
coming from the interior points of the tree generated by the Ward identity (\ref{WIBPZ}) combine exactly 
as in (\ref{middle}), thus corresponding to possibility (3) above, where the additional point $y$ is 
inserted somewhere on one of the interior lines of the tree graph, not connected to one of the endpoints.

Since the argument goes through unchanged if the labels on the leftmost points $z_1, z_2$ and the rightmost 
points $z_{n-1}, z_n$ are replaced by any permutations  of the labels in (\ref{Gamlooptree}), it follows 
that the sum of all subexpressions generated by the Ward identity (\ref{WIBPZ}) correspond to inserting 
$y$ at all $n+1$ possible locations in each tree graph (including the new endpoints), thus generating $n+1$ 
tree graphs with $n+1$ insertions of the holomorphic $T_{zz}$ from every single tree graph with $n$ insertions 
of $T$. Since there are $n!/2$ distinct terms in $\G^{\rm tree}_n$, there are $(n+1)!/2$ such distinct terms
in $\G^{\rm tree}_{n+1}$, again verifying our previous counting. Since we have verified the equivalence
of loops and trees explicitly for $n=2, 3$ and $4$, the proof of the equivalence (\ref{Gamlooptree}) for
all higher $n$ follows by induction. We have checked (\ref{Gamlooptree}) explicitly for the cases of 
$n=5,6$ using Mathematica, the latter case involving $6!/2 =360$ distinct tree contributions. The
algebraic identity relating these $360$ terms to the sum of $5!/2 = 60$ distinct loop contributions would
be difficult to surmise without the guidance of the Ward identity (\ref{WIBPZ}).

\begin{figure}[ht]
\centering
 \includegraphics[width=0.85\columnwidth]{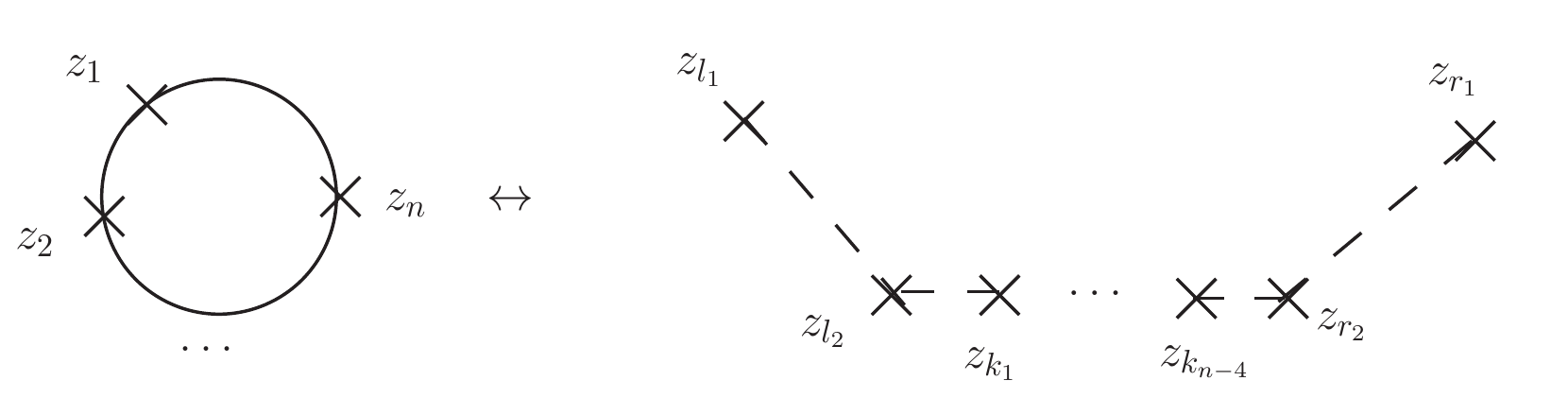}
 \caption{Equivalence of the fermion loop with scalar trees for arbitrary $T$-insertions.}
 \label{fig:nT-illustr}
\end{figure}

\subsection{Trace Insertions and Contact Terms}

So far, we have studied purely holomorphic correlation functions. Their anti-holomorphic counterparts 
are straightforwardly derived in the same way by replacing all $z$ by $\bar z$.
What remains, are all those correlators with insertions of $T_{z\bar z}=\inv4\bar g^{\m\n}T_{\m\n}$.
These can be easily generated by the trace Ward identity \eqref{traceWI-multiple} for $n > 2$, 
with the two-point function already treated separately in (\ref{TTscalar}) and (\ref{2Ttrace}).
Since the stress-energy tensor has only $3$ components in $d=2$ dimensions, the trace
components together with the holomorphic and anti-holomorphic components
already treated provides complete information of any components of arbitrary numbers of
stress-energy tensor correlators in $d=2$.

Starting again with the three point function, with one trace $4T_{z \bar z}$ insertion,
direct computation of the tree graphs using \eqref{TzzbarPhi} yields
\begin{align}
\big\lag 4T_{z_1\bar z_1}T_{z_2z_2}T_{z_3z_3}\big\rag
 &=\frac{N}{6 (2\pi)^2}\d^{(2)}(z_1-z_2)\pa_{z_2}^2\pa_{z_3}^2\ln|z_2-z_3|^2
 +(z_2 \leftrightarrow z_3)\nn
 &=-\frac{N}{ (2\pi)^2} \left[\d^{(2)}(z_1-z_2) + \d^{(2)}(z_1-z_3)\right] \frac{1}{(z_2-z_3)^4}
\label{3T1trace}
\end{align}
which agrees with \eqref{traceWI-multiple} for $\ell =1, n=3$, upon inserting 
\eqref{TTscalar} for the holomorphic two-point function. Notice that only two of the 
distinct three trees in Fig. \ref{fig:TTT-illustr} contribute to (\ref{3T1trace}), namely 
the ones with the trace component at $z_1$ at one of the end points, and this
trace is non-zero because of the explicit non-zero contribution of the linear 
term in the $\bPhi$ stress-energy tensor component (\ref{TzzbarPhi}) canceling
the scalar propagator $G_{\bPhi}$ attached to this point.

\begin{figure}[ht]
\centering
\includegraphics[width=0.8\columnwidth]{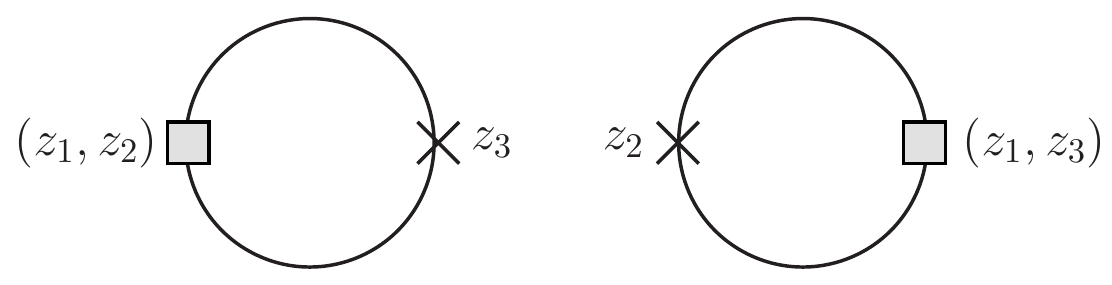}
\caption{The contribution to the fermion loop three point function obtained by
varying one of the vertices, represented by the square.}
\label{fig:TTT-onetrace}
\end{figure}

From the loop point of view the result (\ref{3T1trace}) arises in a different way,
namely from the two contact terms obtained from varying the stress-energy tensor
two-point function $\lag T^{\m_2\n_2}T^{\m_3\n_3} \rag$ with respect to
$g_{\m_1\n_1}$, as in (\ref{Gamall}) with the topology of the bubble diagrams
illustrated in Fig. \ref{fig:TTT-onetrace}.

Looking at two trace insertions of $T_{z \bar z}$ components in either the loop or tree
representation we obtain
\begin{align}
 \langle 4T_{z_1\bar z_1}T_{z_2z_2}4T_{z_3\bar z_3}\rangle
 &=-\frac{N}{3\pi}\d^{(2)}(z_1-z_2)\pa_{z_2}^2\d^{(2)}(z_2-z_3)
 \label{3T-2traces}
\end{align}
which is now clearly coming from the anomaly contribution to the two-point function (\ref{2Ttrace}).
Using \eqref{2Ttrace} we verify that \eqref{3T-2traces} agrees with \eqref{traceand1}, as expected.
In both cases above, one may straightforwardly replace the purely holomorphic components with 
purely anti-holomorphic components.

When $n\ge 4$ contact terms in the tree diagrams arise also from explicit local variations of the
$T^{\m\n}$ vertex at internal points as in (\ref{Gamall}), just as they do in the loop representation
for all $n\ge 3$. For the scalar boson this variation is
\be
2\,\var{T^{\m\n}[\bPhi](y)}{g_{\a\b}(x)}\bigg|_{g=\bar g}=\left(-\bar g^{\m\n}\pa^\a\bPhi\pa^\b \bPhi
+  \bar g^{\a(\m}\bar g^{\n)\b}\bar g^{\kappa\l}\pa_{\kappa} \bPhi\pa_\l \bPhi\right)\d^{(2)}(y-z)
\label{varT}
\ee
which has the trace
\be 
2\,\bar g_{\m\n}\var{T^{\m\n}[\bPhi](y)}{g_{\a\b}(x)}\bigg|_{g=\bar g} =\left(-2\,\pa^{\a}\bPhi\pa^{\b}\bPhi + \bar g^{\a\b}\bar g^{\kappa\l}
\pa_{\kappa}\bPhi\pa_{\l} \bPhi\right)\d^{(2)}(y-z)
\label{varTtrace}
\ee
where only the first term contributes for holomorphic or anti-holomorphic $\a=\b$. Thus for one trace insertion 
this term contributes a factor of $-2\d^{(2)}(y-z)$ times the usual holomorphic $T_{zz}$ vertex quadratic in $\bPhi$ 
from (\ref{TzzbPhi}), which is just what is required to satisfy the single trace insertion Ward identity (\ref{WI2n}).

For the quantum loops we are assured by the equivalence of $N$ bosons with $N$ fermions that the corresponding 
variation of the fermion stress-energy tensor (\ref{SETF}) produces all the contact terms needed to satisfy the multiple
trace insertion Ward identities (\ref{traceWI-multiple}) and (\ref{traceand1}), all such contact terms coming either from
explicit vertex variations analogous to (\ref{varTtrace}), or inherited from the anomalous two-point function trace
(\ref{2Ttrace}). For the tree diagrams the only sources of trace insertions are either the endpoint insertions of 
linear terms in $\bPhi$ (which encode the anomaly through the linear $\bPhi R$ vertex) or explicit contact
variations of the (\ref{varTtrace}) kind at intermediate points of the tree. These latter contributions must
be taken into account for $n\ge 4$ (since a non-endpoint must be varied at least once which is first possible at $n=4$).

Considering this first non-trivial case of $n=4$ for tree diagrams with one trace insertion and three holomorphic $T_{zz}$
insertions, the Ward identity \eqref{traceWI-multiple} with $\ell=1$ and $n=4$ and the explicit expression for the 
three point function \eqref{3Tcomplex} gives
\begin{align}
 &\big\lag T(y)T_{z_1z_1}T_{z_2z_2}T_{z_3z_3}\big\rag=-2\sum_{i=1}^3\d^{(2)}(y-z_i)\big\lag T_{z_1 z_1}T_{z_2z_2}T_{z_3z_3}\big\rag\nn
 &=\frac{2N}{3(2\pi)^3}\Big[\d^{(2)}(y-z_1) + \d^{(2)}(y-z_2) + \d^{(2)}(y-z_3)\Big] \nn
 &\quad\times\left[\frac1{(z_1-z_2)^3(z_2-z_3)^3}+\frac1{(z_2-z_1)^3(z_1-z_3)^3}+\frac1{(z_2-z_3)^3(z_3-z_1)^3}\right]
 \label{4T1trace}
\end{align}
totaling nine terms in all. Six of these terms, namely those with $\d^{(2)}(y-z_i)$ multiplying the trees with $z_i$ at one
of the two endpoints, correspond to correlators with the trace $T(y)$ coming from the linear term trace
contribution (\ref{TzzbarPhi}) at the endpoints canceling the connecting $G_{\bPhi}$ propagator as before. 
For example for the first of the nine terms in (\ref{4T1trace})
\begin{align}
 &-\frac{N}{3\pi}\big\lag \pa_{y}\pa_{\bar y}\bPhi(y)\, :\!\pa_{z_1}\bPhi(z_1)\pa_{z_1}\bPhi(z_1)\!:\ 
 :\!\pa_{z_2}\bPhi(z_2)\pa_{z_2}\bPhi(z_2)\!:\ \pa_{z_3}^2\bPhi(z_3)\big\rag \nn
 &=\frac{N}{3\pi(4\pi)^2}\ \pa_{z_1}\d^{(2)}(y-z_1)\, \pa_{z_1}\pa_{z_2}\ln(z_1-z_2)\, \pa_{z_2}\pa_{z_3}^2\ln(z_2-z_3)\nn
 &=\frac{2N}{3(2\pi)^3}\, \d^{(2)}(y-z_1)\, \frac{1}{(z_1-z_2)^3(z_2-z_3)^3}\,.
\end{align}
The remaining three of nine terms in (\ref{4T1trace}) with $\d^{(2)}(y-z_i)$ multiplying a contribution with that $z_i$ at the
center point of the tree, \textit{viz.}
\begin{align}
 &\frac{2N}{3(2\pi)^3}\bigg[\d^{(2)}(y-z_2)\frac1{(z_1-z_2)^3(z_2-z_3)^3}+\d^{(2)}(y-z_1)\frac1{(z_1-z_2)^3(z_3-z_1)^3}\nn
 &\hspace{1.4cm}+\d^{(2)}(y-z_3)\frac1{(z_2-z_3)^3(z_3-z_1)^3}\bigg]
\end{align}
are reproduced by the explicit variation of that $z_i$ vertex by $g_{\a\b}(y)$ \eqref{varTtrace}. For example
the first of these three terms comes from varying the $z_2$ vertex by $g_{\a\b}(y)$ which gives
\begin{align}
 &\Big\lag\frac{N}{12\pi}\pa_{z_1}^2\bPhi(z_1)\left[-2\pa_{z_2}\bPhi(z_2)\pa_{z_2}\bPhi(z_2)\d^{(2)}(y-z_2)\right]\pa_{z_3}^2\bPhi(z_3)\Big\rag\nn
 &=-\frac{4N}{12\pi(4\pi)^2}\, \d^{(2)}(y-z_2)\, \left[\pa_{z_1}^2\pa_{z_2}\ln(z_1-z_2)\right]\left[\pa_{z_2}\pa_{z_3}^2\ln(z_2-z_3)\right]\nn
 &=\frac{2N}{3(2\pi)^3}\d^{(2)}(y-z_2)\frac{1}{(z_1-z_2)^3(z_2-z_3)^3}
\end{align}
with the correct factors.

In a similar manner multiple trace insertions in either the loop diagram or at intermediate points of the tree
diagrams require additional variations of $T_{\m\n}$ with respect to the metric, which generates products of
$\d$-functions at coincident points, according to the general trace Ward identities (\ref{traceWI-multiple})
and (\ref{traceand1}). Generalizing both the loop and tree diagrams in this way to allow for multiple
variations and contact terms allows us to extend the identity \eqref{Gamlooptree} to the case of mixed 
$T_{z\bar z}$ components, respectively with arbitrary numbers of trace insertions. This completes
the proof of the equivalence of one-loop correlators of arbitrary numbers of stress-energy tensor insertions
of any kind with the linear tree diagrams generated by the effective action \eqref{SanombPhi} of the 
scalar boson field $\bPhi$.

\section{Summary and Conclusions}
\label{Sec:Conclusion}

Our main purpose in this paper has been to demonstrate that fermion pairing 
into a bosonic degree of freedom in relativistic quantum field theory is a general
phenomenon associated with an anomaly. The remarkable features of the Schwinger model 
are a consequence of the chiral anomaly of the fermions, which leads directly to the existence 
of a chiral boson composed of fermion pairs. We showed that with some interesting
differences most of this analysis may be extended to the conformal trace anomaly of the 
stress-energy tensor in $d=2$ spacetime dimensions. In particular, the physical 
Fock space representation of the scalar boson associated with the stress-energy tensor conformal 
anomaly as a certain correlated fermion pair related to the Virasoro generators through (\ref{aLrel})
defines a `bosonization' distinct from that in the Schwinger model and coupling to the electric field.

The quickest path to the boson description is via the functional integral method and the effective actions 
obtained by integration of the anomaly by (\ref{varGam})-(\ref{SeffA}) in the case of the chiral anomaly,
and (\ref{sigvar})-({\ref{Sanom}) in the case of the conformal anomaly. In both cases this leads immediately 
to a gauge (or coordinate) invariant non-local action in terms of field strengths with a propagator $\sq^{-1}$ 
that signals propagation of a massless boson field not present in the classical theory. In both cases
this propagating boson field may be made explicit in the local form of the quadratic effective action (\ref{Seffchi})
or (\ref{Sanomloc}) with a kinetic term. By keeping track of degrees of freedom through the functional determinants 
it is easy to see that the propagating boson is one collective degree of freedom of the original $N$ fermion
theory in the singlet channel, leaving $N-1$ fermion states, so that no net degrees of freedom are gained or
lost by the fermions pairing into an effective boson.

A second clear signature of the fermion pairing phenomenon may be seen in the appearance of a massless
boson $1/k^2$ pole in correlation functions, such as the two-point current polarization tensor (\ref{Pichi}) or the
stress-energy polarization tensor (\ref{VacPolF0}). In each case the existence and residue of this pole
is determined by the anomaly. In each case, deforming the system away from exactly zero mass fermions
shows that corresponding spectral functions obey a UV finite sum rule, which collapses to a $\d (k^2)$
in the massless limit. These behaviors of correlation and spectral functions are closely related to the 
anomalous commutator or Schwinger terms in the current algebras (\ref{Schwcomm}) and (\ref{SchwT}),
which in turn are consequences of proper definition of the fermion vacuum with a filled Dirac sea and
normal ordering definition of the currents. In the Fock space operator description the Schwinger
anomalous commutator terms become nothing else than the canonical commutators of the boson
fields (\ref{andef})-(\ref{phiposn}) and (\ref{Phican})-(\ref{aLrel}) composed of fermion pairs
verifying again that the boson is a \textit{bona fide} propagating quantum field in its own right.

In both the chiral and conformal bosonization schemes there is an interesting connection to the topology
of the field configuration space. The zero mode in the chiral bosonization scheme describe winding
modes in the field configuration space related to the Chern-Simons number, the Wilson-Aharonov-Bohm 
phase and the breaking of chiral symmetry in the Schwinger model. The zero mode in the conformal 
bosonization scheme and analogous Chern-Simons charge derived from the Euler characteristic
describes the conformal mapping of ${\mathbb R}^2$ to the cylinder ${\mathbb R} \times {\mathbb S}^1$
and the background conformal charge (\ref{topcharge}) of the dilation current. The Casimir energy
of the fermions on the cylinder with anti-periodic boundary conditions on the spatial interval $[0,L]$
is thereby related to topology and a condensate (\ref{phibar}) in the bosonic description.

The appearance of the fermion pairing may be seen explicitly in the current and stress-energy
correlation functions in that the sum over arbitrary two-fermion intermediate states in each
coincides with the sum over scalar boson intermediate states which are a specific coherent 
superposition of fermion pairs. This is verified for the two-point correlation functions
in Secs. \ref{sec:JJ} and \ref{sec:IntermedTT}. For the currents the effective action (\ref{SeffA})
is quadratic in the gauge potentials and there are no further connected correlators to
be considered. For the conformal anomaly the effective action (\ref{Sanom}) contains all 
higher metric variations, so that connected correlators of arbitrary numbers of stress-energy
tensors are non-vanishing. 

In \secref{sec:correlation-functions} the equivalence between fermion pairs and bosons
has been extended to the one-loop correlation functions of arbitrary numbers of fermion 
stress-energy tensors, which are mapped precisely to the set of linear tree diagrams, 
generated by the boson effective action (\ref{SanombPhi}): \textit{cf.}\ \eqnref{Gamlooptree}
and \figref{fig:nT-illustr}. This proof involves some non-trivial combinatoric identities among 
polynomials, which are made transparent only by use of the general $n$-point Ward identities, 
including the general anomalous trace Ward identities in (\ref{WardIds}). Since all the propagator 
lines in the tree diagrams are precisely those of the same boson field $\bPhi$ in  (\ref{SanombPhi}),
the equivalence (\ref{Gamlooptree}) of the fermion loop and boson tree diagrams shows that all 
intermediate states of these arbitrary higher order $n$-point stress-energy tensor amplitudes
involve exactly the same correlated fermion pair states of the boson field as that of the basic
two-point $\lag T^{\m\n}T^{\a\b}\rag$ amplitude.

The close association between the phenomenon of fermion pairing into bosons and quantum
anomalies has been studied here exclusively in $d=2$ dimensions, in order to keep matters
as simple as possible and all formulae explicit. However, it should be clear that many of
the same features of this close association carry over to higher even dimensions, with
appropriate modifications. In particular the bosonic effective action for both the conformal
and chiral anomalies in $d=4$ have been discussed in \cite{Mottola:2006ew} and
\cite{Giannotti:2008cv}, with the massless $1/k^2$ pole and UV finite sum rule
explicitly exhibited in the latter case. The principal differences in higher dimensions
are that the anomalous amplitudes appear first in higher $n$-point correlation
functions than $n=2$ in $d=2$, and the effects of the anomaly and pairing
phenomenon are present only in special tensor combinations, in particular
kinematic limits, the other tensor combinations being non-anomalous and 
determined by other model dependent considerations.

\subsection*{Acknowledgements}
\vspace{-1mm}
D. N. Blaschke is a recipient of an APART fellowship of the Austrian Academy of Sciences,
and is also grateful for the hospitality of the theory division of LANL and its partial financial support.
R. Carballo-Rubio acknowledges support from CSIC through the JAE-predoc program, cofunded by FSE.
E. M. gratefully acknowledges helpful discussions with Federico R. Urban at the early stages of this work.

\appendix
\Appendix{Commutator Algebra of Fermion Charge Density}
\label{sec:commutators}
In this appendix, we compute explicitly the commutator relations for the fermion density operator defined in Eq.\ (\ref{rhodef}).
It is clear that the mixed commutator of left and right movers $[\rho^{(\mp)}_n,\rho^{(\pm)}_{n'}]= 0$,
while $[\rho^{(+)}_n,\rho^{(+)}_{n'}] = [\rho^{(-)}_n,\rho^{(-)}_{n'}]$, so that we can drop the chirality index in 
order to compute this commutator. To that end let us first evaluate
\bes
\begin{align}
[c_q,c_{q'}c^{\dagger}_{q'-n}]&=c_qc_{q'}c^\dagger_{q'-n}-c_{q'}c^{\dagger}_{q'-n}c_q=-c_{q'}\,\delta_{q,q'-n}\\
[c^\dagger_q,c_{q'}c^{\dagger}_{q'-n}]&=c^\dagger_qc_{q'}c^{\dagger}_{q'-n}-c_{q'}c^{\dagger}_{q'-n}c^\dagger_q
=  \delta_{q,q'}\,c^\dagger_{q'-n}
\end{align}
\ees
so that
\bes
\be
[c_q,\r_n]=c_{q+n}\,, \qquad\qquad
[c_q^\dag,\r_n]=-c^\dagger_{q-n}\,.
\ee
\ees
Therefore for $n> 0, n' <0$
\begin{align}
[\r_n, \r_{n'}] &= \bigg(\sum_{q\geq -n' + \inv2} c^\dag_{q-n} c_{q+n'}  
- \sum_{q\geq n + \inv2} c^\dag_{q-n- n'} c_q
+ \sum_{\inv2 \leq q\leq n - \inv2} c_qc^\dag_{q-n - n'} \bigg.\nn
&\quad\qquad\bigg.- \sum_{\inv2 \leq q\leq -n' - \inv2} c_{q+n'}c^\dag_{q-n}
+ \sum_{q\leq - \inv2} c_qc^\dag_{q-n - n'} - \sum_{q\leq - \inv2} c_{q+n'}c^\dag_{q-n}\bigg)\nn
& =   \sum_{\inv2 \leq q < n + \inv2} \left(c^\dag_{q - n - n'} c_q + c_qc^\dag_{q - n - n'}\right)
=  \sum_{\inv2 \leq q < n + \inv2} \d_{n,-n'}\  =\  n\, \d_{n,-n'}
\label{rhonm}
\end{align}
which is also valid for $n >0 , n' >0$, when it vanishes. Using then also (\ref{rhodag})
we find that (\ref{rhonm}) is valid for all $n, n' \neq 0$. 
In order to obtain this result it is crucial to use the correctly fermion normal ordered definition (\ref{rhodef}),
which gives rise to a finite range of $n$ values of $q$ contributing in  (\ref{rhonm}).
 
This non-zero commutator simply expresses
the anomalous commutator of currents (\ref{j0j1comm}), showing it to be an exact result, valid at the
operator level, since
\begin{align}
[j^0(t,x), j^1(t, x')] &= \frac{1}{L^2} \sum_{n\neq 0} n e^{ik_n (x - x')} - \frac{1}{L^2} \sum_{n\neq 0} n e^{-ik_n (x - x')}\nn
&= \frac{1}{\pi L} \sum_{n \in \Z} k_n e^{ik_n (x - x')} = -\frac{i}{\pi}\, \pa_x \delta(x-x')\,.
\end{align}

\Appendix{Virasoro Algebra of Fermion Energy Density}
\label{sec:virasoro}

For completeness we also provide a direct computation of the commutator of two Virasoro generators of the fermions,
highlighting the importance of the definition of the fermion vacuum by the normal ordering prescription.
Suppressing the $\pm$ chirality indices for notational simplicity and using
\begin{align}
 [c_n,\bL_{n'}]&=\left(n+\frac{n'}{2}\right)c_{n+n'} \,, &
 [c_n^\dagger,\bL_{n'}]&=-\left(n-\frac{n'}{2}\right)c^\dagger_{n-n'}
\end{align}
yields
\begin{align}
[\bL_n,\bL_{n'}]&=-\sum_{q=-\infty}^{-\inv2}\left(q-\frac{n}{2}\right)[c_qc^\dagger_{q-n},\bL_{n'}]+\sum_{q=\inv2}^\infty\left(q-\frac{n}{2}\right)[c^\dagger_{q-n}c_q,\bL_{n'}]\nonumber\\
&=-\sum_{q=-\infty}^{-\inv2}\left(q-\frac{n}{2}\right)\left(c_q[c^\dagger_{q-n},\bL_{n'}]+[c_q,\bL_{n'}]c^\dagger_{q-n}\right)\nn
&\quad +\sum_{q=\inv2}^\infty\left(q-\frac{n}{2}\right)\left(c^\dagger_{q-n}[c_q,\bL_{n'}]+[c^\dagger_{q-n},\bL_{n'}]c_q\right)\nonumber\\
&=\sum_{q=-\infty}^{-\inv2}\left(q-\frac{n}{2}\right)\left(q-n-\frac{n'}{2}\right)c_qc^\dagger_{q-n-n'}-\sum_{p=-\infty}^{n'-\inv2}\left(p-n'-\frac{n}{2}\right)\left(p-\frac{n'}{2}\right)c_{p}c^\dagger_{p-n-n'}\nonumber\\
&\quad +\sum_{p=n'+\inv2}^\infty\left(p-n'-\frac{n}{2}\right)\left(p-\frac{n'}{2}\right)c^\dagger_{p-n-n'}c_{p}-\sum_{q=\inv2}^\infty\left(q-\frac{n}{2}\right)\left(q-n-\frac{n'}{2}\right)c^\dagger_{q-n-n'}c_q
\end{align}
where $p=q+n'$.
Using the identity
\begin{align}
 \left(q-n'-\frac n2\right)\left(q-\frac{n'}2\right)-\left(q-n-\frac{n'}2\right)\left(q-\frac n2\right)
 &=(n-n')\left(q-\frac{n'+n}{2}\right)
\end{align}
we find
\begin{align}
[\bL_n,\bL_{n'}]&=-(n-n')\sum_{q=-\infty}^{-\inv2}\left(q-\frac{n+n'}{2}\right)c_qc^\dagger_{q-n-n'} -\sum_{p=\inv2}^{n'-\inv2}\left(p-n'-\frac{n}{2}\right)\left(p-\frac{n'}{2}\right)c_{p}c^\dagger_{p-n-n'} \nonumber\\
&\quad +(n-n')\sum_{q=n'+\inv2}^\infty\left(q-\frac{n+n'}{2}\right)c^\dagger_{q-n-n'}c_{q}
-\sum_{p=\inv2}^{n'-\inv2}\left(p-\frac{n'}{2}\right)\left(p-n'-\frac{n}{2}\right)c^\dagger_{p-n-n'}c_p\nonumber\\
&=(n-n')\bL_{n+n'}-\delta_{n,-n'}\sum_{p=\inv2}^{n'-\inv2}\left(p-n'-\frac{n}{2}\right)\left(p-\frac{n'}{2}\right)
\,.
\end{align}
Finally, one needs to evaluate the sum
\begin{align}
\sum_{p=\inv2}^{n'-\inv2}\left(p-\frac{n'}{2}\right)^2=\frac{1}{12}n'\left(n'^2-1\right)
\end{align}
to obtain the algebra
\begin{align}
 \co{\bL_n}{\bL_{n'}}&=(n-n')\bL_{n+n'}+\frac{n(n^2-1)}{12}\d_{n,-n'}
 \,.
\end{align}




\end{document}

%% file: Figures/rho-jj.pdf_tex
\begingroup%
  \makeatletter%
  \providecommand\color[2][]{%
    \errmessage{(Inkscape) Color is used for the text in Inkscape, but the package 'color.sty' is not loaded}%
    \renewcommand\color[2][]{}%
  }%
  \providecommand\transparent[1]{%
    \errmessage{(Inkscape) Transparency is used (non-zero) for the text in Inkscape, but the package 'transparent.sty' is not loaded}%
    \renewcommand\transparent[1]{}%
  }%
  \providecommand\rotatebox[2]{#2}%
  \ifx\svgwidth\undefined%
    \setlength{\unitlength}{461.00789063bp}%
    \ifx\svgscale\undefined%
      \relax%
    \else%
      \setlength{\unitlength}{\unitlength * \real{\svgscale}}%
    \fi%
  \else%
    \setlength{\unitlength}{\svgwidth}%
  \fi%
  \global\let\svgwidth\undefined%
  \global\let\svgscale\undefined%
  \makeatother%
  \begin{picture}(1,0.66326617)%
    \put(0,0){\includegraphics[width=\unitlength]{Figures/rho-jj.pdf}}%
    \put(0.09909334,0.07989458){\makebox(0,0)[rb]{\smash{0}}}%
    \put(0.09909334,0.15052244){\makebox(0,0)[rb]{\smash{0.1}}}%
    \put(0.09909334,0.2211503){\makebox(0,0)[rb]{\smash{0.2}}}%
    \put(0.09909334,0.29177815){\makebox(0,0)[rb]{\smash{0.3}}}%
    \put(0.09909334,0.36257955){\makebox(0,0)[rb]{\smash{0.4}}}%
    \put(0.09909334,0.4332074){\makebox(0,0)[rb]{\smash{0.5}}}%
    \put(0.09909334,0.50383526){\makebox(0,0)[rb]{\smash{0.6}}}%
    \put(0.09909334,0.57446312){\makebox(0,0)[rb]{\smash{0.7}}}%
    \put(0.09909334,0.64509098){\makebox(0,0)[rb]{\smash{0.8}}}%
    \put(0.11349657,0.04865867){\makebox(0,0)[b]{\smash{0}}}%
    \put(0.28807058,0.04865867){\makebox(0,0)[b]{\smash{2}}}%
    \put(0.4626446,0.04865867){\makebox(0,0)[b]{\smash{4}}}%
    \put(0.63739215,0.04865867){\makebox(0,0)[b]{\smash{6}}}%
    \put(0.81196617,0.04865867){\makebox(0,0)[b]{\smash{8}}}%
    \put(0.98654019,0.04865867){\makebox(0,0)[b]{\smash{10}}}%
    \put(0.01926824,0.37021499){\rotatebox{90}{\makebox(0,0)[b]{\smash{$m^2 \rho(s)$}}}}%
    \put(0.54993161,0.00180481){\makebox(0,0)[b]{\smash{$s/m^2$}}}%
  \end{picture}%
\endgroup%

%% file: Figures/rho-TT.pdf_tex
\begingroup%
  \makeatletter%
  \providecommand\color[2][]{%
    \errmessage{(Inkscape) Color is used for the text in Inkscape, but the package 'color.sty' is not loaded}%
    \renewcommand\color[2][]{}%
  }%
  \providecommand\transparent[1]{%
    \errmessage{(Inkscape) Transparency is used (non-zero) for the text in Inkscape, but the package 'transparent.sty' is not loaded}%
    \renewcommand\transparent[1]{}%
  }%
  \providecommand\rotatebox[2]{#2}%
  \ifx\svgwidth\undefined%
    \setlength{\unitlength}{461.00789063bp}%
    \ifx\svgscale\undefined%
      \relax%
    \else%
      \setlength{\unitlength}{\unitlength * \real{\svgscale}}%
    \fi%
  \else%
    \setlength{\unitlength}{\svgwidth}%
  \fi%
  \global\let\svgwidth\undefined%
  \global\let\svgscale\undefined%
  \makeatother%
  \begin{picture}(1,0.66326617)%
    \put(0,0){\includegraphics[width=\unitlength]{Figures/rho-TT.pdf}}%
    \put(0.14230301,0.07989458){\makebox(0,0)[rb]{\smash{0}}}%
    \put(0.14230301,0.1741229){\makebox(0,0)[rb]{\smash{0.0005}}}%
    \put(0.14230301,0.26835122){\makebox(0,0)[rb]{\smash{0.001}}}%
    \put(0.14230301,0.36257955){\makebox(0,0)[rb]{\smash{0.0015}}}%
    \put(0.14230301,0.45663433){\makebox(0,0)[rb]{\smash{0.002}}}%
    \put(0.14230301,0.55086266){\makebox(0,0)[rb]{\smash{0.0025}}}%
    \put(0.14230301,0.64509098){\makebox(0,0)[rb]{\smash{0.003}}}%
    \put(0.15670624,0.04865867){\makebox(0,0)[b]{\smash{0}}}%
    \put(0.32260362,0.04865867){\makebox(0,0)[b]{\smash{10}}}%
    \put(0.48867453,0.04865867){\makebox(0,0)[b]{\smash{20}}}%
    \put(0.6545719,0.04865867){\makebox(0,0)[b]{\smash{30}}}%
    \put(0.82064281,0.04865867){\makebox(0,0)[b]{\smash{40}}}%
    \put(0.98654019,0.04865867){\makebox(0,0)[b]{\smash{50}}}%
    \put(0.01926824,0.37021499){\rotatebox{90}{\makebox(0,0)[b]{\smash{$m^2 \rho(s)$}}}}%
    \put(0.57162321,0.00180481){\makebox(0,0)[b]{\smash{$s/m^2$}}}%
  \end{picture}%
\endgroup%